\documentclass[aps,pre,groupedaddress]{revtex4-2}

\usepackage{fullpage}
\usepackage{todonotes,color}
\usepackage[centertags]{amsmath}
\usepackage{amscd,amsfonts,amssymb,latexsym}
\usepackage{graphicx}
\usepackage{theorem}
\usepackage{enumerate}
\usepackage[toc,page]{appendix}
\usepackage{float}
\usepackage{subcaption,overpic}
\usepackage{todonotes,color}

\newcommand{\bn}{\mathbf{n}}
\newcommand{\Ca}{a}

\newcommand{\eq}[2]{\del^2 #1 + V_0^2(#2 + 2 #1) - #1}
\newcommand{\eqr}[1]{\del^2 #1 + 3 V_0^2 #1 - #1}

\newcommand{\tst}{\mbox{T.S.T.}}
\newcommand{\RHS}{\rho}
\newcommand{\Vexp}{V_{\mathrm{exp}}}
\newcommand{\bxi}{{\boldsymbol{\xi}}}
\newcommand{\bX}{{\bf X}}
\newcommand{\bD}{{\boldsymbol{\Delta}}}
\newcommand{\brho}{{\boldsymbol{\rho}}}

\newcommand{\Uexp}{U_{\mathrm{exp}}}
\newcommand{\Wexp}{W_{\mathrm{exp}}}

\newcommand{\del}{\nabla}

\newcommand{\dd}{\partial}

\newcommand{\la}{\lambda}
\newcommand{\non}{\nonumber}
\newcommand{\eps}{\epsilon}
\newcommand{\beqa}{\begin{eqnarray}}
\newcommand{\eeqa}{\end{eqnarray}}
\newcommand{\beqas}{\begin{eqnarray*}}
\newcommand{\eeqas}{\end{eqnarray*}}
\newcommand{\ba}{\begin{align}}
\newcommand{\ea}{\end{align}}
\newcommand{\bas}{\begin{align*}}
\newcommand{\eas}{\end{align*}}
\newcommand{\beq}{\begin{equation}}
\newcommand{\eeq}{\end{equation}}
\newcommand{\re}{\mathrm{Re}}
\newcommand{\im}{\mathrm{Im}}
\newcommand{\ra}{\rightarrow}
\newcommand{\al}{\alpha}

\newcommand{\R}{{\mathbb R}}

\newcommand{\ee}{\mathrm{e}}
\newcommand{\ii}{\mathrm{i}}
\newcommand{\pdhfrac}[2]{\mathchoice{\frac{#1}{#2}}{#1/#2}{#1/#2}{#1/#2}}
\newcommand{\fdd}[2]{\pdhfrac{\mathrm{d}#1}{\mathrm{d}#2}}
\newcommand{\sdd}[2]{\pdhfrac{\mathrm{d}^2#1}{\mathrm{d}#2^2}}
\newcommand{\pd}[2]{\pdhfrac{{\partial}#1}{{\partial}#2}}

\renewcommand{\d}[1]{\mathrm{d}#1}

\newcommand{\Ve}{V_G}

\newcommand{\Vs}{V_{s}}

\newcommand{\Vn}{V}
\newcommand{\V}{V}

\newcommand{\A}{A}

\newcommand{\bx}{{\bf x}}

\graphicspath{{./}{./Figures/}}
\makeatletter
\def\input@path{{./}{./Figures/}}
\makeatother

\begin{document}

\title{Blowup of Multi-Peaked Waveforms in the Two-Dimensional Nonlinear
Schr{\"o}dinger Model}

\author{S. Jon Chapman}
\affiliation{Mathematical Institute, University of Oxford, AWB, ROQ, Woodstock Road, Oxford OX2 6GG}

\author{M. Kavousanakis}
\affiliation{School of Chemical Engineering, National Technical University of Athens, 15780, Athens, Greece}

\author{E.G. Charalampidis}
\affiliation{Department of Mathematics and Statistics, and Computational Science Research Center, San Diego State University, San Diego, CA 92182-7720, USA}

 \author{I.G. Kevrekidis}
\affiliation{Department of Chemical and Biomolecular Engineering \& \\
Department of Applied Mathematics and Statistics, Johns Hopkins University, Baltimore, MD 21218, USA}

\author{P.G. Kevrekidis}
\affiliation{Department of Mathematics and Statistics, University of
  Massachusetts, Amherst MA 01003-4515, USA}

\affiliation{Department of Physics, University of Massachusetts Amherst, Amherst, MA 01003, USA}

\affiliation{Department of Mechanical Engineering, Seoul National University, 1 Gwanak-ro, Gwanak-gu, Seoul 08826, South Korea}

\date{\today}

\begin{abstract}
In the present work, we explore the self-focusing and resulting
collapse of two-dimensional waveforms involving multiple pulses
in a nonlinear Schr{\"o}dinger equation with a general power-law
nonlinearity. We find that a wide range of multi-peaked states 
bifurcate from the critical threshold of the cubic nonlinearity, 
thus representing ``bifurcations from 
infinity'', i.e., the relevant pulses start at infinite distance
in the critical limit and draw nearer, as the nonlinearity
exponent increases past that threshold. We identify the resulting ``interacting 
particle system'' as amounting to a force balance between 
the exponentially interacting tails (modulated by a suitable
power law) and a linear phase-induced force. The equilibria
emerging from this force balance are found to be in
{\it excellent} agreement with the identified steady states
of the 
partial differential equation.
~The spectral stability of multi-peaked configurations is analyzed, leading to the conclusion that
all the relevant states are less stable than the single-peak
collapsing solution whose stability was analyzed earlier.~Indeed, we reveal both symmetry-breaking, as well as 
motion-inducing destabilizing dynamics, with the former ones
among them being dominant and ultimately leading to a
single dominant collapse spot.~Moreover, we characterize systematically
both the real and imaginary 
eigenvalues of multi-peaked configurations, partitioning
them in groups of different sizes, 
described by powers
of the solution's blowup rate $G$.
\end{abstract}

\maketitle

\section{Introduction}

The nonlinear Schr{\"o}dinger (NLS) equation holds a prominent role in the theory of nonlinear dispersive partial differential equations, both as a fundamental mathematical model and as an effective envelope description of slowly varying wave packets in nonlinear media~\cite{ablowitz1981solitons,ablowitz2,sulem,AblowitzPrinariTrubatch}. Its physical relevance spans numerous areas, including nonlinear optics where it governs the propagation of optical pulses and beams in fibers and photonic media~\cite{hasegawa:sio95,Kivshar2003}, ultracold atomic gases where it arises as the Gross--Pitaevskii equation describing Bose--Einstein condensates~\cite{Pitaevskii2003,Pethick2008,siambook}, as well as plasma physics and water-wave dynamics~\cite{kono,ablowitz2}. Because of this universality, the NLS equation has become a paradigmatic framework for studying nonlinear wave propagation, coherent structures, and singularity formation under suitable conditions.

A significant body of work on the NLS model has focused on the existence and dynamics of solitary wave solutions. These include bright solitons in focusing media~\cite{Kivshar2003}, dark solitons arising in defocusing settings~\cite{Frantzeskakis_2010}, as well
as on higher-dimensional structures such as topological/vortical
ones~\cite{siambook}. Also of wide interest is the phenomenon of finite-time singularity formation or {\em collapse}. In focusing NLS systems, a sufficiently strong  self-focusing nonlinearity can overcome dispersive spreading, leading to the finite-time blowup of the solution amplitude. In this regime the (formerly ground-state) solitary wave becomes orbitally unstable and the dynamics may evolve toward singular solutions~\cite{sulem,fibich2015}. The mathematical structure and physical implications of collapse in NLS-type models have been extensively investigated over the past several decades~\cite{zakharov1972collapse,weinstein1983sharp,lemesurier,fibich,sulem,fibich2015}.

From a modeling perspective, collapse may arise through several mechanisms that enhance the nonlinear interaction. In equations with power-law nonlinearities of the form $|u|^{2\sigma}u$ where $\sigma$ represents their strength, 
increasing 
its value can lead to regimes in which self-similar collapsing solutions appear, a feature that has led to related bifurcation studies~\cite{siettos,jon1,jon2}. Alternatively, collapse can be triggered by increasing the spatial dimensionality while keeping the nonlinear exponent fixed.~Indicatively, 
the Laplacian for radially symmetric configurations assumes the form $\Delta u = u_{rr} + (d-1)u_r/r$, and sufficiently large dimension $d$ promotes focusing dynamics and singularity formation~\cite{sulem,lemesurier,fibich2015,budd:1999,eva}, a theme
that was extensively explored in earlier studies. Analytical and numerical works have revealed a rich structure of blowup dynamics in such regimes, including self-similar solutions and universal blowup profiles near criticality~\cite{merle1993blowup,merle2005blowup}. Recently, the study
of the relevant collapse and the proof of existence of associated solutions
has been the source of a novel vein of interest, through the use of
``rigorous numerics'' and the realm of computer-assisted proofs,
certifying the existence of single-~\cite{donninger,dahne2024selfsimilarsingularsolutionsnonlinear}
and multi-bump~\cite{dahne2024selfsimilarsingularsolutionsnonlinear}
collapse waveforms.

Experimental investigations have provided valuable insights into these phenomena as well. Early nonlinear optics experiments demonstrated the formation of the celebrated Townes soliton and its associated collapse dynamics~\cite{moll}. Subsequent work examined collapse processes involving beams carrying topological charge~\cite{gaeta2}. In recent years, advances in ultracold atomic gases have enabled complementary explorations in Bose--Einstein condensates. In particular, two independent experiments have reported realizations of Townes solitons in quasi-two-dimensional condensate systems~\cite{dalibard,chenlung}. Additional directions have also emerged, including the collapse of co-propagating beams with different wavelengths in two-color optical systems~\cite{twocolor}, as well as collapse phenomena induced by rapidly quenching atomic interactions from repulsive to attractive in vortical condensate states~\cite{banerjee2024collapse}.

The latter experiment~\cite{banerjee2024collapse} provides a particularly intriguing scenario motivating the present work. There, a vortex condensate initially forms a ring-shaped density distribution which subsequently breaks into a necklace-like configuration of localized peaks. In this setting,
the phase singularity prevents collapse at the center of the condensate.
Accordingly, the fragmentation of the ring produces several localized density maxima that collapse along the periphery. Similar mechanisms have previously been discussed in the context of azimuthal modulational instabilities of vortex solitons~\cite{CAPLAN20091399}. Such instabilities naturally generate multiple localized structures that may undergo simultaneous collapse, a situation that has been theoretically~\cite{Nawa1998}, as well
as numerically~\cite{REN2000246} investigated in earlier studies. From a theoretical standpoint, a particularly simple setting in which to analyze the interaction and collapse of multiple localized peaks is the one-dimensional case ($d=1$). Related configurations have been explored in earlier work by treating the spatial dimension $d$ as a bifurcation parameter~\cite{budd:1999}, an approach that provides useful mathematical insight but is less directly connected to physical realizations.
It is worthwhile to note here that there exists a complementary thread
of activity that has examined the self-similar dynamics of
singular ring-like profiles (as well as of multi-ring generalizations; see 
also~\cite{dahne2024selfsimilarsingularsolutionsnonlinear})
in critical and supercritical NLS equations~\cite{FIBICH2005193,FIBICH200755},
as well as in a variety of related (e.g., biharmonic, nonlinear heat,
nonlinear biharmonic heat etc.)~\cite{BARUCH20101968} models.
As we will observe these rings will be subject to the azimuthal
symmetry-breaking instability that was observed in~\cite{banerjee2024collapse}
and as such are intimately connected also with the multi-peaked
solutions presented herein.

Motivated by our earlier investigations~\cite{jon1,jon2}, which considered the mathematically equivalent  but more analytically tractable formulation (in
comparison to the ``varying-dimension'' problem) involving a variable nonlinear exponent $\sigma$, we revisit here the problem of multi-pulse self-similar blowup solutions but in a two-dimensional setting. 
This naturally generalizes and substantially extends ---given
the wealth of additional corresponding 2D solutions--- our one-dimensional
analysis of such multi-peaked states.
In particular, we formulate the computation of these states as a bifurcation problem emanating from the critical condition $\sigma d = 2$ 
at which the NLS equation
is conformally invariant.
~Then, collapsing solutions exhibit a blowup rate $G$ that vanishes in the dynamical evolution at this critical condition leading to the well established log-log blowup law~\cite{sulem,fibich2015}. 
The basic mathematical and numerical framework enabling the computation of these solutions is presented in \ref{sec:theor_numer_setup}.

Fixing the spatial dimension to $d=2$ and varying the nonlinear exponent $\sigma$ away from this critical threshold, we compute the bifurcation diagram
of a wide range of 
multi-peaked collapsing solutions. Our results indicate that all of these solution branches emerge from infinity at the bifurcation point, 
{i.e., the inter-soliton distance tends to infinity, as we
approach the bifurcation point.}
As the relevant nonlinearity exponent
increases, the associated pulses approach each other. 
We reveal an effective ``force balance'' between the
inter-pulse tail-tail interaction force, and an effective onsite force 
{---acting individually on each pulse, rather than depending on
the inter-pulse separation---}
emerging from the (complex) solution nontrivial phase structure.
We find it remarkable that this interplay is strongly reminiscent
of a so-called Toda lattice with a mass, an effective one-dimensional
variant of which appeared, e.g., in a fundamentally different setting 
of dark solitons interacting in an external trap potential in the
work of~\cite{Coles_2010}.~The numerical findings reported herein are explained through an asymptotic 
description for the spacing between pulses as a function of the deviation from criticality. 

Furthermore, extending the stability analysis of the single-pulse collapsing branch developed in~\cite{jon1,jon2}, we conduct a 
spectral study of the multi-pulse branches and determine how the number of unstable eigenmodes depends on the blowup rate $G$.
The complexity of the relevant system precludes us from a full
characterization of the entirety of the eigenvalue spectrum 
(since each of the multiple peaks of the two-dimensional
configuration is associated with 4 pairs of eigenvalues,
leading to the need to account for $4 N$ eigenvalue pairs).
Nevertheless, we provide a full description of simpler configurations,
such as the two-peaked setting, and for {\it all} possible 
states, we analytically characterize their dominant
(scaling with $G^{1/2}$, where $G$ is the blowup rate) eigenvalue pairs
both along the real and along the imaginary axis. 
It is important to also indicate that we find that the dominant
eigenvalue of such configurations is one associated with symmetry breaking,
and eventually favors a single peak to grow in comparison to the
remaining ones and lead to an isolated focusing event.
Our final count
indicates that for a configuration of $N$-peaks, there exist
$2N-2$ both real and imaginary eigenvalues proportional to 
$G^{1/2}$, $2N$ real eigenvalues proportional to $G$, while $2$
more will be at the origin, reflecting the symmetries of the
configuration. 

This paper is organized as follows.~In Sec.~\ref{sec:theor_numer_setup} we present the theoretical formulation of the problem and key analytical findings whereas 
Sec.~\ref{sec:numerics} discusses 
our numerical results.%
~In Sec.~\ref{asymptotics}, we derive the general formula for the equilibrium position of the peaks of an arbitrary $N$-peaked configuration together with its leading-order eigenvalues and normal form by bringing forth exponential asymptotics.~These analytical predictions are further validated in Appendix A through large-scale, state-of-the-art finite-element computations, hence providing an independent numerical confirmation of the bifurcation structure, equilibrium peak locations, and leading-order eigenvalues.%
~Finally, Sec.~\ref{sec:conclusions} summarizes our conclusions and outlines a number of promising directions for future research. 

\section{Theoretical Formulation and Principal Analytical Findings}
\label{sec:theor_numer_setup}

Our starting point is the two-dimensional nonlinear Schr\"odinger (NLS) equation with a power-law nonlinearity given by~\cite{sulem}
\begin{align}
\ii \pd{\psi}{t} + \del^2\psi + |\psi|^{2 \sigma} \psi = 0, \quad \psi(\bx,t)\in\mathbb{C},
\label{eq:NLS_original}
\end{align}
where $\sigma$ is the nonlinearity exponent.~The Hamiltonian of the NLS reads
\[ H = \int_{-\infty}^\infty \left(
\left|\del \psi\right|^2 -\frac{1}{\sigma+1} |\psi|^{2\sigma+2}
\right) \, \d \bx, 
\]
and satisfies
\[ \ii \pd{\psi}{t} = \frac{\delta H}{\delta \psi^*}, \qquad \ii 
 \pd{\psi^*}{t} = -\frac{\delta H}{\delta \psi}.\]

We note that on an infinite domain, we require that $H$ is finite.~To set up the stage for our subsequent analysis, we introduce 
the well-established stretched variables~\cite{sulem,fibich,fibich2015} 
\[  \boldsymbol{\xi} = \frac{\bx}{L}, \quad \tau = \int_0^t
  \frac{\d t'}{L^2(t')}, \quad \psi(\bx,t) = L^{-1/\sigma}
  u(\boldsymbol{\xi},\tau) \]
to obtain
\begin{align}
  \ii \pd{u}{\tau} + \del_{\boldsymbol{\xi}}^2u +  |u|^{2 \sigma} u
  - \ii L L_t \boldsymbol{\xi} \cdot \del_{\boldsymbol{\xi}}u- \frac{\ii L L_t}{\sigma} u= 0,
\label{eq:NLS_coexplod_cart}  
\end{align}
  and
  \[ H = L^{-2/\sigma-2}\int_{-\infty}^\infty \left(
\left|\del_{\boldsymbol{\xi}}u\right|^2 -\frac{1}{\sigma+1} |u|^{2\sigma+2}
\right) \, \d \bx.
\]
On a large finite domain, we impose the boundary conditions
\[ {\bf n}\cdot\del_{\boldsymbol{\xi}}u = 0,
\]
where $\bf n$ is the outward normal vector to the domain. We also
impose the pinning condition 
\[ \im(u(0,\tau)) = 0\]
{(see \S \ref{pinning})}.
We write 
\[ G = -L L_t, \qquad u(\boldsymbol{\xi},\tau) =
  \V(\boldsymbol{\xi},\tau) \ee^{\ii \tau-\ii G(\tau) |\boldsymbol{\xi}|^2/4}\] 
to obtain
\begin{equation}
  \ii \pd{\V}{\tau} + \frac{G' |\boldsymbol{\xi}|^2}{4} \V+
  \del_{\boldsymbol{\xi}}^2\V+ |\V|^{2 \sigma}\V - \V  
- \frac{\ii (\sigma-1) G}{ \sigma} \V
+ \frac{G^2 |\boldsymbol{\xi}|^2 }{4}\V= 0,\label{maineqn}
\end{equation}
with $G' = \fdd{G}{\tau}$, 
boundary conditions
\beq
{\bf n} \cdot  \del_{\boldsymbol{\xi}}V = \frac{\ii G }{2}
{\bf n} \cdot \boldsymbol{\xi} V,\label{bc}
\eeq
and pinning condition
\[\im(\V(0,\tau)) = 0.\] 
As $G \ra 0$ the equilibrium positions of the peaks, $\mathbf{X}_i$, 
will be found to satisfy (see 
\S\ref{asymptotics} for the derivation)  
\begin{equation} \label{eq:peakasymptotics}
    b_0 \mathbf{X}_i = 8 \sqrt{2 \pi} \frac{A_2^2}{G^2} \sum_{j \neq i}^N \mathrm{e}^{-|\mathbf{X}_i-\mathbf{X}_j|} \frac{\mathbf{X}_i-\mathbf{X}_j}{|\mathbf{X}_i-\mathbf{X}_j|^{3/2}}, \quad i=1,\dots N,
\end{equation}
where 
\begin{equation}
    b_0 = \int_{\mathbb{R}^2} V_0^2 \textrm{d}V \approx 11.7, 
\end{equation}
\begin{equation}
    V_0(|\boldsymbol{\xi}|) \sim \frac{A_2 \textrm{e}^{-|\boldsymbol{\xi}|}}{|\boldsymbol{\xi}|^{1/2}} \quad \textrm{as } |\boldsymbol{\xi}| \rightarrow \infty, \quad A_2 \approx  3.518,  
\end{equation}
where $V_0$ is two-dimensional soliton solution
(i.e., the radially symmetric solution with
a single peak at the origin and $G=0$; see \S\ref{asynear}).
In what follows,
these asymptotic predictions will be compared with the numerically
calculated positions of the peaks. 

\section{Numerical Results: Existence and Stability of Multi-peaked States}
\label{sec:numerics}

\subsection{Existence Results}
\label{sec:steady}

\subsubsection{Computation of radially-symmetric multi-peaked self-similar solutions}\label{sec:multiradial}

The radially-symmetric version of Eq.~\eqref{eq:NLS_original}
reads
\begin{equation}
\label{eq:NLSradial}
    \mathrm{i} \frac{\partial \psi}{ \partial t}+\frac{\partial^2 \psi}{\partial r^2}+\frac{1}{r} \frac{\partial \psi}{\partial r}+|\psi|^{2\sigma} \psi=0, \quad \psi=\psi(r,t)\in\mathbb{C},
\end{equation}
%
and self-similar blowup solutions are sought by using the stretched variables mentioned before.~Upon using the latter, Eq.~\eqref{eq:NLSradial} is written as
%

%
%

\begin{equation}
\label{eq:nlscoexp}
     \mathrm{i} \frac{\partial v}{\partial \tau}+\frac{\partial^2 v}{\partial \rho^2}+\frac{1} {\rho} \frac{\partial v}{\partial \rho}+|v|^{2 \sigma}v - v + \mathrm{i} G(\tau) \left( \rho \frac{\partial v}{\partial \rho} + \frac{v}{\sigma} \right)  =0,
\end{equation}
%
where $G(\tau):=-L L_t$
represents the blowup rate as before. 

When $\sigma>1$, i.e., supercritical case for the 2D NLS, self-similar solutions to Eq.~\eqref{eq:nlscoexp} 
exist as stationary solutions, i.e.,
$v(\rho,\tau) \mapsto u(\rho)$ as $\tau\mapsto \infty$.
%
We compute such solutions numerically by posing Eq.(\ref{eq:nlscoexp}) on a finite computational domain, $ \rho \in [0, K]$ with homogeneous Neumann boundary conditions, $\frac{\partial v}{\partial \rho} \Big{|}_{\rho=0,K}=0$.
The spatial domain is discretizated uniformly with resolution $\delta \rho$, and spatial derivatives with respect to $\rho$ are approximated using a centered, fourth-order accurate finite difference scheme. 
The identification of stationary solutions proceeds in two stages. 
First, we perform time integration using a backward Euler scheme so that the self-similar dynamics relax toward a stationary state, 
$v(\rho,\tau) \mapsto v(\rho) $.
Because the blowup rate $G(\tau)$ is an additional unknown, the system must be closed by selecting a unique solution from the one-parameter family of degenerate solutions.
\label{pinning}
This is achieved by imposing a template (or pinning) condition of the form: $\mathrm{Im} \left[ v(\rho=0,\tau)\right]=C \equiv \mathrm{const.}$, which originates from the phase condition: 

\begin{equation}
    \int_{0}^K \mathrm{Im} \left[ v(\rho,\tau)\right] T(\rho) \mathrm{d}\rho=C,
\end{equation}

\noindent where $T(\rho)$ is a template function \cite{rowley_2003}.
By choosing $T(\rho)=\delta(\rho)$, we obtain the pointwise (pinning) condition discussed above.
In our computations we set $C=0$, thereby seeking solutions satisfying $\mathrm{Im} \left[ v(\rho=0,\tau)\right]=0$.
Alternative phase conditions yield identical results for the amplitude $|v|$ and the blowup rate $G$, although the real and imaginary parts of $v$ may differ. 
Once the dynamics approaches a stationary profile $v(\rho)$ and a constant blowup rate $G$, the exact solution is obtained using Newton's method. 
The waveform produced by the time integration serves as the initial guess for the Newton solver, which typically converges within 2-3 iterations with high accuracy. 
Finally, the resulting self-similar solution allows us to perform spectral stability analysis in the co-exploding frame. 
For the stability computations presented below, we use MATLAB's $\mathrm{eigs}$ routine for sparse eigenvalue problems.

In this work, however, our primary focus is on multi-peaked self-similar solutions of the NLS for $\sigma>1$.
These solutions are dynamically unstable, and therefore the two-stage procedure described above fails to converge to them;
specifically, the self-similar dynamics do not converge to unstable solutions. 
To compute these solutions, we adopt an approach similar to that of \cite{FIBICH200755} and to the method used in our previous work on the one-dimensional NLS \cite{multi1dpaper}.
The stationary problem that we solve is:

\begin{equation}
    \label{eq:formulti}
    \frac{\mathrm{d}^2v_+}{\mathrm{d} \rho^2} + \frac{1}{\rho} \frac{\mathrm{d}v_+}{\mathrm{d}\rho} -v_+ + \mathrm{i} G_+ \left( \frac{1}{\sigma} v_+ + \rho \frac{\mathrm{d}v_+}{\mathrm{d} \rho}\right) + |v_+|^{2\sigma}v_+=0,
\end{equation}

\noindent subject to boundary conditions:

\begin{equation}
    \label{eq:multibc}
    \frac{\mathrm{d}v_+}{ \mathrm{d}\rho} (0)=0, \quad \im[v_+(0)]=0, \quad \left| K \frac{\mathrm{d}v_+}{\mathrm{d}\rho}(K) + \left( 1+ \mathrm{i}/G_+\right)v_+(K) \right|=0.
\end{equation}

\begin{figure}
    \centering
    \includegraphics[width=0.5\linewidth]{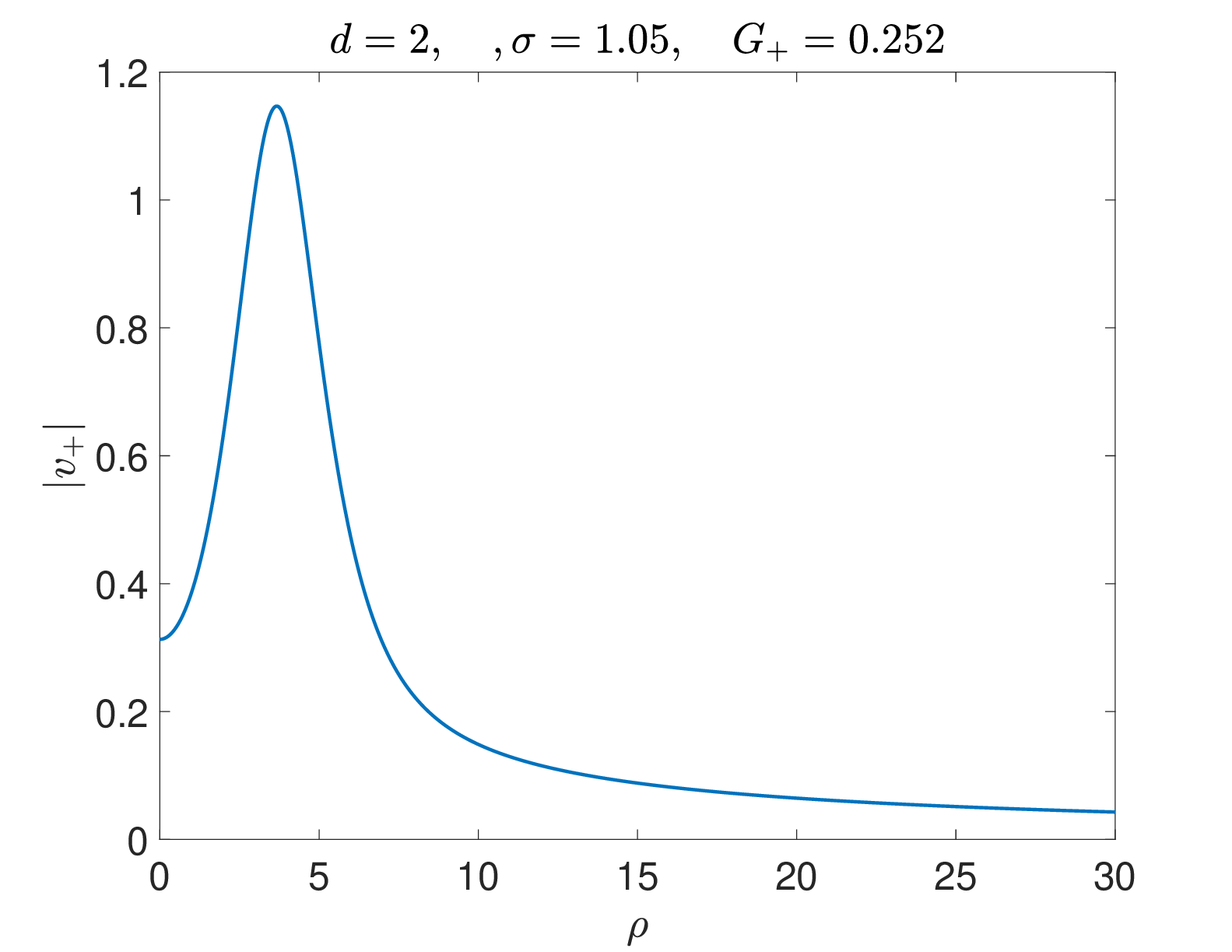}
    \caption{Amplitude $|v_+|$ of the ring radially symmetric solution to the boundary value problem of Eq.~(\ref{eq:formulti})-(\ref{eq:multibc}) for $\sigma=1.05$.
    The hump position is located at  $\rho \approx 3.68 $.
    }
    \label{fig:two_humped_example}
\end{figure}

The BVP above is solved using a shooting method combined with a gradient-free minimization procedure to determine the unknown parameters $\re{[v_+(0)]}$ and $G_+$.
Starting from an initial guess $\re{[v_+(0)]^{(0)}}, G_+^{(0)}$, the latter is treated as an initial condition for Eq.~(\ref{eq:formulti}).
The equation is then integrated forward in $\rho$ up to $\rho=K$, where $K$ typically lies between 200 to 1000.
The boundary condition at $\rho=K$ in Eq.~(\ref{eq:multibc}) is used to define our objective function that is minimized. 
In our implementation, we employ MATLAB's $\mathrm{fminsearch}$ function together with the $\mathrm{ode23t}$ integrator.
Upon convergence, this procedure yields $\re{[v_+(0)]}, G_+$, and the profile $v_+(\rho)$.
The resulting solution is then used as an initial guess for a Newton solver applied to the steady-state of Eq.~(\ref{eq:nlscoexp}), with the pinning condition $\im[v(0)]=0$.
Figure~\ref{fig:two_humped_example} illustrates an example of a ring solution to the boundary value problem defined by Eqs.~(\ref{eq:formulti})-(\ref{eq:multibc}) computed on the interval $[0,100]$ for $\sigma=1.05$.
One can perform a systematic scan over the initial guesses for $\re{[v_+(0)]}, G_+$, which yields a wide variety of multi-humped solutions. 
The ring solution shown in Fig.~\ref{fig:two_humped_example} is obtained using the initial guess: $\re{[v_+(0)]}=0.3, G_+=0.3$.
The localized bump is located around $\rho \approx 3.68 $ and the corresponding converged parameters are: 
$\re{[v_+(0)]}=0.313$ and $G_+=0.252$.
Upon convergence, the computed values of $v_+$ and $G_+$ are used as input for our Newton solver in order to obtain the steady-state numerical solution of Eq.~(\ref{eq:nlscoexp}), subject to Neumann boundary conditions and to the pinning condition $\im[v(\rho=0)]=0$.
This procedure yields the radially symmetric, ring self-similar solution of the 2D NLS equation, in line with earlier works in this setting, such as, e.g.,
the ones of~\cite{FIBICH200755,fibich2015}.

\subsubsection{Computation of non-radially-symmetric multi-peaked self-similar solutions}\label{sec:nonradial}






To identify non-radially-symmetric self-similar solutions to the NLS of Eq.~\eqref{eq:NLS_original}, we consider Eq.~\eqref{eq:NLS_coexplod_cart} with $\boldsymbol{\xi}=(\xi,\eta)$ that is written as:
\begin{equation}
\label{eq:nlscoexp2d}
     \mathrm{i} \frac{\partial v}{\partial \tau}+\frac{\partial^2 v}{\partial \xi^2}+\frac{\partial^2 v}{\partial \eta^2}+|v|^{2 \sigma}v - v + \mathrm{i} G(\tau) \left( \xi \frac{\partial v}{\partial \xi} + \eta \frac{\partial v}{\partial \eta} + \frac{v}{\sigma} \right)  =0.
\end{equation}

\begin{figure}[pt!]
\centering
\begin{tabular}{cc}
\includegraphics[width=0.49\linewidth]{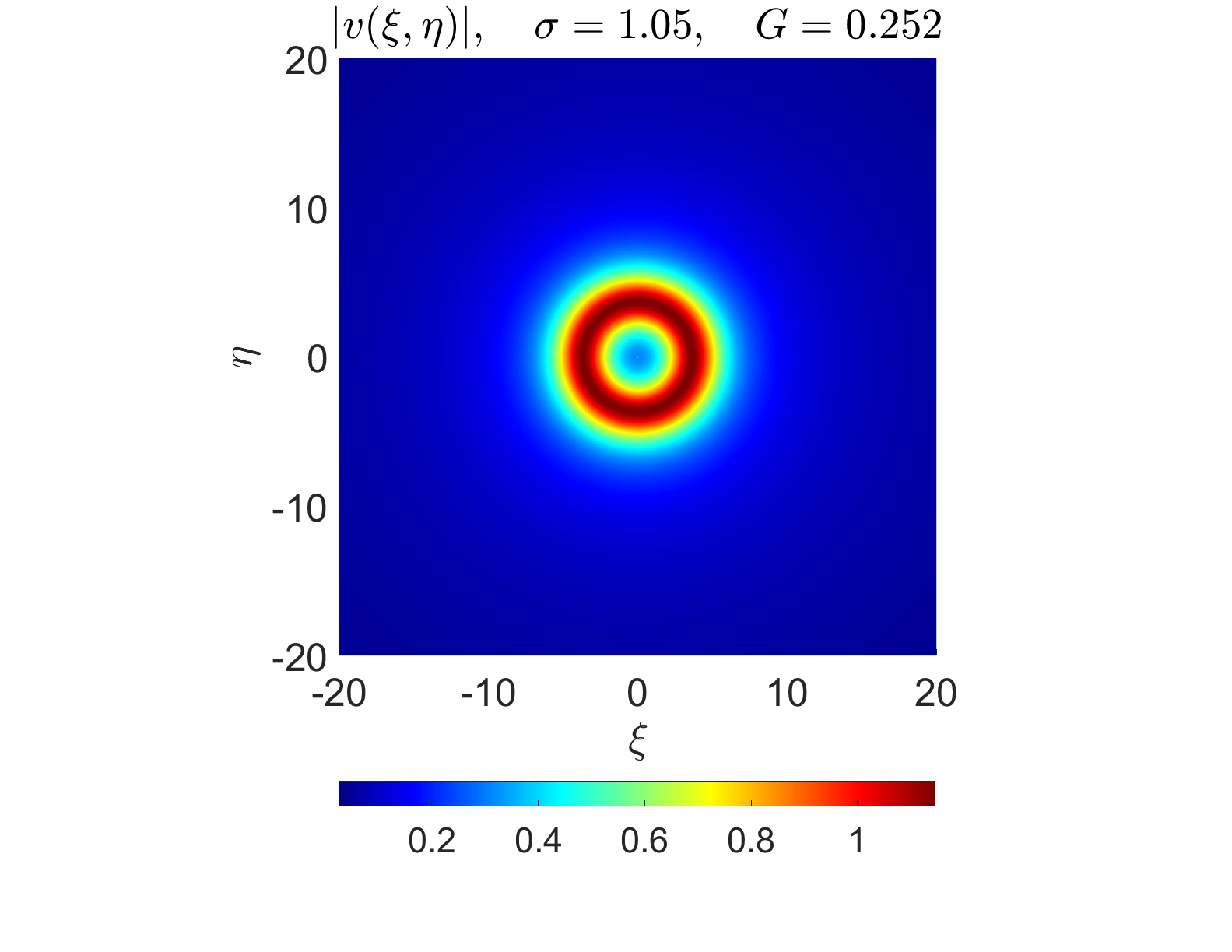}     &  \includegraphics[width=0.49\linewidth]{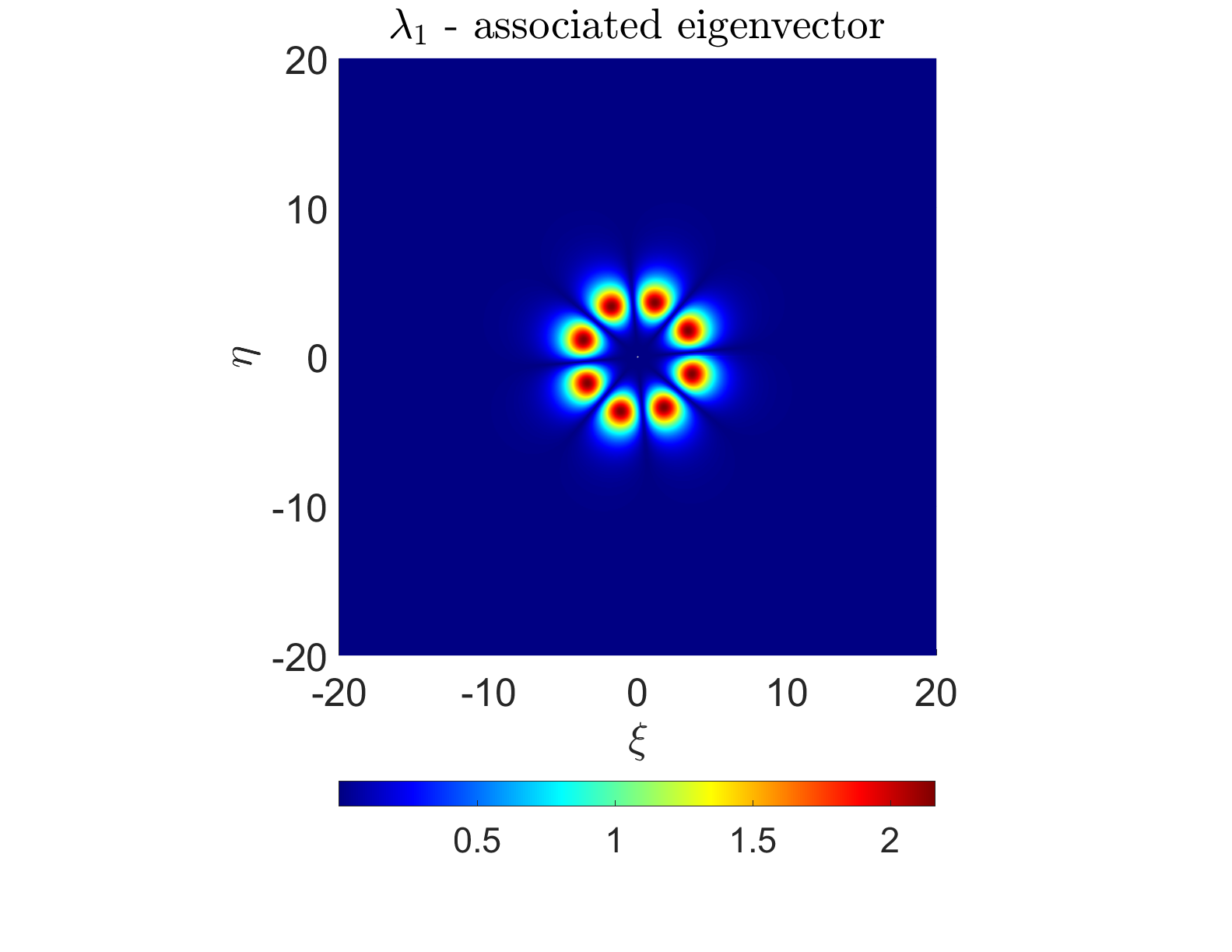}\\
 (a)    & (b) \\
 \includegraphics[width=0.49\linewidth]{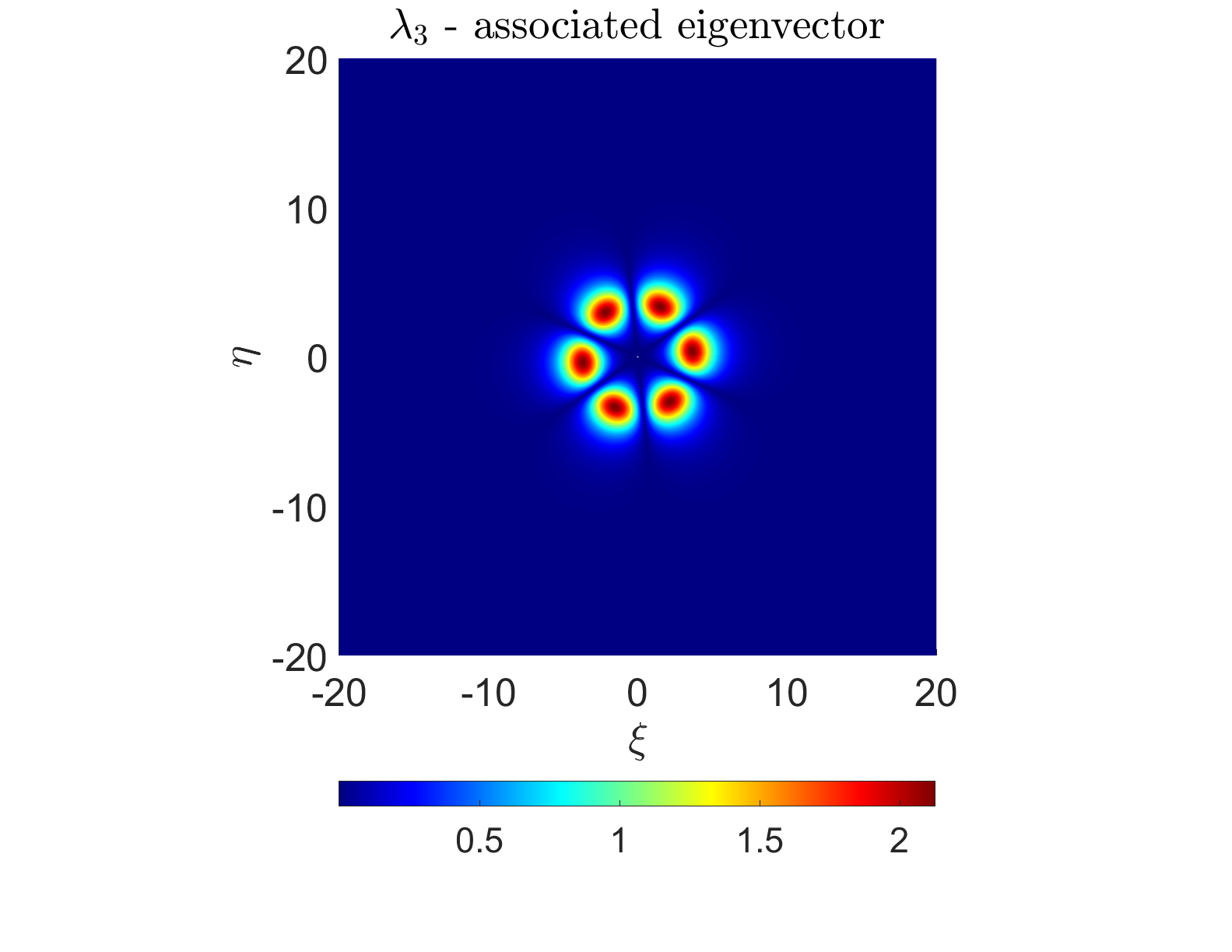}     &  \includegraphics[width=0.49\linewidth]{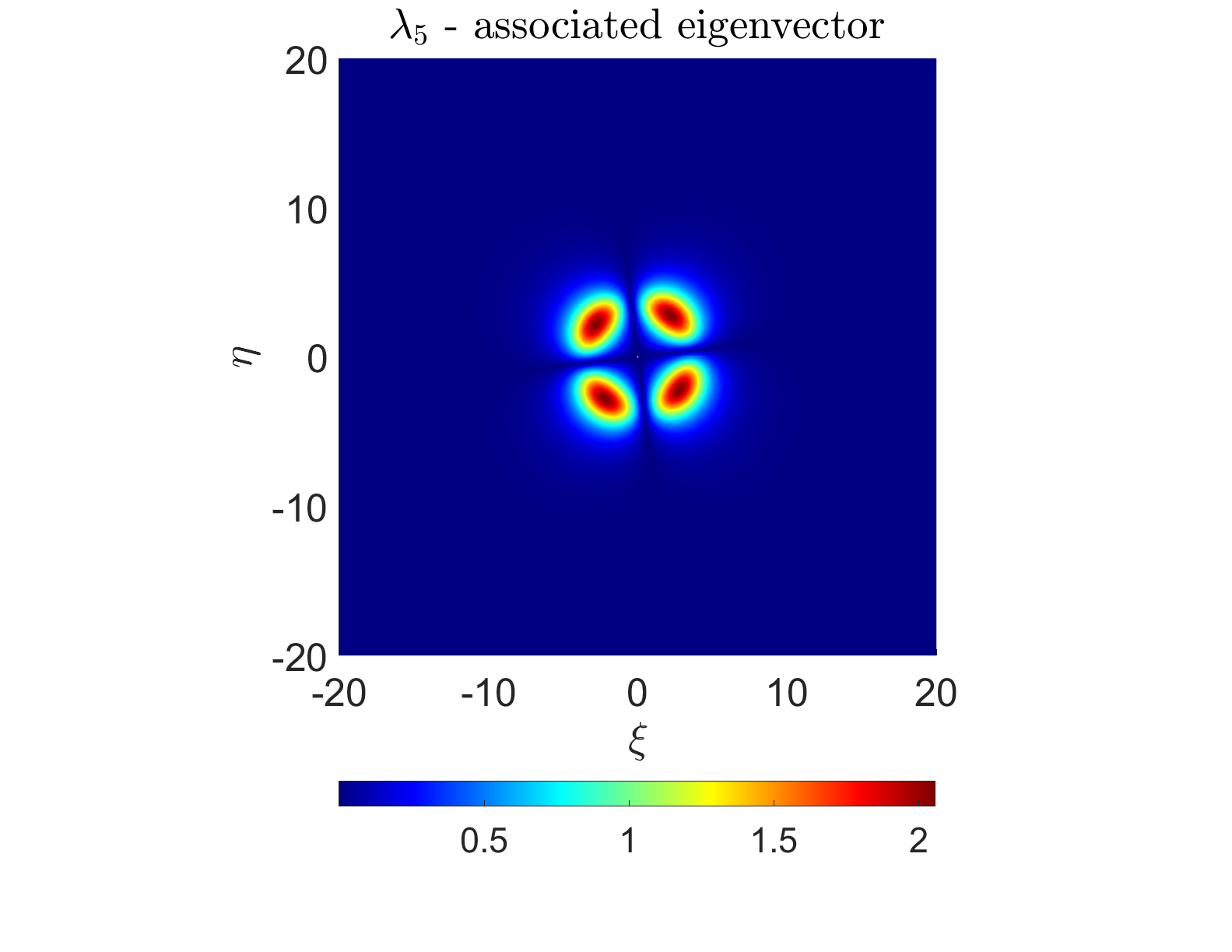}\\
 (c) & (d)
\end{tabular} 
\caption{
\label{fig:two_humped_2D}
(a) Unstable ring self-similar solution (here we show the amplitude) of the 2D NLS equation at $\sigma=1.05$.
(b)-(d) Representative unstable eigenmodes that break the radial symmetry and can lead to multi-peaked structures. From (b) to (d) the modes correspond to eigenvalues: $\lambda_1=1.06$ (8 peaks), $\lambda_{3}=1.05$ (6 peaks), and $\lambda_{5}=0.88$ (4 peaks).
}
\end{figure}

\noindent We compute stationary solutions to Eq.~(\ref{eq:nlscoexp2d}) on 
$(\xi,\eta)\in\Omega=(-K,K) \times (-K,K)$ with $\partial v/\partial {\bf n}=0$ on $\partial\Omega$,
and employ a pinning condition (e.g., $\im[v(\xi=0,\eta=0)]=0$), which enables the determination of the blowup rate $G$.
Self-similar blowup solutions of the radially symmetric NLS equation are also expected to be solutions of the non-radially symmetric formulation.
Indeed, when the ring self-similar solution described in the previous subsection is used as an initial guess (suitably adapted to the Cartesian grid), the numerical solver converges to the profile illustrated in Fig.~\ref{fig:two_humped_2D}(a).
To assess the stability of the computed ring solution, we evaluate its spectrum using MATLAB's $\mathrm{eigs}$ sparse eigensolver. We find three positive eigenvalues associated with continuous symmetries:  $\lambda=2G$ (rescaling) and a double eigenvalue $\lambda=G$ (translations in $x$ and $y$). In addition, there is an eigenvalue at $\lambda=0$ (due to the gauge invariance of the NLS model~\cite{sulem}) and further positive eigenvalues that correspond to symmetry-breaking perturbations in the case of  the radially symmetric ring state.
The number of such positive eigenvalues increases as $\sigma$ approaches its critical value ($\sigma_{crit}=1$).
Figure~\ref{fig:two_humped_2D}(b)-(d) shows representative unstable eigenmodes that break radial symmetry and can trigger the formation of multi-peaked structures.

\begin{figure}[pt!]
\centering
\begin{tabular}{cc}
\includegraphics[width=0.49\linewidth]{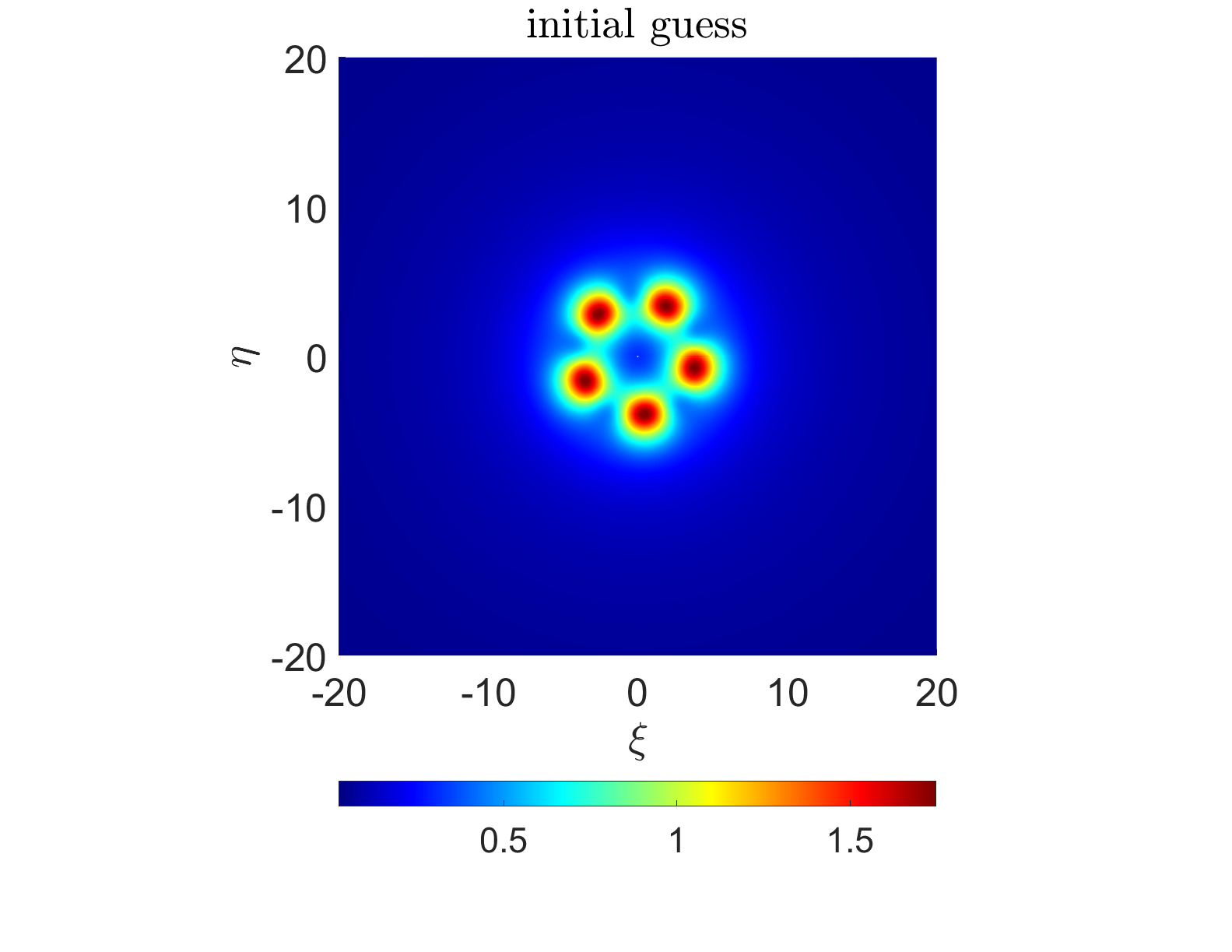}     &  \includegraphics[width=0.49\linewidth]{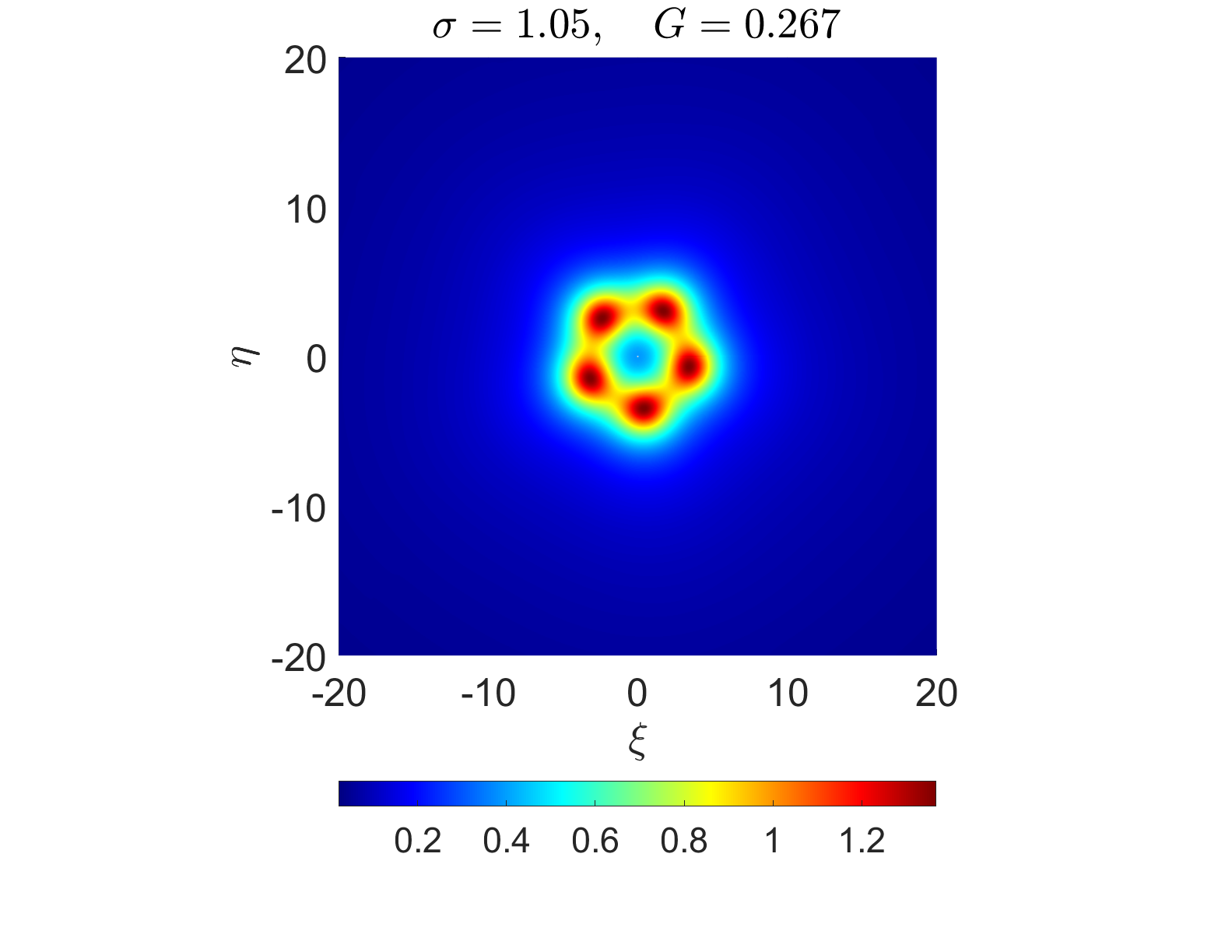}\\
 (a)    & (b)
\end{tabular} 
\caption{
(a) Initial guess for damped Newton iterations, obtained by perturbing the ring solution at $\sigma=1.05$ along the eigenvector associated with $\lambda_7$. 
(b) Converged five-peaked solution; the blow-up rate is $G=0.267$. 
Both panels illustrate the amplitude $|v|$ of the solution.
}
\label{fig:perturbexample}
\end{figure}

For the particular ring solution shown, we compute 11 unstable eigenvalues beyond the symmetry-related ones ($\lambda \approx 2G$, $\lambda \approx G$ and $\lambda \approx 0$): $\lambda_{1,2}\approx 1.06, \lambda_{3,4}\approx1.05, \lambda_{5,6}\approx0.88, \lambda_{7,8}\approx0.77, \lambda_{9,10}\approx0.62$, and $\lambda_{11}\approx0.13$.
%
%
This suggests that suitably perturbing the ring self-similar state can steer Newton iterations toward potential multi-peaked solutions. Convergence 
thereto may require many iterations; in several cases we employed damping strategies inspired by damped Newton methods. 
As an illustration, Fig.~\ref{fig:perturbexample}(a) shows an initial guess obtained by perturbing the ring solution at $\sigma=1.05$ along the eigenvector associated with $\lambda_7$.
After a few Newton steps, the method converges to the five-peaked profile in Fig.~\ref{fig:perturbexample}(b), with blow-up rate $G=0.267$.

An alternative route to multi-peaked solutions leverages the asymptotic predictions 
of Eq.~(\ref{eq:peakasymptotics}) 
%
which provides the peak locations $\bX_i$ as functions of the blow-up rate $G$, 
while Eq.~(\ref{eq:sigmaG}) below relates $\sigma$ to $G$.
We form an initial guess for damped Newton iterations by shifting the origin-centered, single-peaked, stable self-similar solution to each predicted position $X_i$. 
For example, a good two-peaked initial guess $v^{(0)}$ is obtained by superimposing two shifted copies at $-\bX_1=\bX_2=(X,0)$ with $X$ given by Eq.~(\ref{eq:peakasymptotics})  (which in this case
results in Eq.~(\ref{eq:twohumpspos}) discussed below). The resulting
guess profile then reads:

\begin{equation}
\label{superposition}
    v^{(0)}(\xi,\eta;\sigma)=v_1(\xi+X,\eta;\sigma)+v_1(\xi-X,\eta;\sigma),
\end{equation}
\noindent where $v_1$ denotes the single-peaked self-similar solution. 

\begin{figure}[ht!]
\centering
\begin{tabular}{cc}
\includegraphics[width=0.49\linewidth]{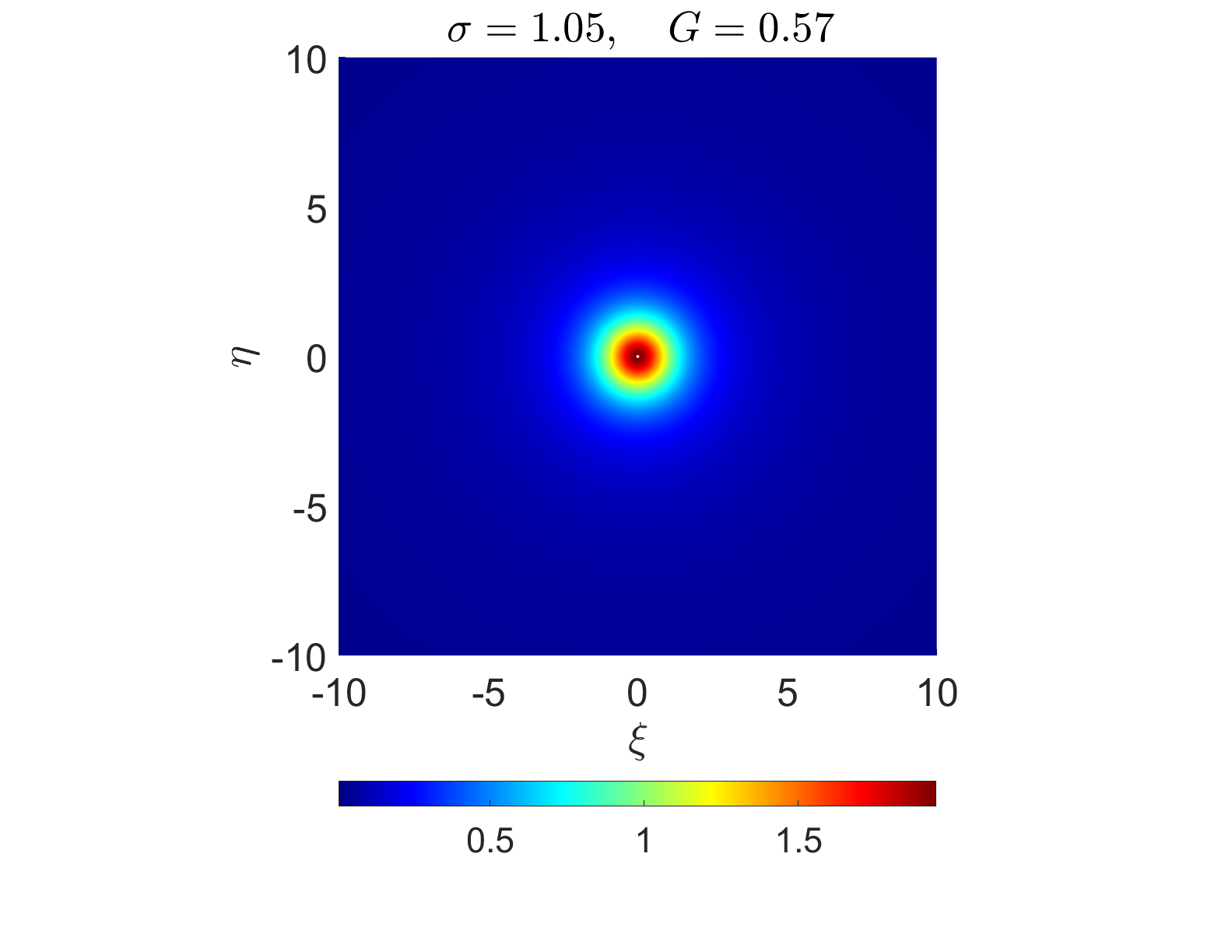}     &  \includegraphics[width=0.49\linewidth]{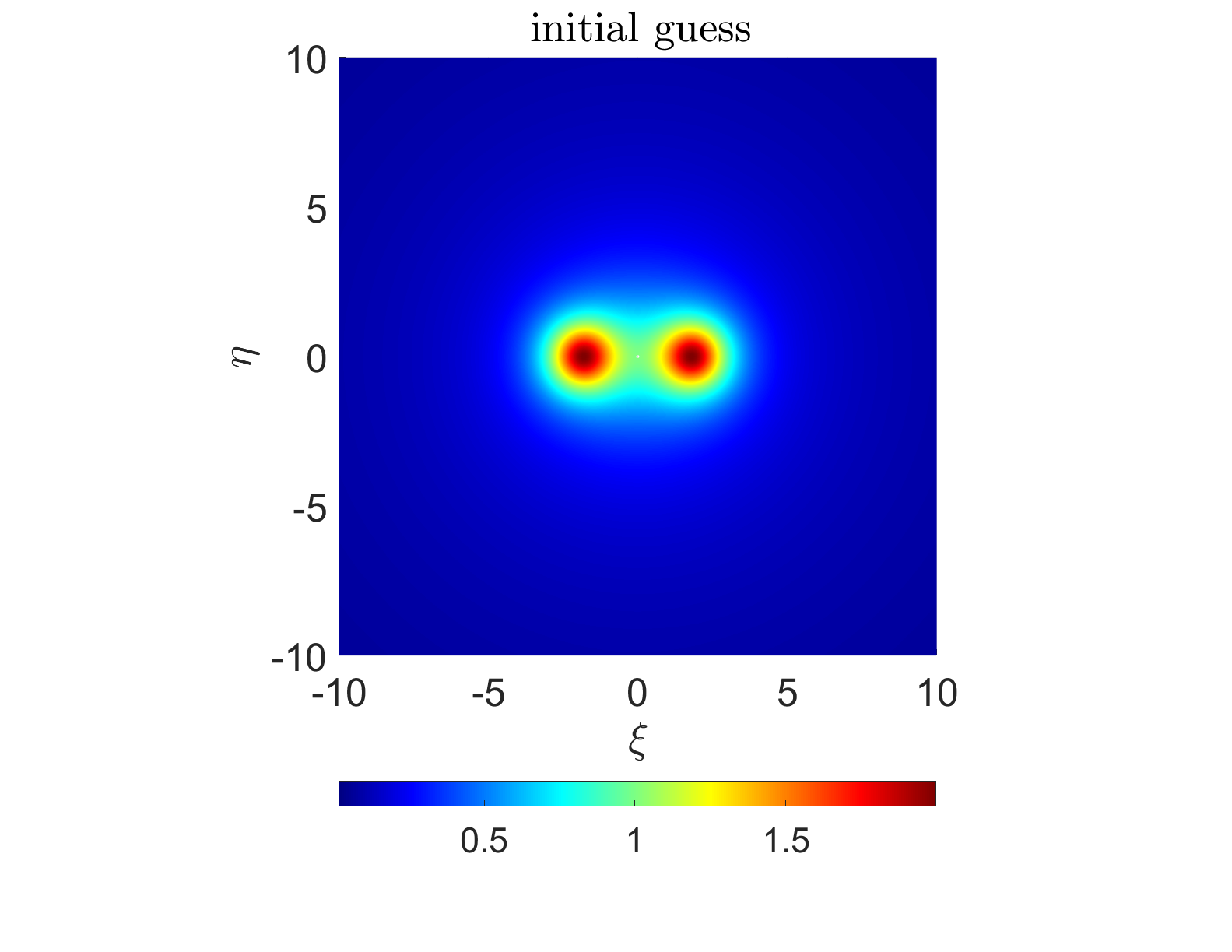} \\
 (a)    & (b)
\end{tabular} 
\begin{tabular}{c}
\includegraphics[width=0.49\linewidth]{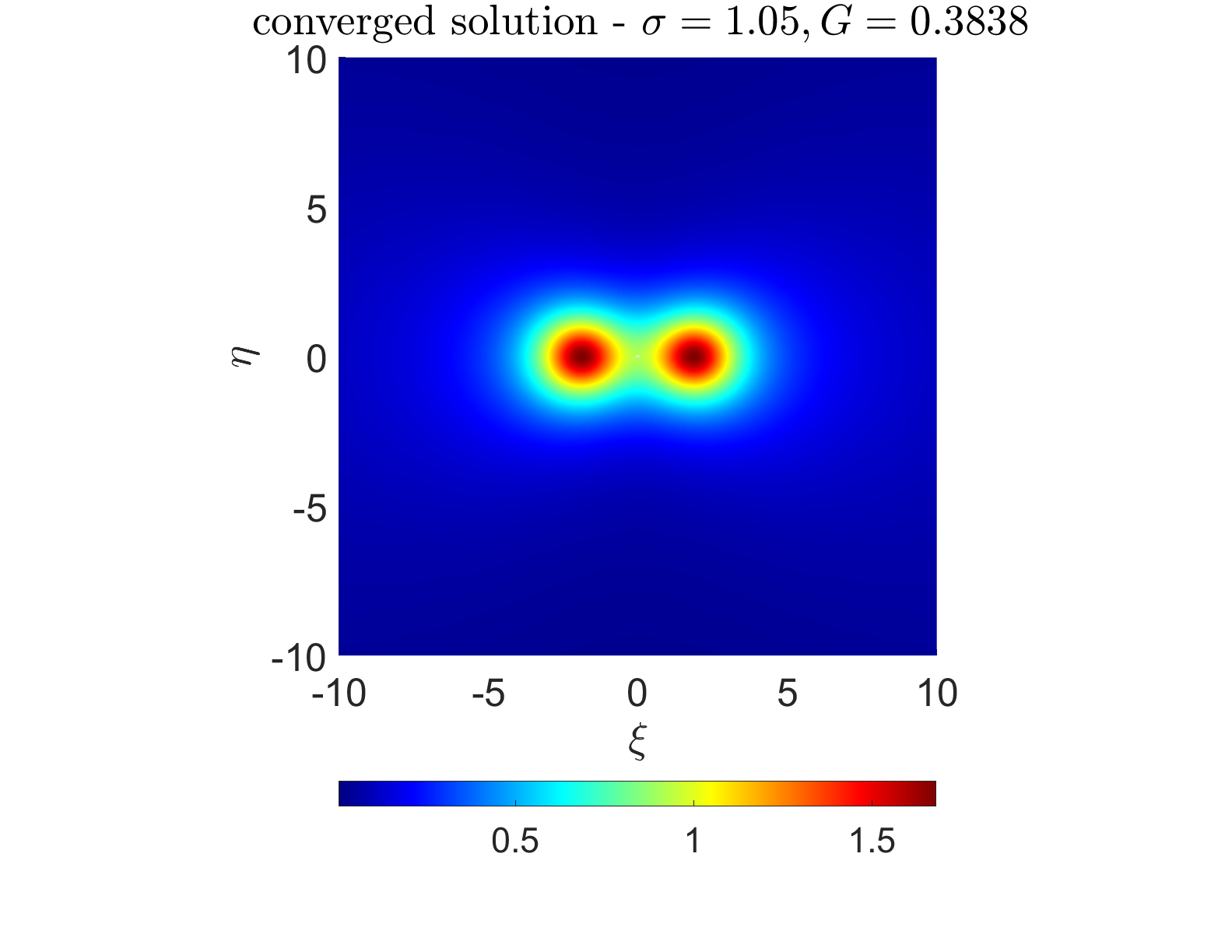}\\
 (c)
\end{tabular} 
\caption{
(a) Single-peaked (stable) self-similar solution (amplitude) at $\sigma=1.05$.
(b) Superposition of $\pm X \approx 1.836$ shifts of the single-peaked solution along the horizontal axis, used as an initial guess for damped Newton iterations.
(c) Converged two-peaked solution; the blow-up rate is $G=0.3838$.
}
\label{fig:shiftsolution}
\end{figure}

In Fig.~\ref{fig:shiftsolution}, we present (a) the single-peaked solution at $\sigma=1.05$ (with corresponding blow-up rate $G=0.57$), (b) the superposition of shifts by $\pm X \approx \pm 1.836$ (from Eq.~(\ref{eq:peakasymptotics}) with $G = 0.3855$, which yields $\sigma=1.05$ via Eq.~(\ref{eq:sigmaG})) used as an initial guess for the damped Newton iterations, and (c) the converged two-peaked self-similar solution at $\sigma=1.05$ with blow-up rate $G=0.3838$.
We also note that the computed value $G$ deviates only by $0.5\%$ from the asymptotics prediction $G=0.3855$.

\begin{figure}[ht!]
\centering
\begin{tabular}{ccc}
\includegraphics[width=0.33\linewidth]{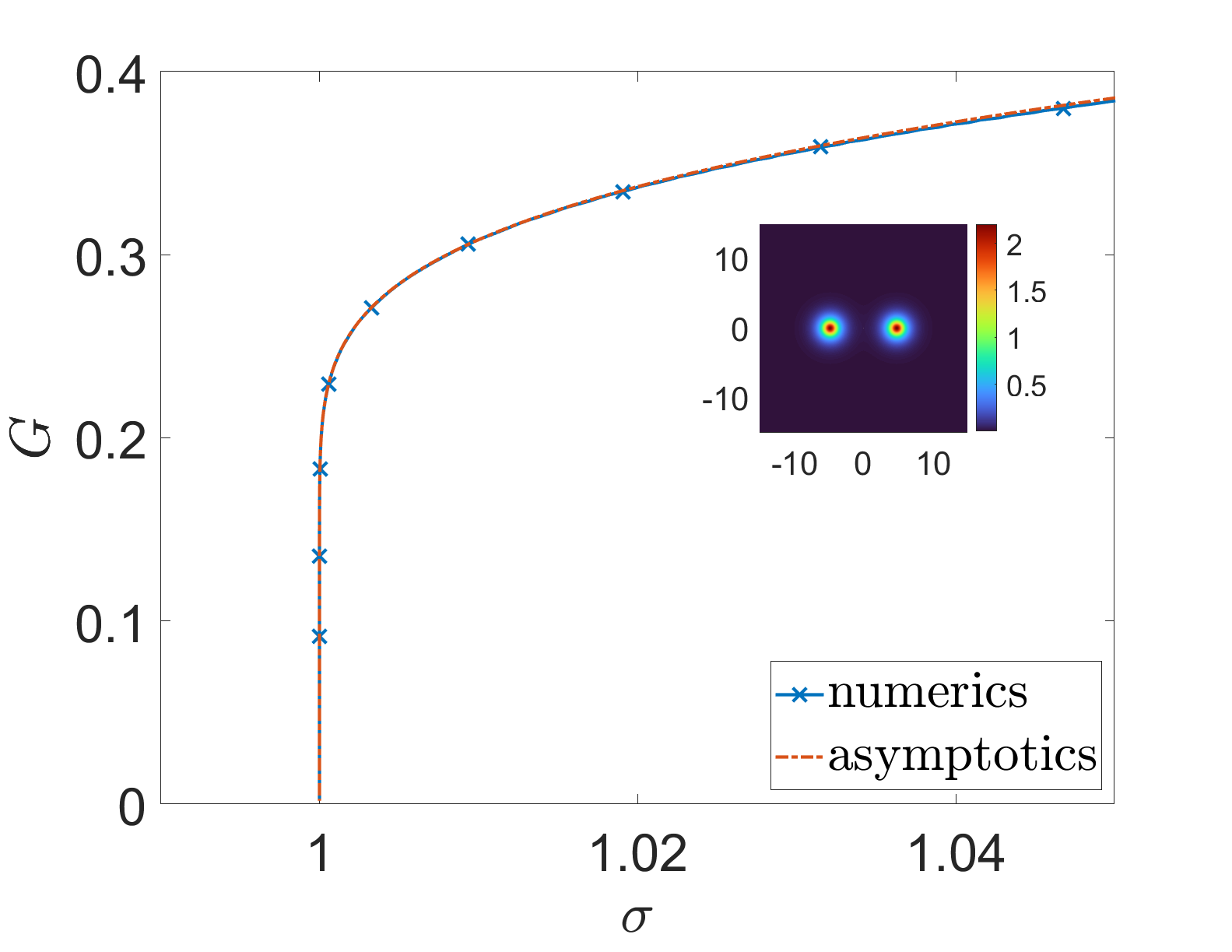}     &  \includegraphics[width=0.33\linewidth]{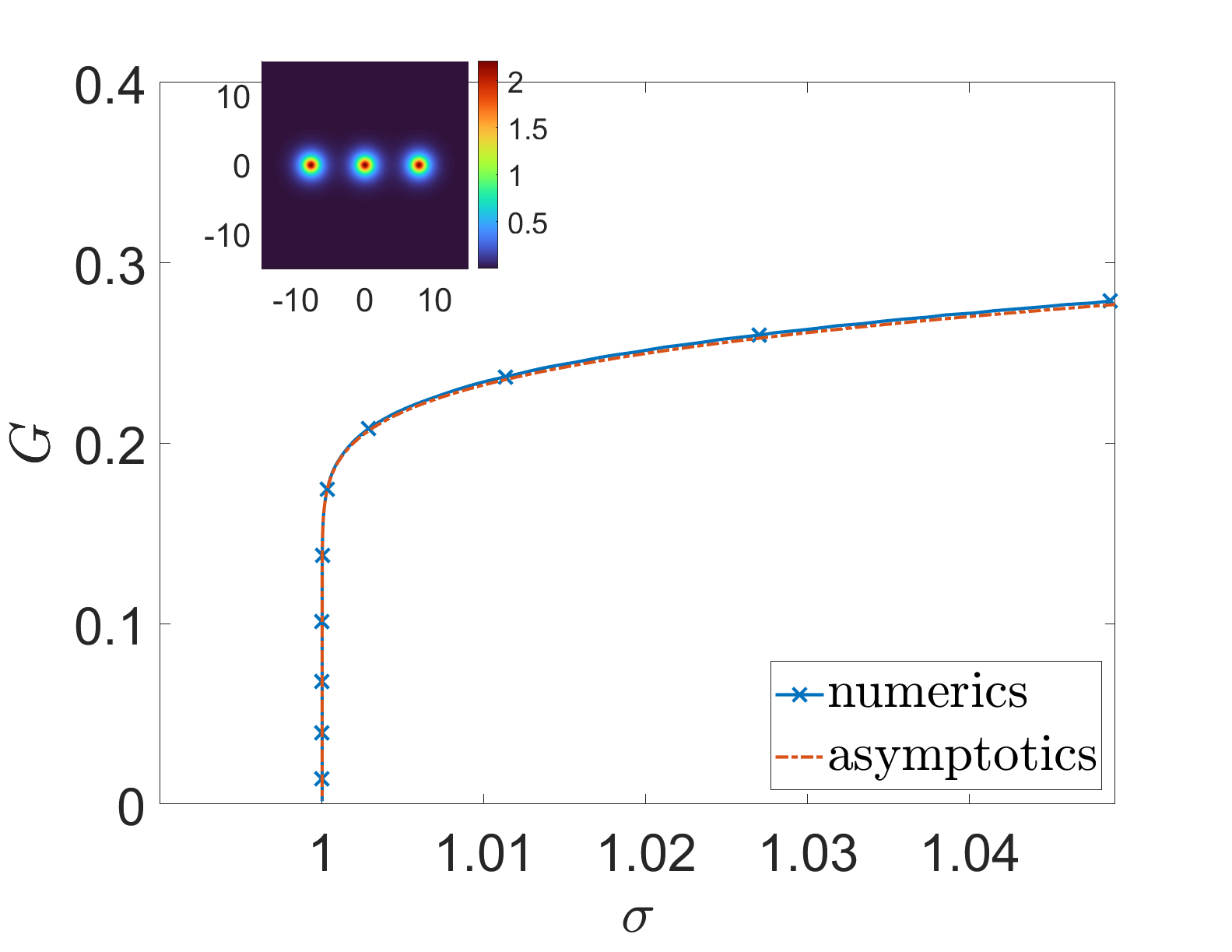} & \includegraphics[width=0.33\linewidth]{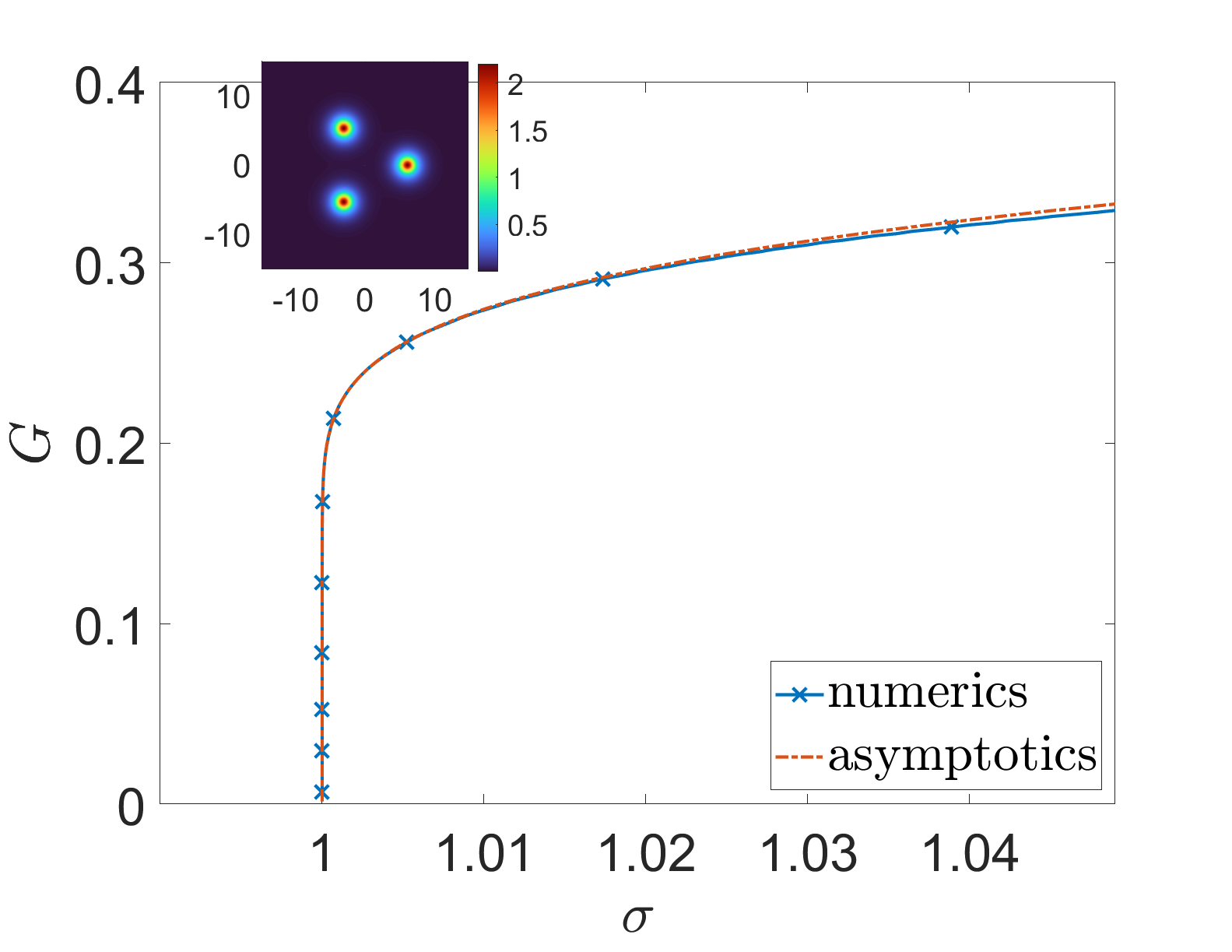} \\
(i)     & (ii) & (iii) \\
    \includegraphics[width=0.33\linewidth]{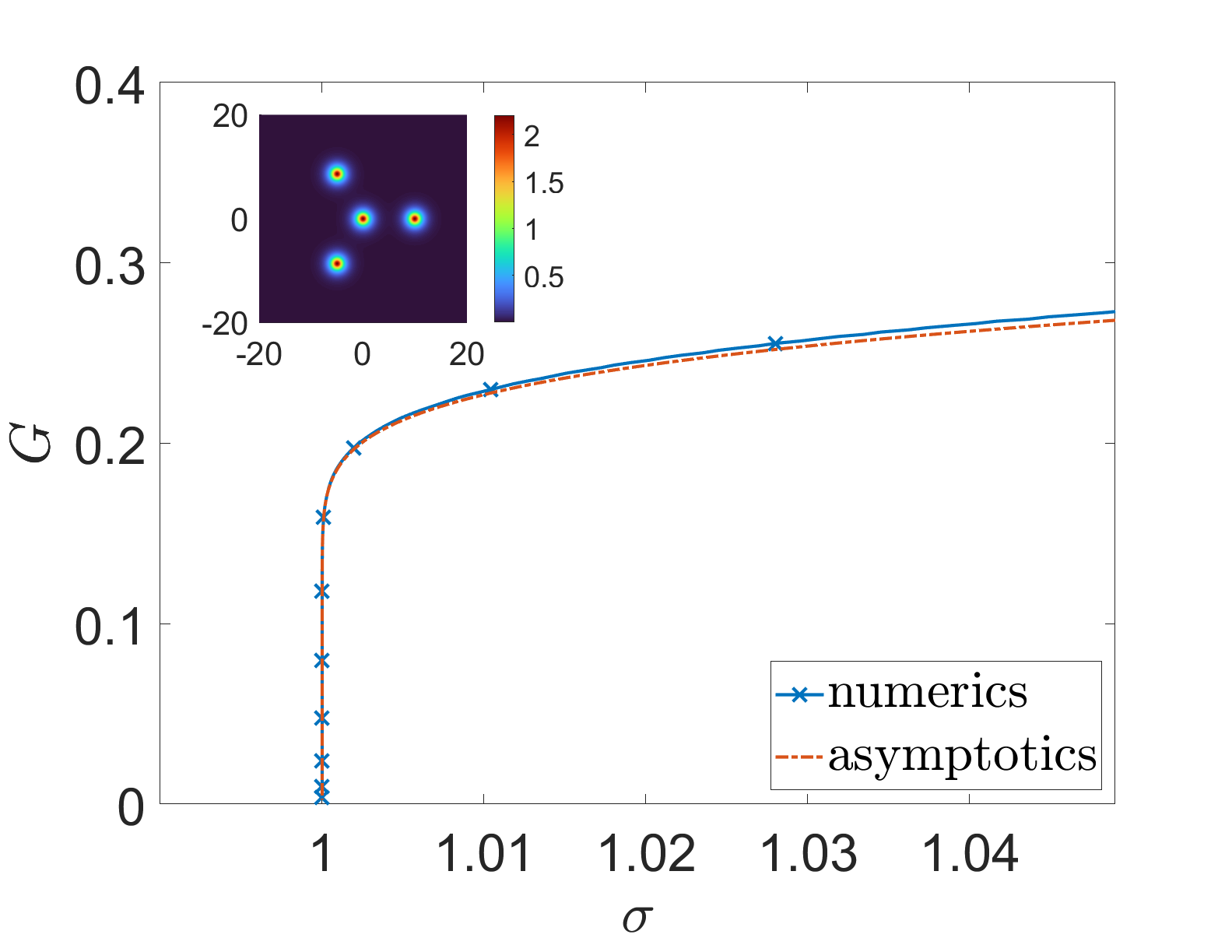} & \includegraphics[width=0.33\linewidth]{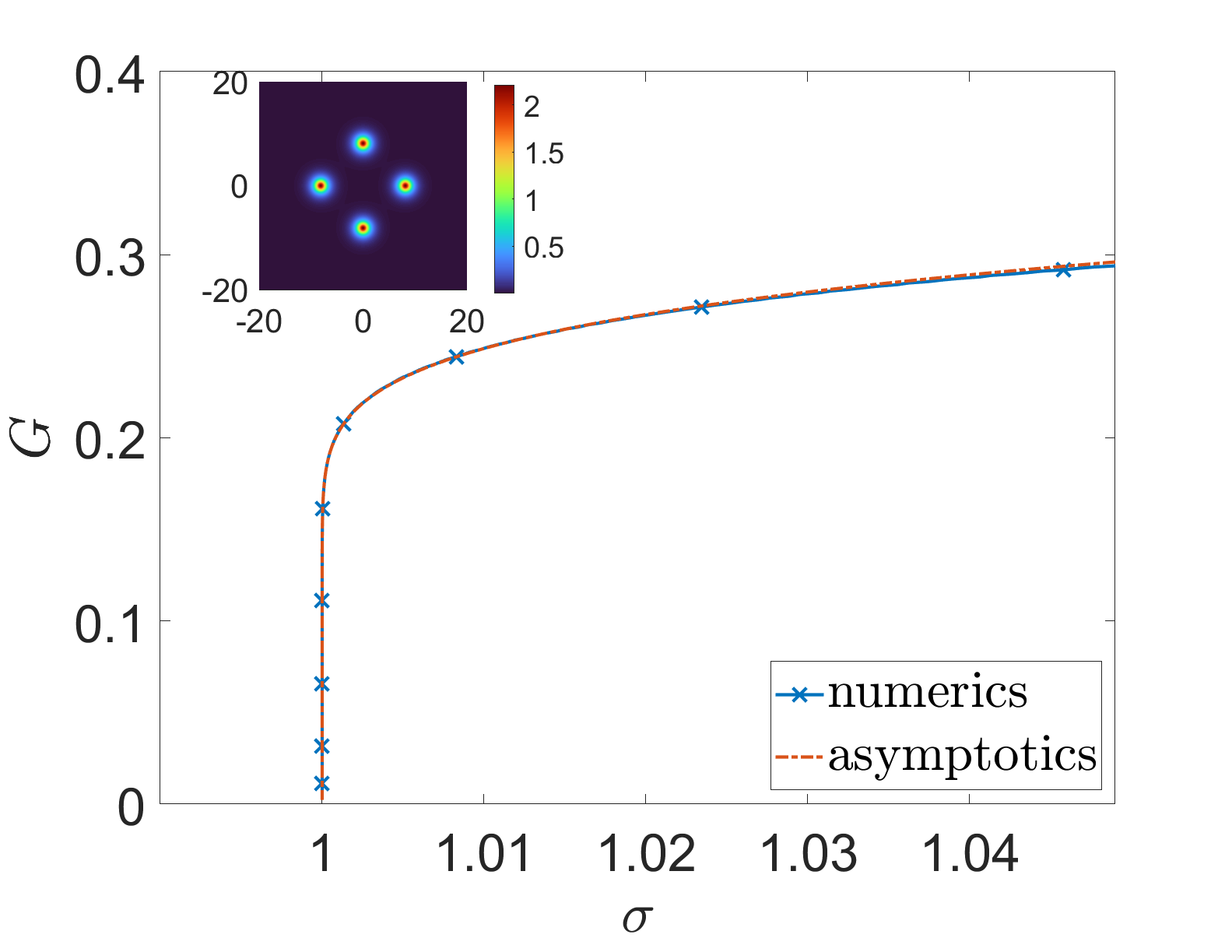} & \includegraphics[width=0.33\linewidth]{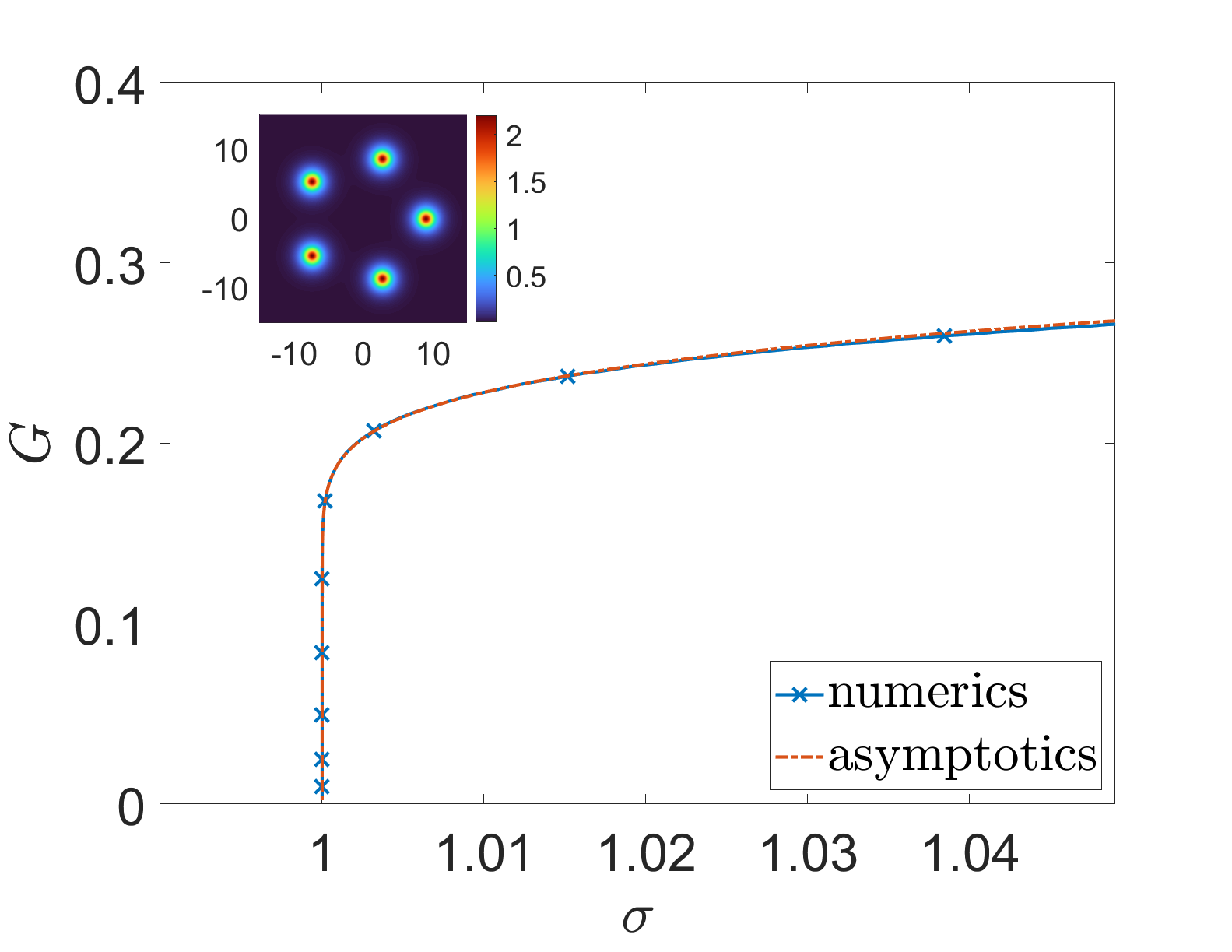} \\
(iv)     & (v) & (vi) \\
\end{tabular}
\begin{tabular}{c}
 \includegraphics[width=0.31\linewidth]{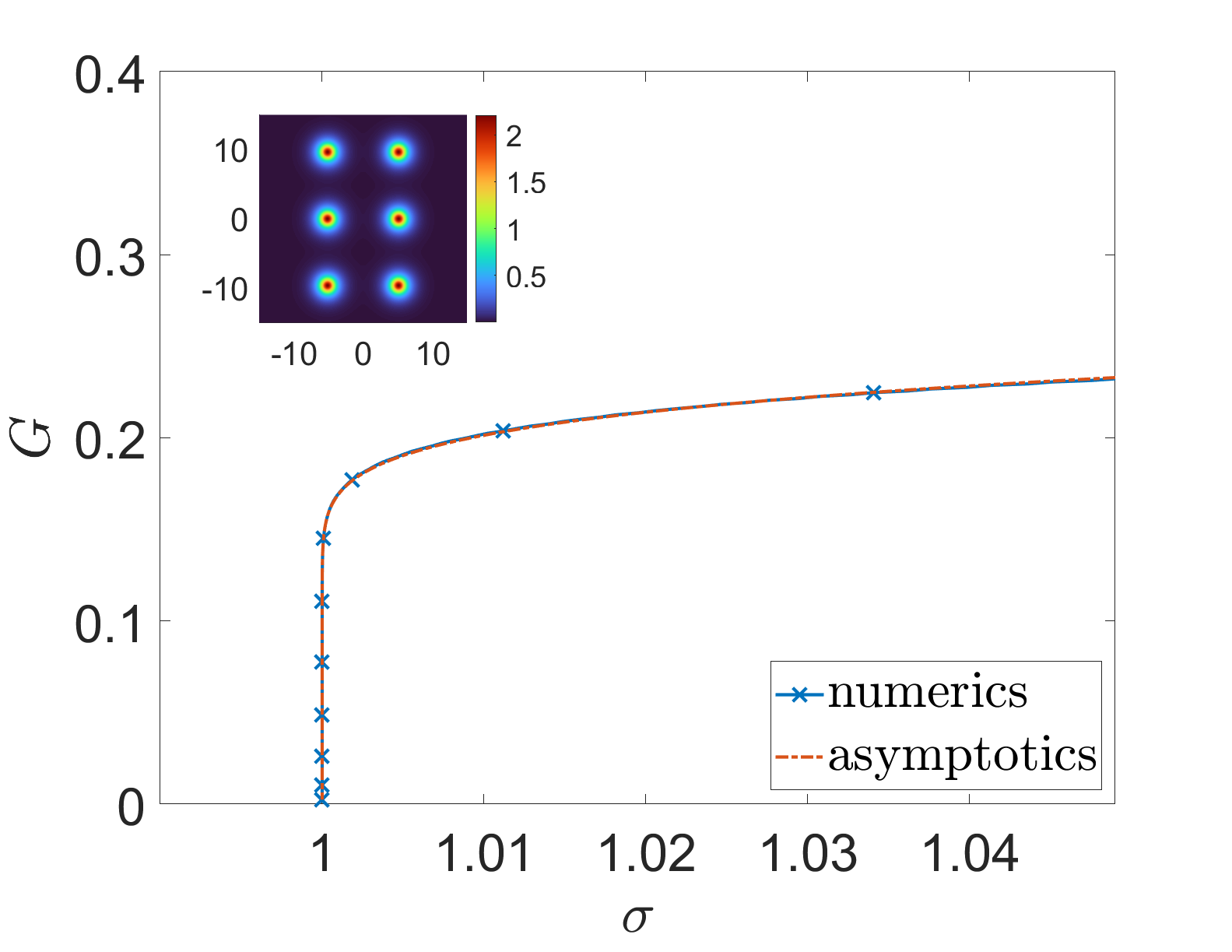}    \\
 (vii)     
\end{tabular}
\caption{
\label{fig:bifdiagrammulti}
Bifurcation diagrams for self-similar solutions with: (i) two peaks; (ii) three aligned peaks; (iii) three peaks at the vertices of an equilateral triangle; (iv) four peaks with three at the vertices of an equilateral triangle and one at the barycenter; (v) four peaks at the vertices of a square; (vi) five peaks at the vertices of a pentagon; and (vii) six peaks aligned in two parallel lines. Dash-dotted red curves indicate asymptotic predictions, and blue x-marks show numerical results. Insets display representative solutions with blow-up rates: (i) $G=1.8 \times 10^{-3}$, (ii) $G=1.6 \times 10^{-3}$, (iii) $G=1.5 \times 10^{-3}$, (iv) $G=2.3 \times 10^{-3}$, (v) $G=2.3 \times 10^{-3}$, (vi) $G=2.4 \times 10^{-3}$, and (vii) $G=2.4 \times 10^{-3}$. 
}
\end{figure}

Up to this point, we have demonstrated two complementary strategies for achieving convergence with Newton's method: (i) perturbing the ring solution along unstable eigenmodes that break radial symmetry, and (ii) using asymptotic formulas to predict peak locations and the value $\sigma$ given blow-up rate $G$; then
we superpose individual self-similar profiles, as suggested, e.g.,
by Eq.~(\ref{superposition}). 
We now apply parametric continuation \cite{kuznetsov_book} in the nonlinearity exponent $\sigma$, which allows us to trace entire branches of multi-peaked self-similar solutions.
Figure~\ref{fig:bifdiagrammulti} presents numerically computed branches corresponding to two-, three, four-, five- and six-peaked configurations. 
Notice that for configurations involving more than two peaks, multiple
possibilities arise. E.g., for three-peaked states, we consider
an aligned and an equilateral triangle one, for four, we examine 
a square, as well as an equilateral triangle with a peak at its
barycenter, and so on. For six peaks, we show an aligned configuration
with $2 \times 3$ aligned peaks, while obviously a hexagonal configuration
exists as well.
Each panel shows the dependence of the blow-up rate $G$ on $\sigma$, revealing an almost vertical bifurcation near the critical value $\sigma_{critical}=1$.
The dashed-dotted red curves depict the asymptotic predictions from Eq.~(\ref{eq:peakasymptotics}) and Eq.~(\ref{eq:sigmaG}), which agree closely with the numerical result, even for higher-complexity multi-peaked structures.  

\begin{figure}[ht!]
    \centering
    \begin{tabular}{cc}
         \includegraphics[width=0.49\linewidth]{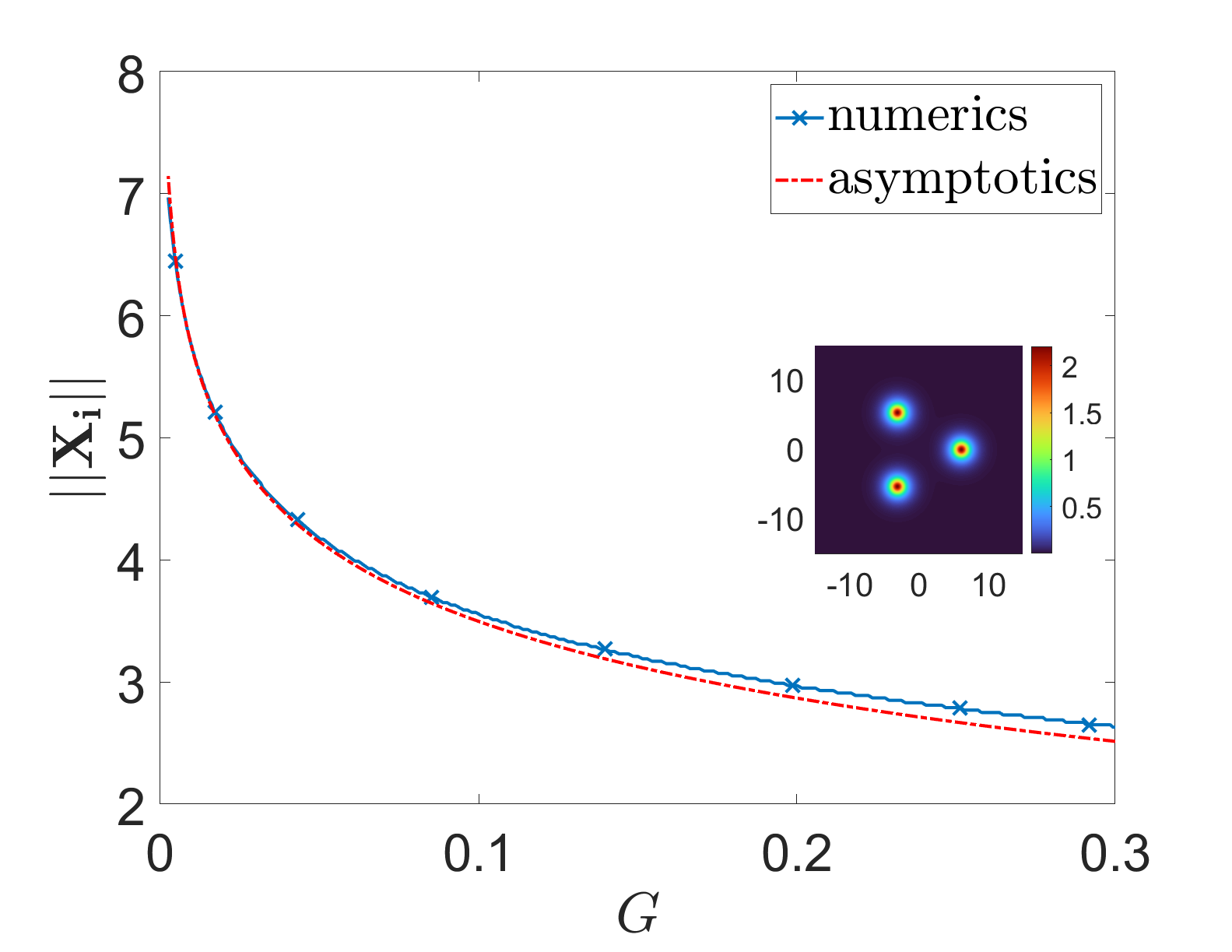}    &      \includegraphics[width=0.49\linewidth]{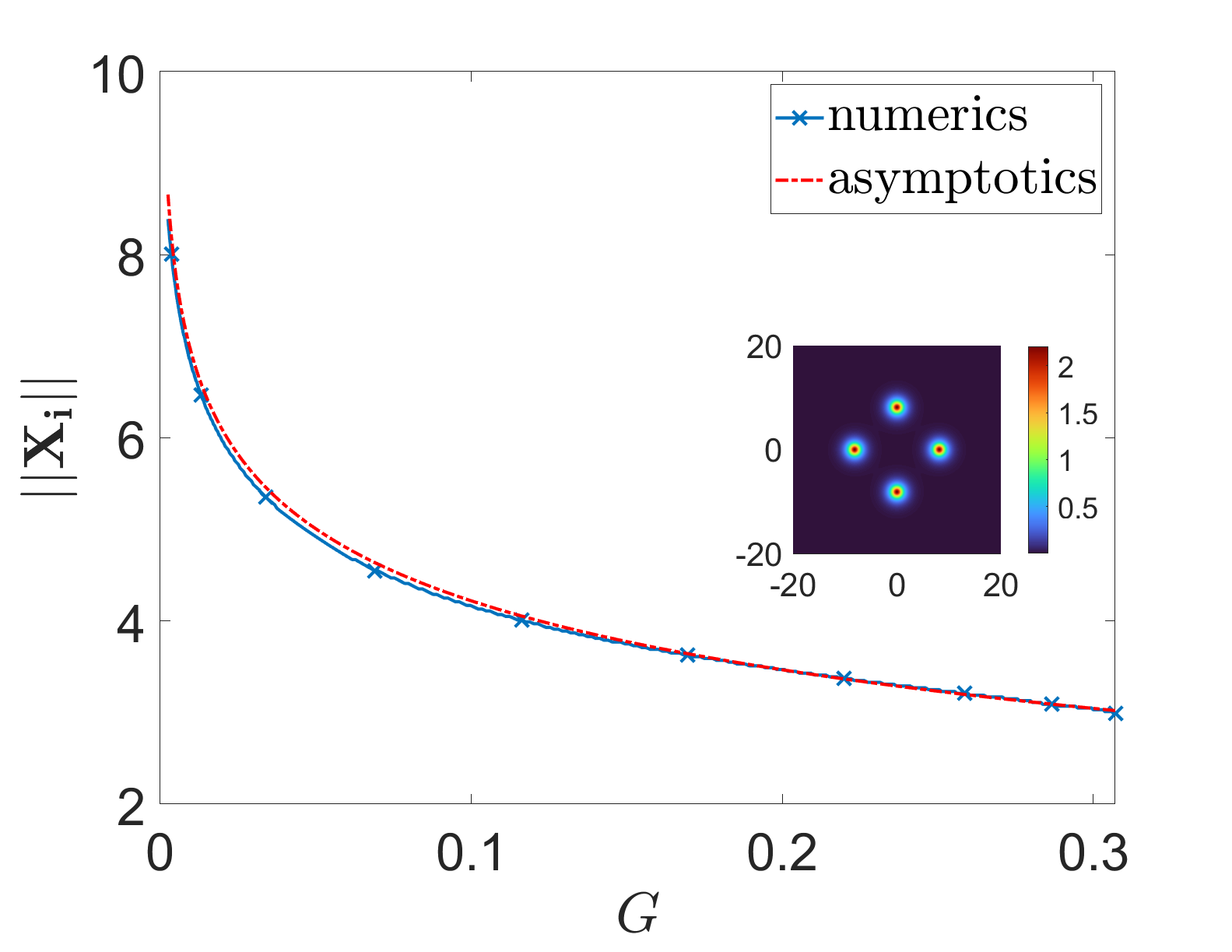} \\
         (i) & (ii) 
    \end{tabular}
    \begin{tabular}{c}
     \includegraphics[width=0.49\linewidth]{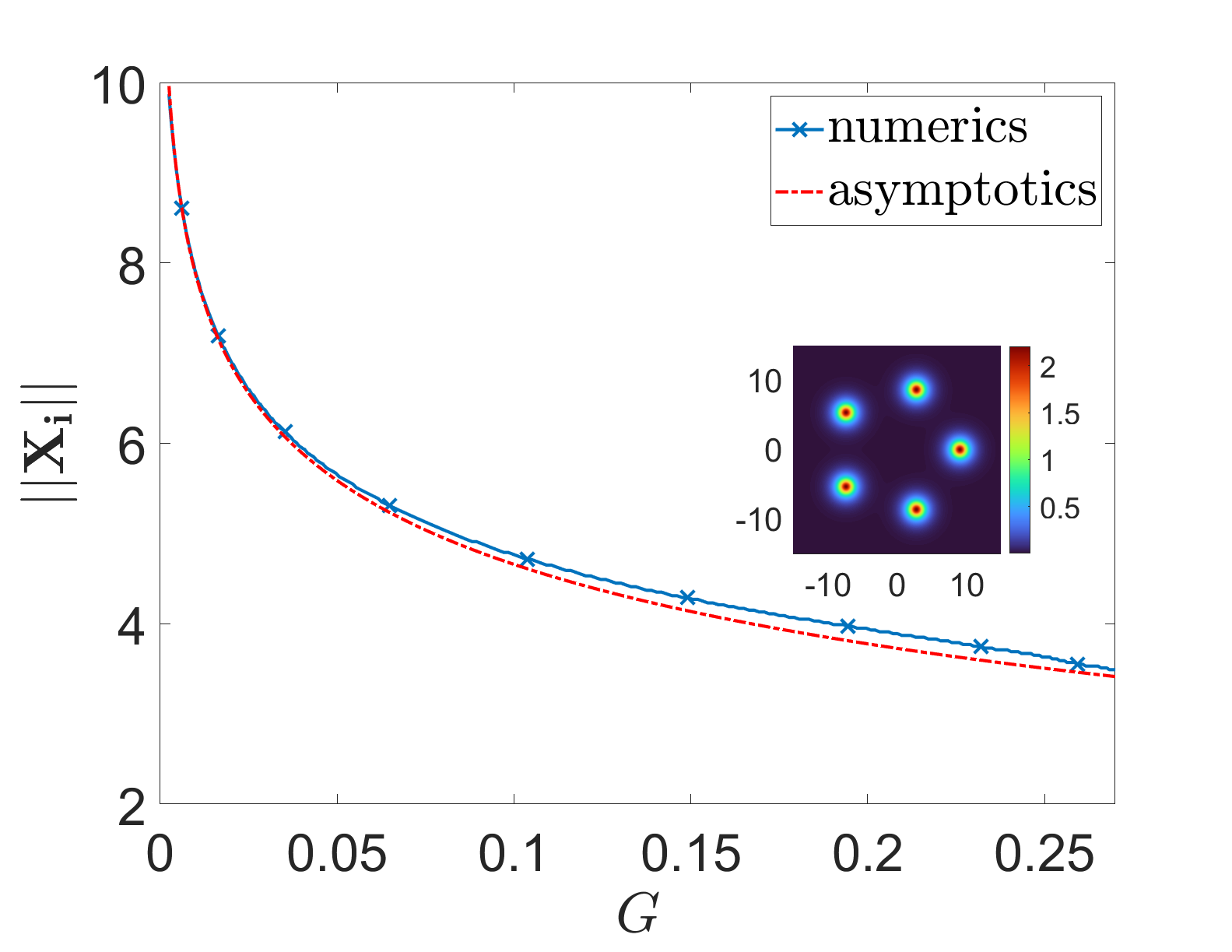}      \\
        (iii)  
    \end{tabular}
    \caption{Comparison of peak positions (radial distance $||\bf{X}_i||$ from the domain center)  between numerical results and the asymptotic prediction (Eq.~(\ref{asypositions})) for (i) three-, (ii) four- and (iii) five-peaked solutions as functions of the blow-up rate $G$. The large distances near the critical  point ($\sigma \rightarrow1, G\rightarrow0$) indicate that these solution branches emerge from infinity at the bifurcation.}
    \label{fig:peak_comparison}
\end{figure}

In Fig.~\ref{fig:peak_comparison}, we plot the peak locations of the computed solutions as functions of the blow-up parameter $G$.
In all cases presented (including additional multi-peaked configurations not shown), the peaks lie far from the origin as the nonlinearity exponent $\sigma$ approaches the critical value $\sigma=1$ (equivalently, $G\rightarrow0$), and they move closer together as $\sigma$ increases (with a concomitant increase in $G$).
This behavior indicates that these multi-peaked branches originate at ``infinity'' at the bifurcation point $\sigma=1$. This is natural, as we will
illustrate below, as the phase profile (absent in the soliton, but present
in the self-similar profiles considered herein) introduces a balancing
``force'' equilibrating the inter-peak interaction effect and spontaneously
giving rise to the stationary configurations we compute.
The comparison with the asymptotic predictions shows excellent agreement near the bifurcation and only small deviations further away.

\subsection{Spectral Analysis}
\label{sec:stab}
\begin{figure}[pt!]
    \centering
    \begin{tabular}{ccc}
\includegraphics[width=0.33\linewidth]{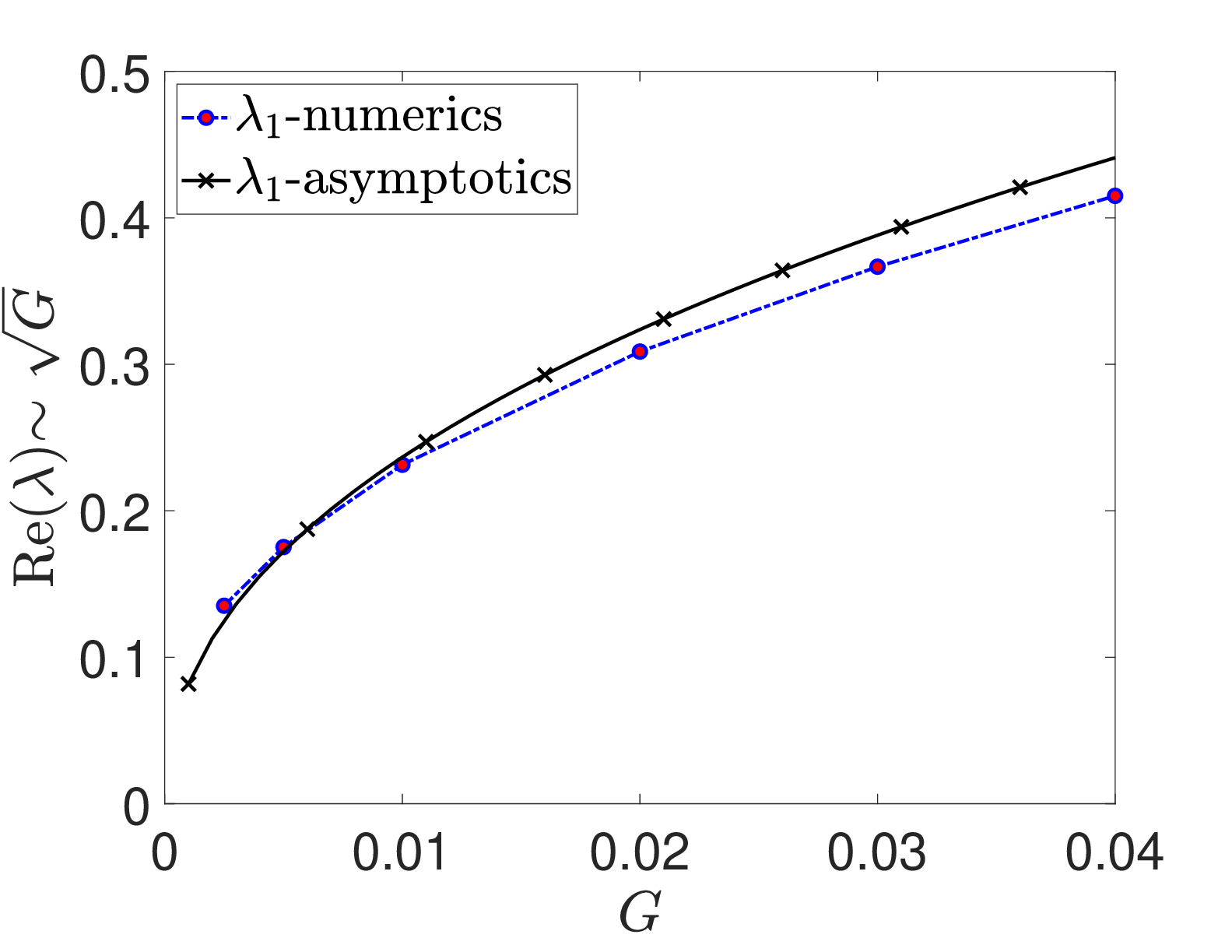}     &  \includegraphics[width=0.33\linewidth]{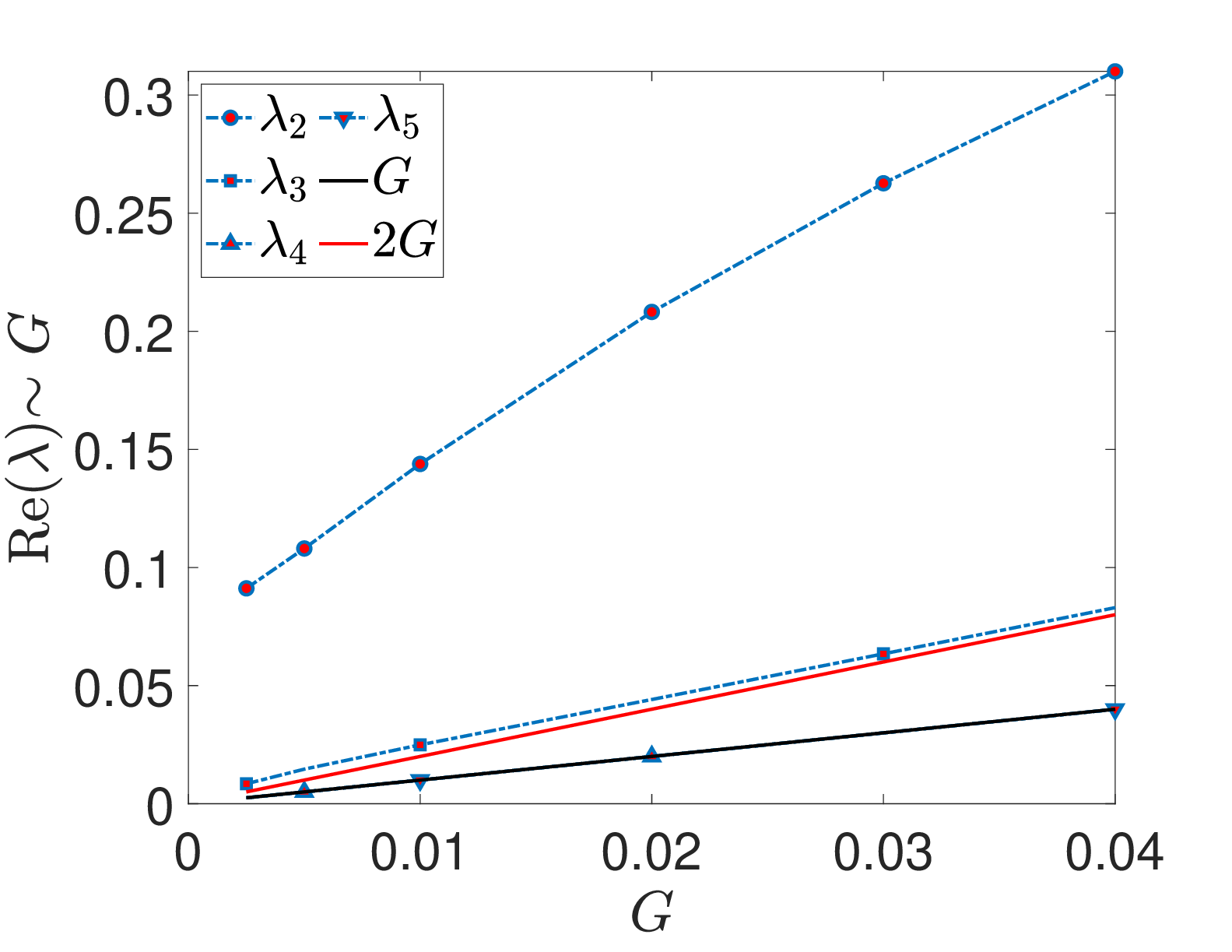} & \includegraphics[width=0.33\linewidth]{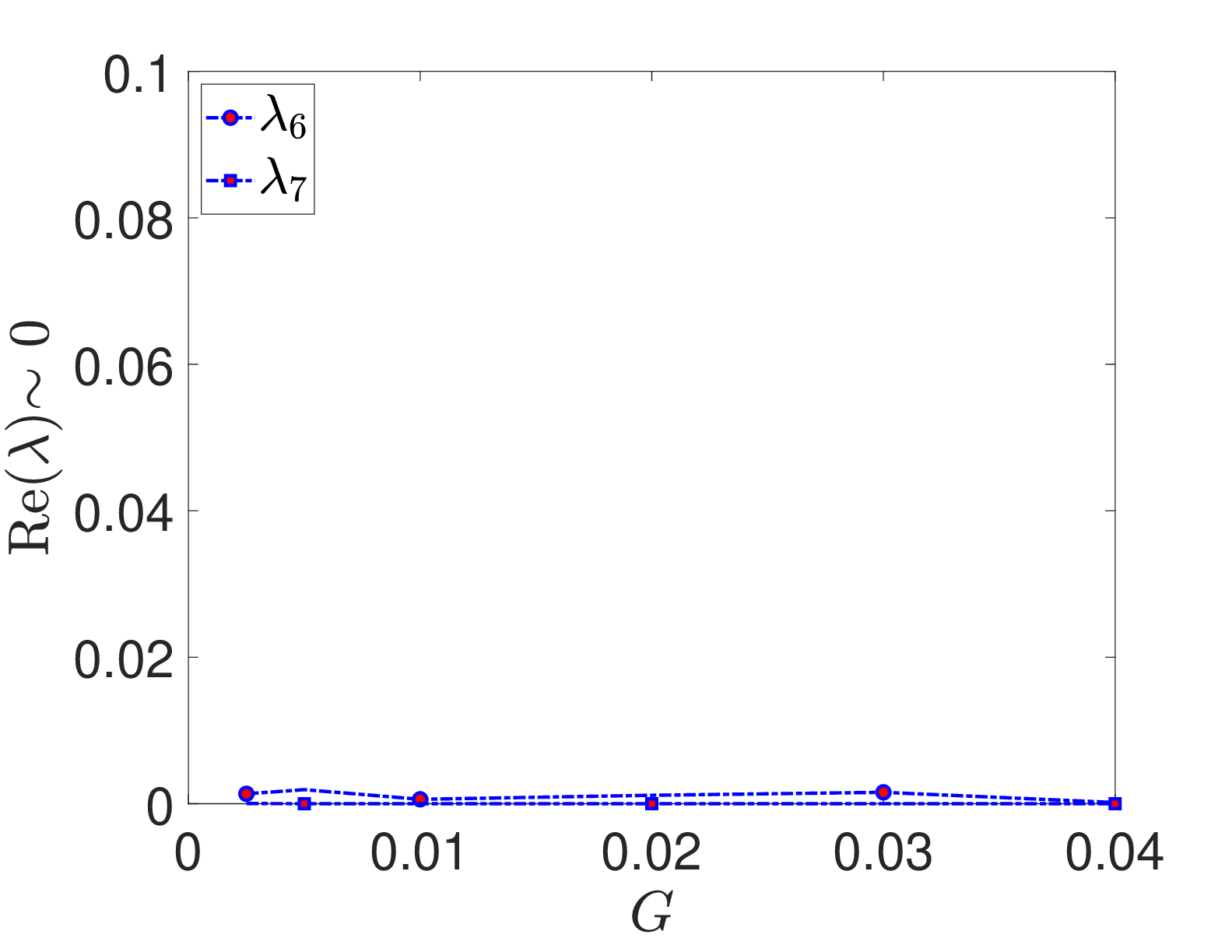} \\
     (i)    & (ii) & (iii) \\
     \end{tabular}
     \begin{tabular}{c}
     \includegraphics[width=0.33\linewidth]{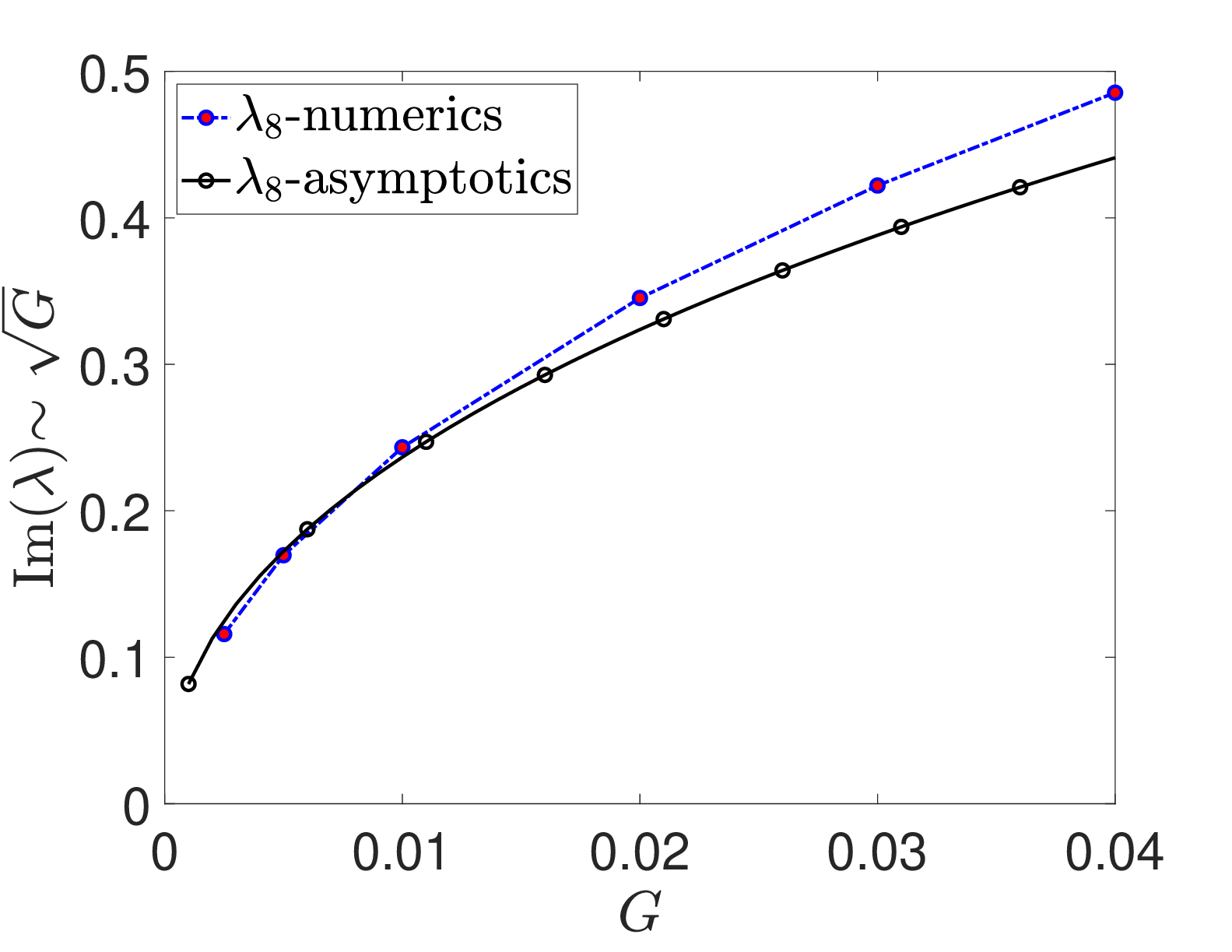}\\
     (iv)\\     
     \end{tabular}
    \caption{Dependence of the dominant eigenvalues on the blowup rate $G$ for two-peaked solutions.
    (i) Dominant real eigenvalue with $\sqrt{G}$ scaling (black lines: asymptotic predictions).
    (ii) Real eigenvalues with $G$ scaling;
    black and red lines mark $G$ and $2G$ lines, respectively. 
    (iii) Eigenvalues $\lambda_6, \lambda_{7}$ are the ones closest to the origin.
    (iv) Purely imaginary eigenvalue $\lambda_{8}$ with $\sqrt{G}$ scaling (black lines: asymptotic predictions).}
    \label{fig:domeigen2peaks}
\end{figure}

To assess the stability of the computed multi-peaked solutions, we compute their spectra using MATLAB's eigs eigenvalue solver. 
We present results for several configurations and focus on dominant eigenvalue pairs on the real and imaginary axes, and on how these depend on the blow-up rate $G$.
Our count, summarizing observations across the configurations considered, indicates that for $N$-peaked configurations, there exist: $N-1$ real 
eigenvalues (and their ``nearly opposites''--- contrary to the
Hamiltonian case, eigenvalues do not come in pairs here) and 
$N-1$ purely imaginary ones (counting the ones with positive imaginary part); both of these are of $O(\sqrt{G})$.
Then, there are $2\times N$ eigenvalues 
(real and purely imaginary eigenvalues) of $O(G)$.
Finally, there  are 2 eigenvalues at the origin, reflecting the
phase invariance and the rotational invariance of our multi-peaked
configurations. Each of these has its own ``former Hamiltonian pair'' which
is typically slightly negative, as was the case, e.g., also in~\cite{jon2}.
Figure ~\ref{fig:domeigen2peaks} summarizes these dependencies for two-peaked solutions. 
%
Specifically, we find one real eigenvalue with $\sqrt{G}$ scaling (panel (i)), one purely imaginary eigenvalue with $\sqrt{G}$ scaling (panel (iv)), and four real eigenvalues with $G$ scaling (panel (ii)).
Among these, three eigenvalues are associated with symmetries: $\lambda_3 \approx 2G$ (scaling), and $\lambda_{4,5} \approx G$ (translations in the original $x$ and $y$ coordinate directions).
Finally, there are two eigenvalues near the origin reflecting the 
symmetries of the configuration, as mentioned above (panel (iii)). 
We thus count $2\times(N=2)-2=2$ eigenvalues with $\sqrt{G}$ scaling and $2 \times (N=2)=4$ eigenvalues with $G$ scaling, plus another $2$ eigenvalues due to
symmetry for a total of $4N$ eigenvalues, originating from the $4N$
eigenvalue pairs of the Hamiltonian solitary wave problem.
Recall that in the latter, each solitary wave bears an eigenvalue pair
due to the conformal invariance, 2 pairs due to translations in the two
directions and 1 pair due to the phase invariance, giving rise to $4$ pairs
for each peak and hence $4N$ pairs for $N$ peak states. Our enumeration
above accounts for each one of these eigenvalues as the non-conservative
setting of the self-similar frame modifies the position (but preserves
the count) of the eigenvalues of the (pre-bifurcation) Hamiltonian setting.
Asymptotic predictions for eigenvalues with $\sqrt{G}$ scaling (Eq.~(\ref{Ghalfeigs}) for $N=2$) are also available and depicted in Figs. ~\ref{fig:domeigen2peaks}(i) and (iv) showing good comparison with numerics. 

\begin{figure}[ht!]
        \includegraphics[width=0.99\linewidth]{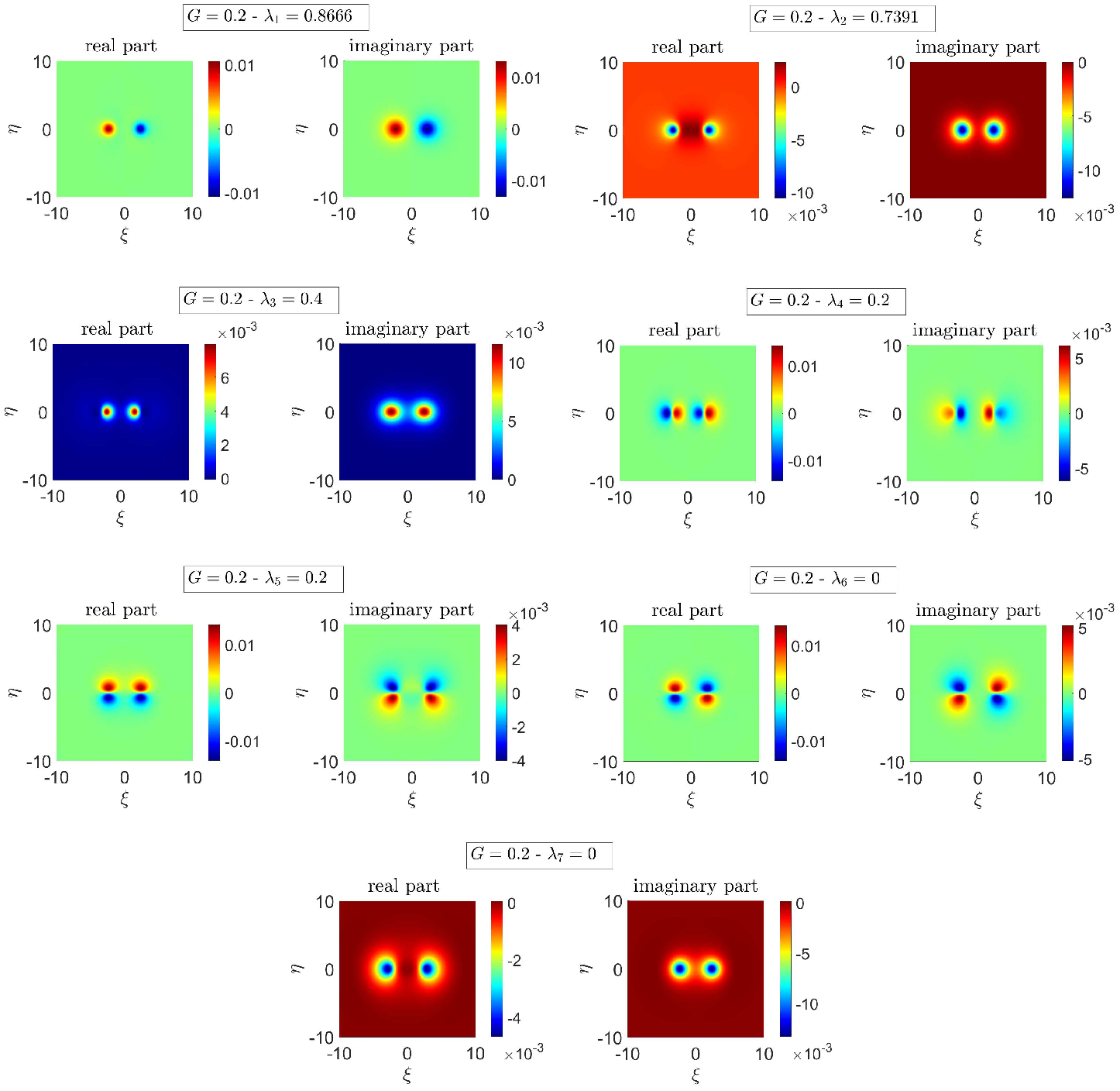}\\
    \caption{Eigenvectors of the first seven dominant real eigenvalues for a two-peaked self-similar solution with blow-up rate, $G=0.2$ ($\sigma=1.001$).
    For an analysis of the origin of the relevant eigenvectors, see the
details in the text.    }
    \label{fig:eigenvectors}
\end{figure}

Figure~\ref{fig:eigenvectors} illustrates the corresponding eigenvectors (real and imaginary parts) for a two-peaked solution with $G=0.2$. It is interesting to seek an interpretation of the corresponding
eigenvectors. The largest real part is anti-symmetric illustrating that its
addition to the relevant two-peak solution should give rise to a symmetry-breaking
effect. Indeed, we expect that the symmetry of the peaks will break, favoring one
(or the other) to be led to a single-peak collapse in the system, as the single
peak is the predominantly robust self-similar manifestation of the system's dynamics. 
Then, there are 4 eigenvalues of O$(G)$,  of which 3 pertain to symmetries:
2 at $\lambda_{4,5}=G$ are clearly associated with the derivative of the solution
and translations in $\xi$ ($\lambda_{4}$) and $\eta$ ($\lambda_{5}$), respectively. On the other hand, the 
conformal invariance, the self-similar rescaling of the solution (which, hence,
resembles the solution itself) is associated with the eigenvector of $\lambda_3$
and eigenvalue $2G$, as discussed previously in~\cite{jon2}. Additionally, $\lambda_6$
is a neutral mode associated with $O(2)$ rotational invariance of the two-peak configuration,
while $\lambda_7$ represents
the phase mode, again emulating the solution itself. The imaginary mode is
not shown here, nevertheless, we can see a concrete and complete account
of the eigenvectors associated with the symmetries, stability and possible
evolution of the two-peak state. Representing such a complete account becomes
considerably more elaborate of an effort for larger $N$, yet the mathematical
principles remain the same as in the above prototypical example.

\begin{figure}[ht!]
    \centering
    \begin{tabular}{ccc}
\includegraphics[width=0.33\linewidth]{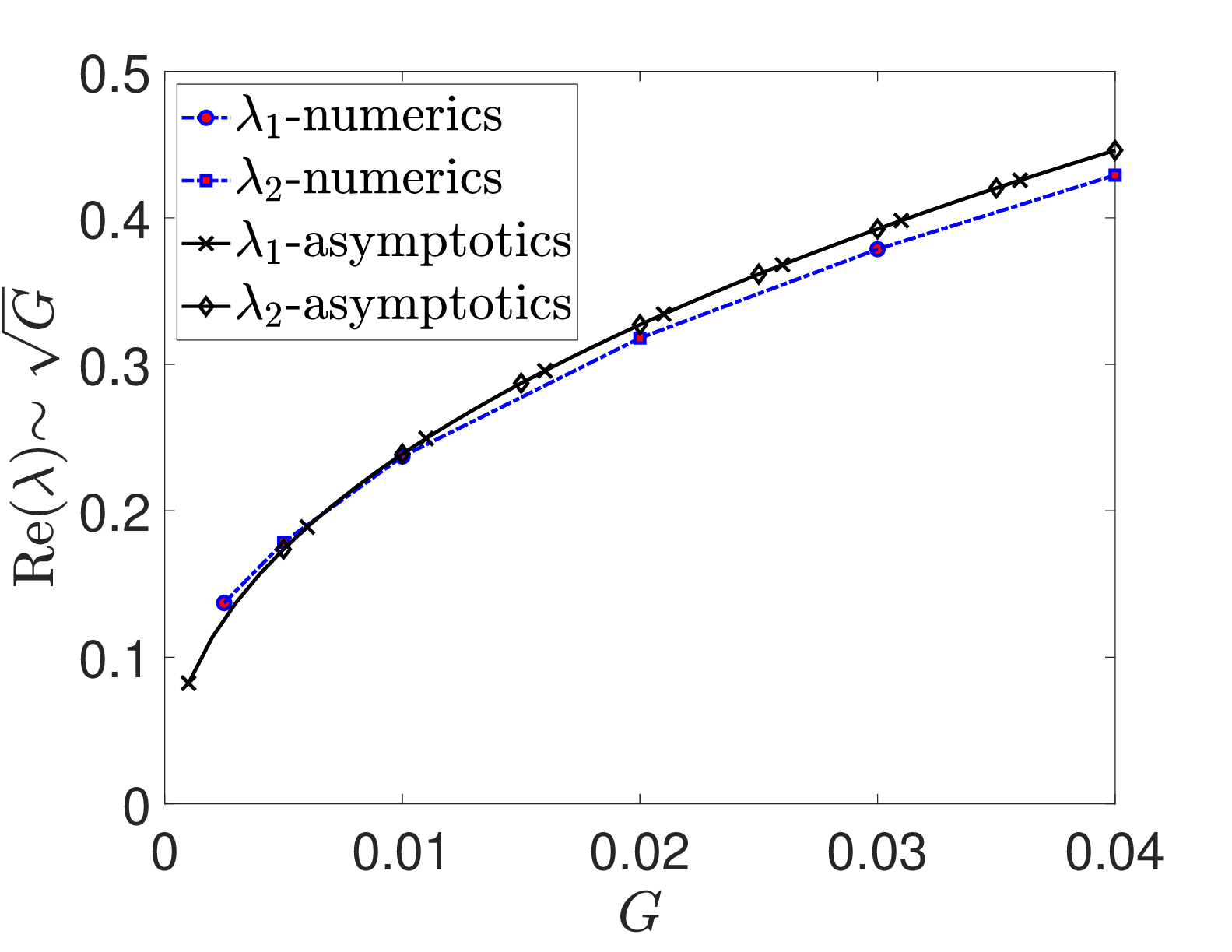}     &  \includegraphics[width=0.33\linewidth]{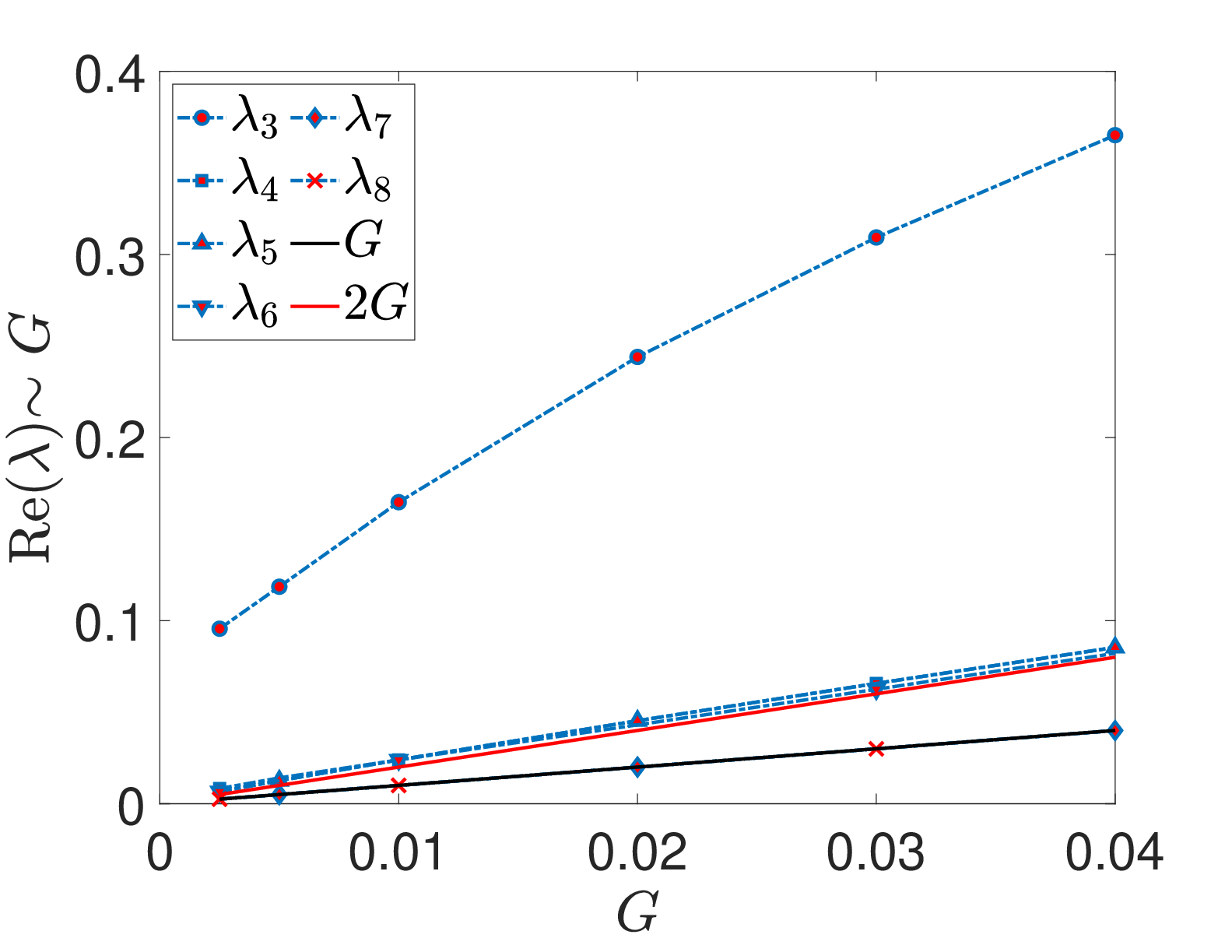} & \includegraphics[width=0.33\linewidth]{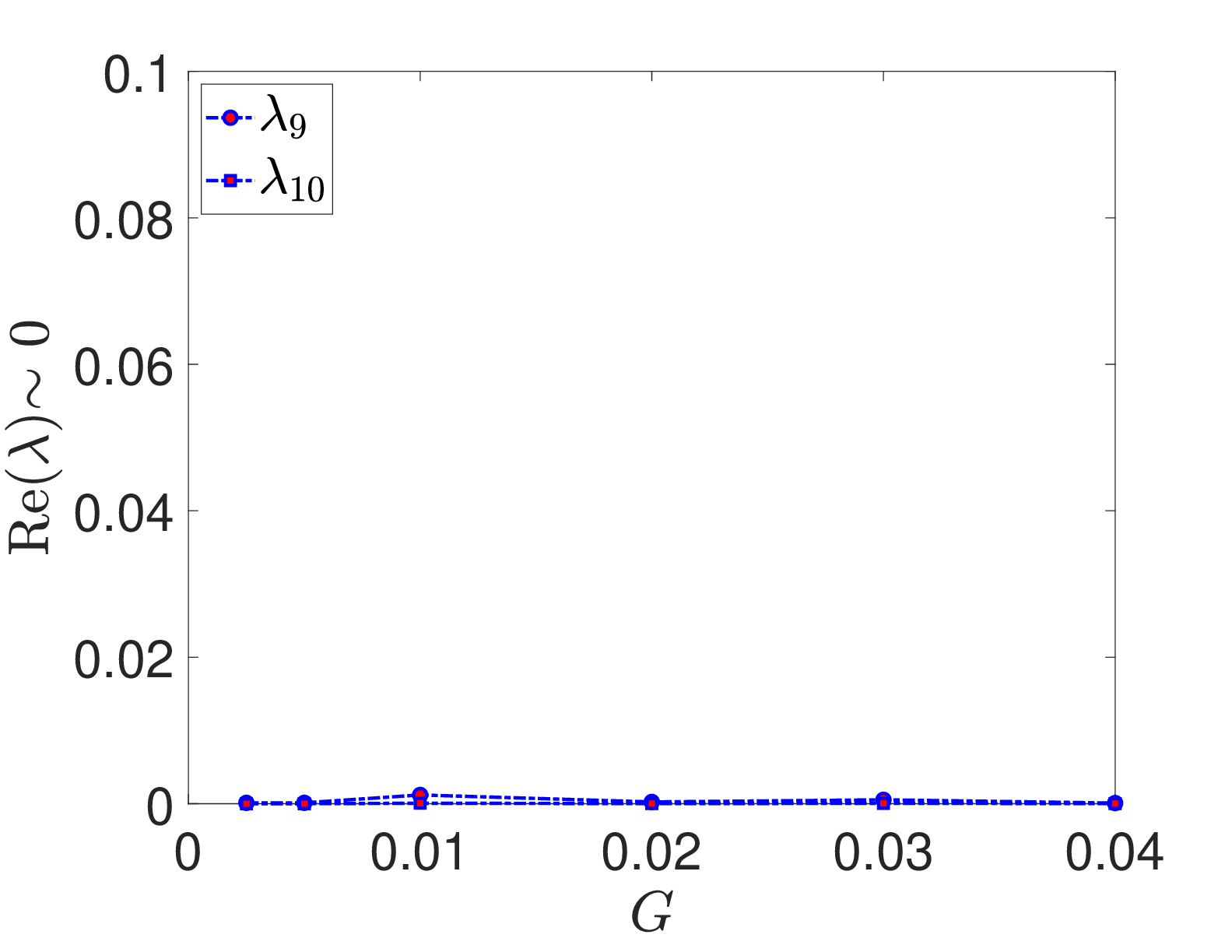} \\
     (i)    & (ii) & (iii) \\
     \end{tabular}
     \begin{tabular}{c}
            \includegraphics[width=0.33\linewidth]{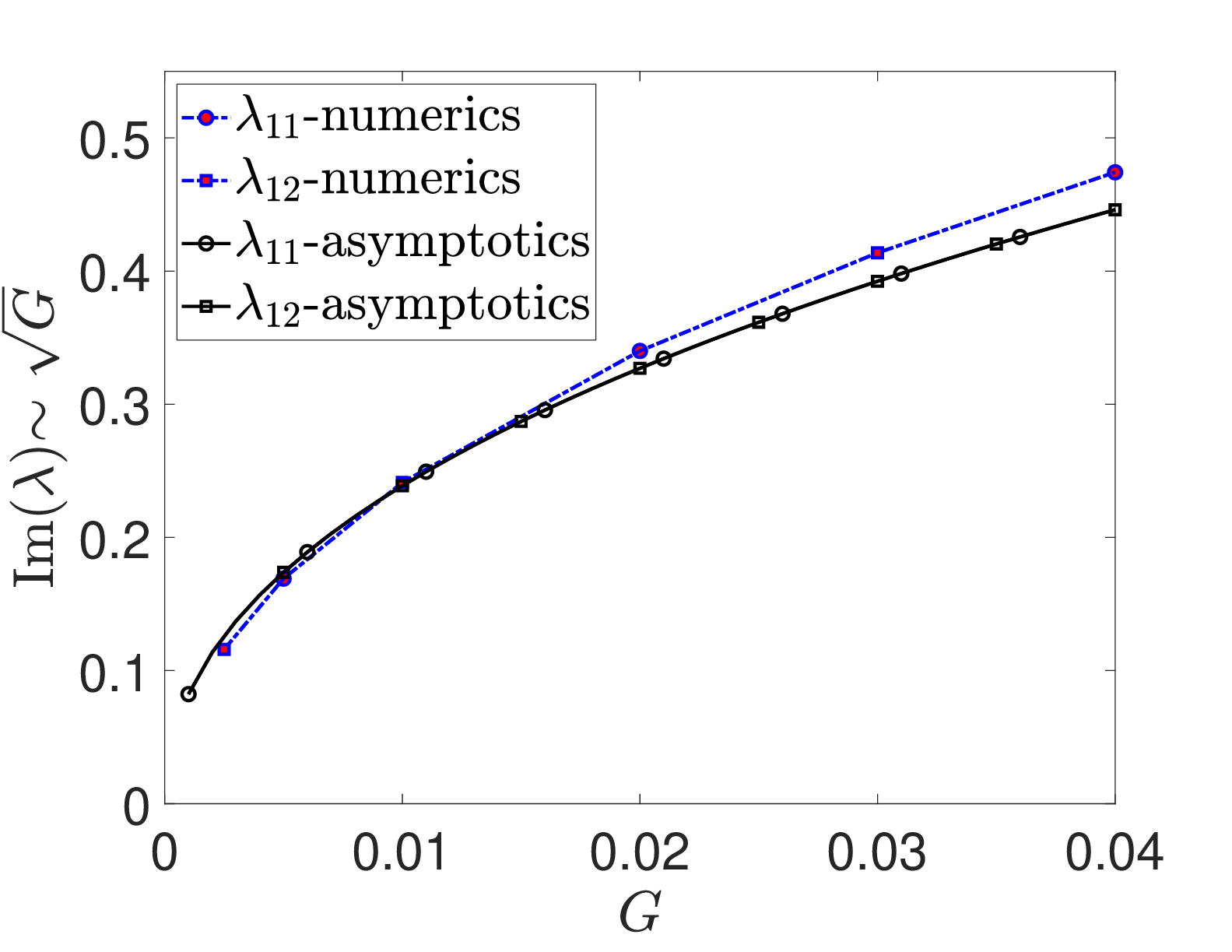} \\
     (iv) \\
    \end{tabular}
    \caption{Dependence of the dominant eigenvalues on the blowup rate $G$ for three-peaked solutions (equilateral triangle configuration).
    (i) Two dominant real eigenvalues with $\sqrt{G}$ scaling (black curves: asymptotic predictions).
    (ii) Dominant real eigenvalues with $G$ scaling (black and red lines mark $G$ and $2G$). 
    (iii) Eigenvalues $\lambda_9, \lambda_{10}$ closest to the origin.
    (iv) Purely imaginary eigenvalues $\lambda_{11}$ and $\lambda_{12}$  with $\sqrt{G}$ scaling (black lines: asymptotic predictions).}
    \label{fig:domeigen3peaks}
\end{figure}

We next present the dominant eigenvalues for three-peaked solutions with peaks at the vertices of an equilateral triangle. 
Figure~\ref{fig:domeigen3peaks}(i) shows two real eigenvalues ($\lambda_{1,2}$) with $\sqrt{G}$ scaling; panel (ii) shows six eigenvalues with $G$ scaling (including the symmetry-related eigenvalues $\lambda_6 \approx 2G$, $\lambda_{7,8} \approx G$); panel (iii) shows $\lambda_{9,10} \approx 0$; and panel (iv) shows two purely imaginary eigenvalues ($\lambda_{11,12}$) with $\sqrt{G}$ scaling. 
Thus, we count $2 \times (N=3)-2=4$ eigenvalues with $\sqrt{G}$ scaling and $2\times(N=3)=6$ eigenvalues with $G$ scaling, in addition to the two
symmetry modes (due to rotation and phase invariance). 


\begin{figure}[ht!]
    \centering
    \begin{tabular}{ccc}
\includegraphics[width=0.33\linewidth]{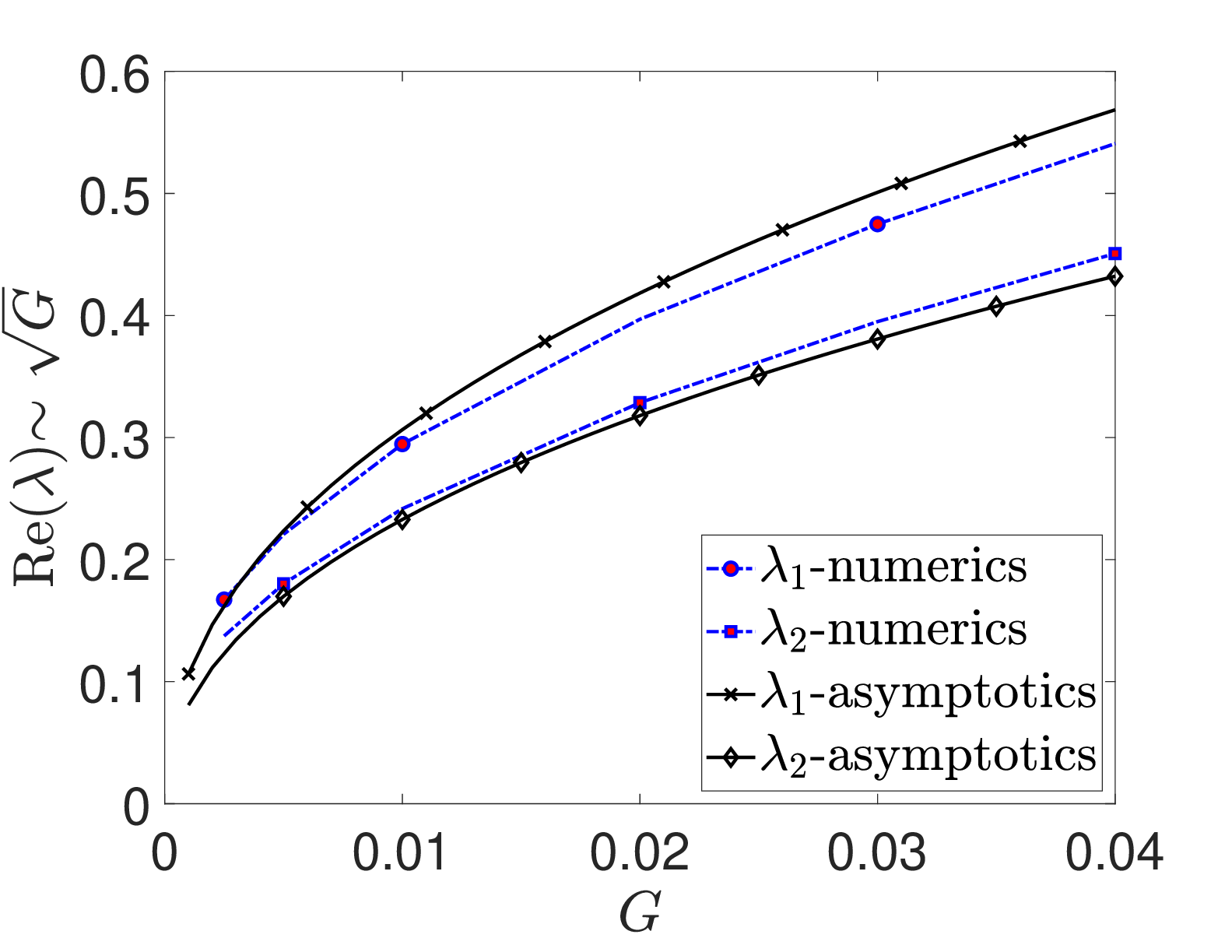}     &  \includegraphics[width=0.33\linewidth]{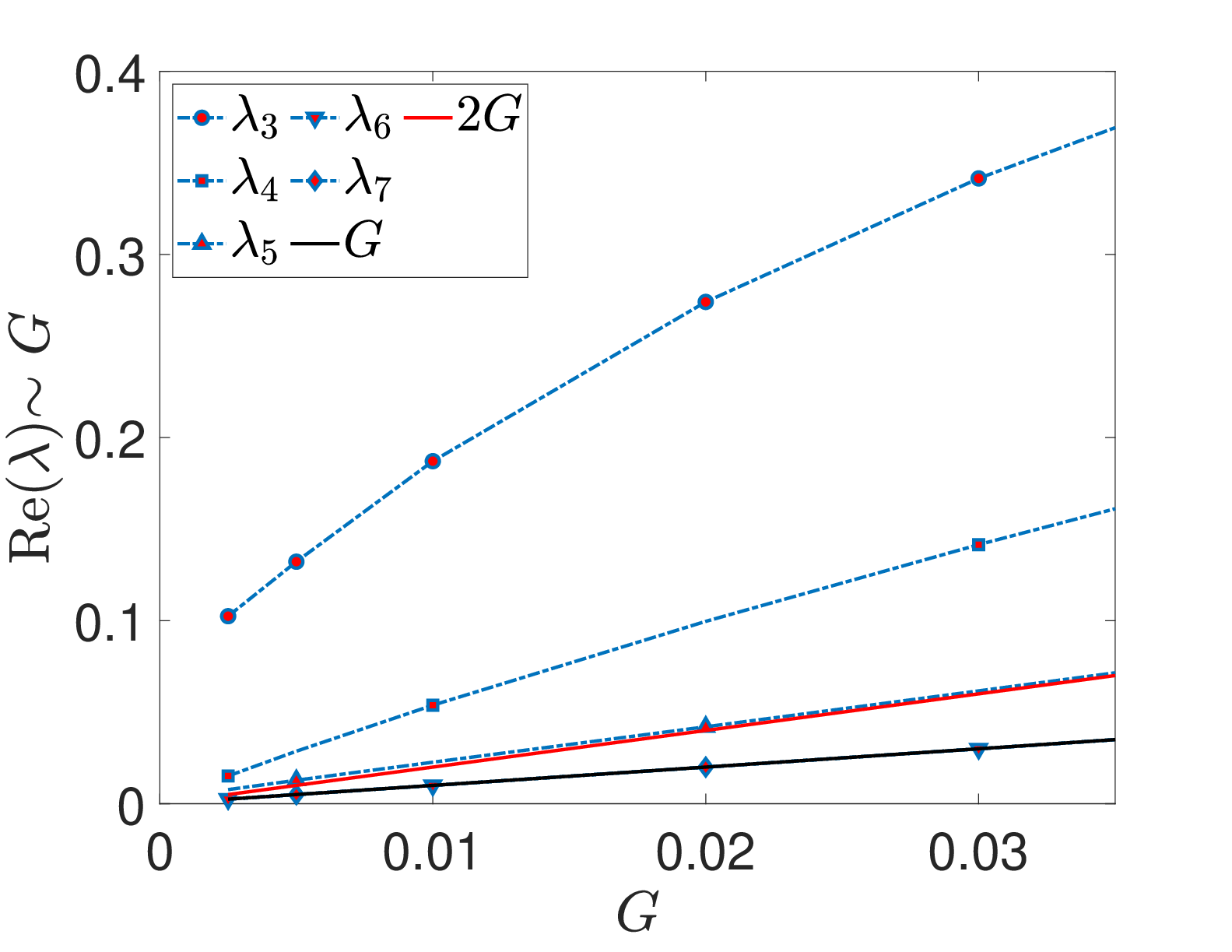} & \includegraphics[width=0.33\linewidth]{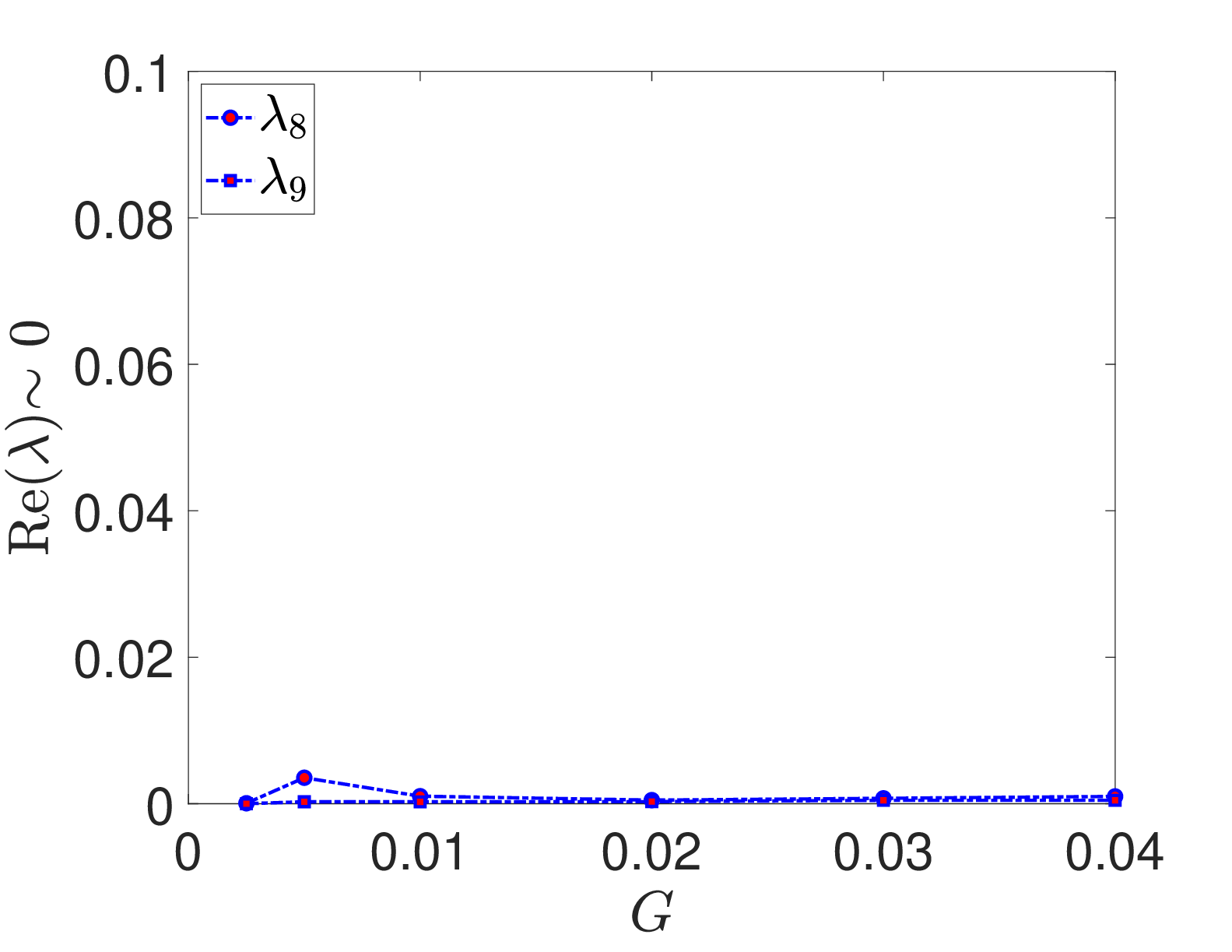} \\
     (i)    & (ii) & (iii) 
     \end{tabular}
     \begin{tabular}{cc}
     \includegraphics[width=0.33\linewidth]{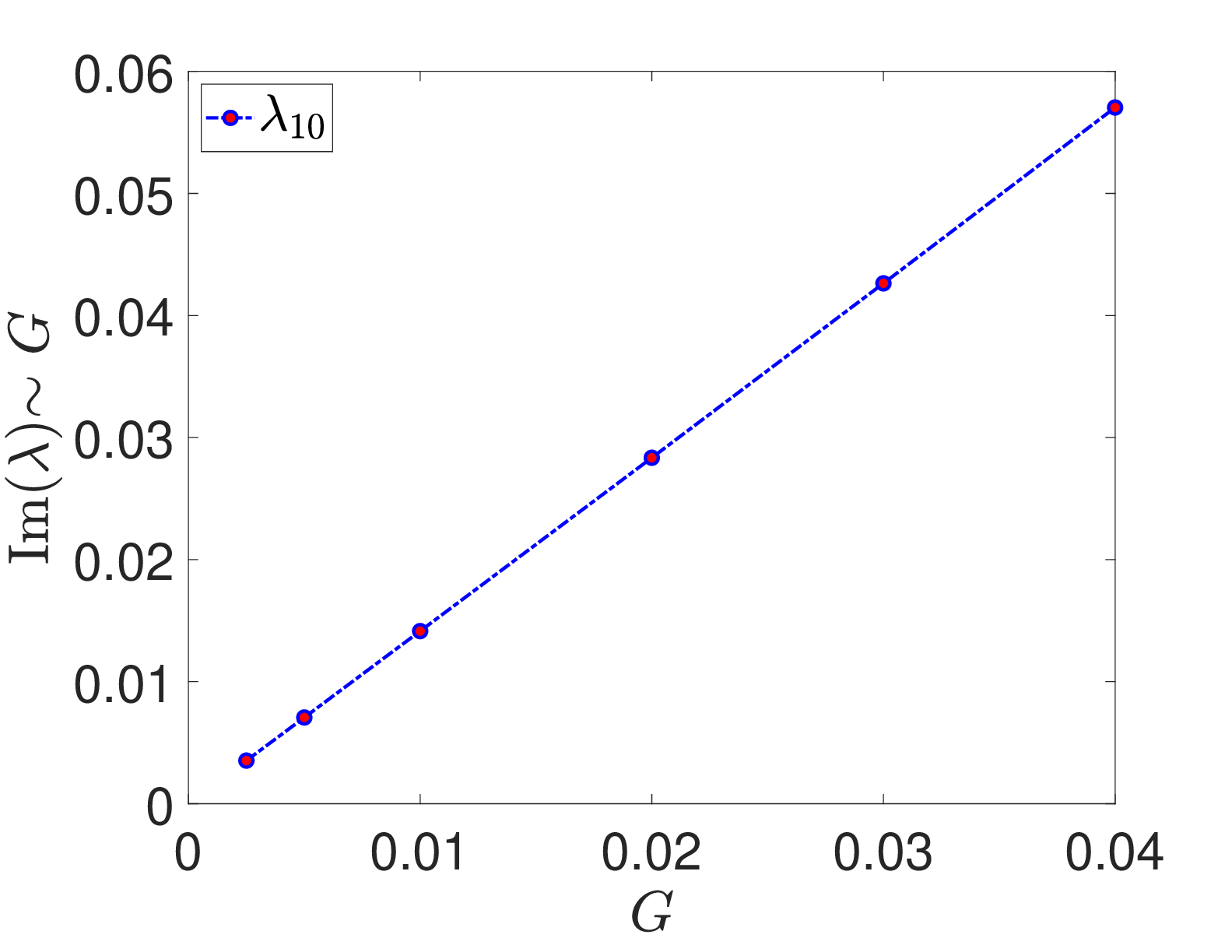} & \includegraphics[width=0.33\linewidth]{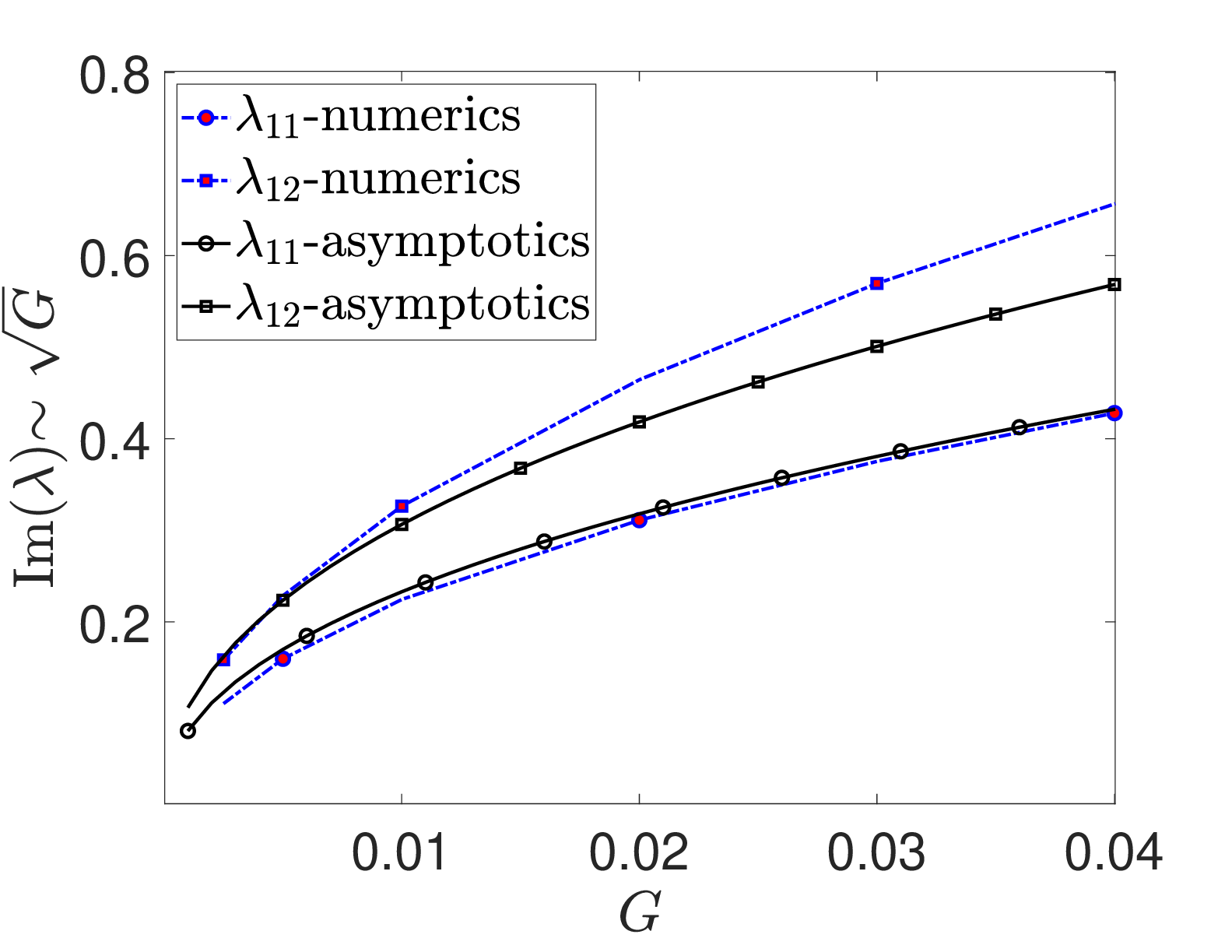}\\
     (iv) & (v) \\     
     \end{tabular}     
    \caption{Dependence of dominant eigenvalues on the blowup rate $G$ for three-peaked solutions with one peak at the domain center.
    (i) Dominant real eigenvalues with $\sqrt{G}$ scaling (black curves: asymptotic predictions)
    (ii) Real eigenvalues with $G$ scaling (black and red lines indicate $G$ and $2G$). 
    (iii) Eigenvalues $\lambda_8, \lambda_{9}$ closest to the origin.
    (iv) One purely imaginary eigenvalue $\lambda_{10}$ with $G$ scaling.
    (v) Purely imaginary eigenvalues $\lambda_{11}$, $\lambda_{12}$ with $\sqrt{G}$ scaling (black lines: asymptotic predictions).}
    \label{fig:domeigen3peaks_inline}
\end{figure}

For the case of three peaks
with one peak at the domain center, Fig.~\ref{fig:domeigen3peaks_inline} yields the dominant eigenvalues.
We find two real eigenvalues ($\lambda_{1,2}$) with $\sqrt{G}$ scaling, five real eigenvalues with $G$ scaling, one purely imaginary eigenvalue ($\lambda_{10}$) with $G$ scaling and two purely imaginary eigenvalues with $\sqrt{G}$ scaling ($\lambda_{11,12}$). 
Relative to the equilateral case, the number of real eigenvalues with $G$ scaling is reduced to five, but including the imaginary eigenvalue with $G$ scaling still gives six eigenvalues with $G$ scaling in total.
(in total: $2\times(N=3)-2=4$ eigenvalues with $\sqrt{G}$ scaling, and  $2\times(N=3)=6$ with $G$ scaling).

\begin{figure}[ht!]
    \centering
    \begin{tabular}{ccc}
\includegraphics[width=0.33\linewidth]{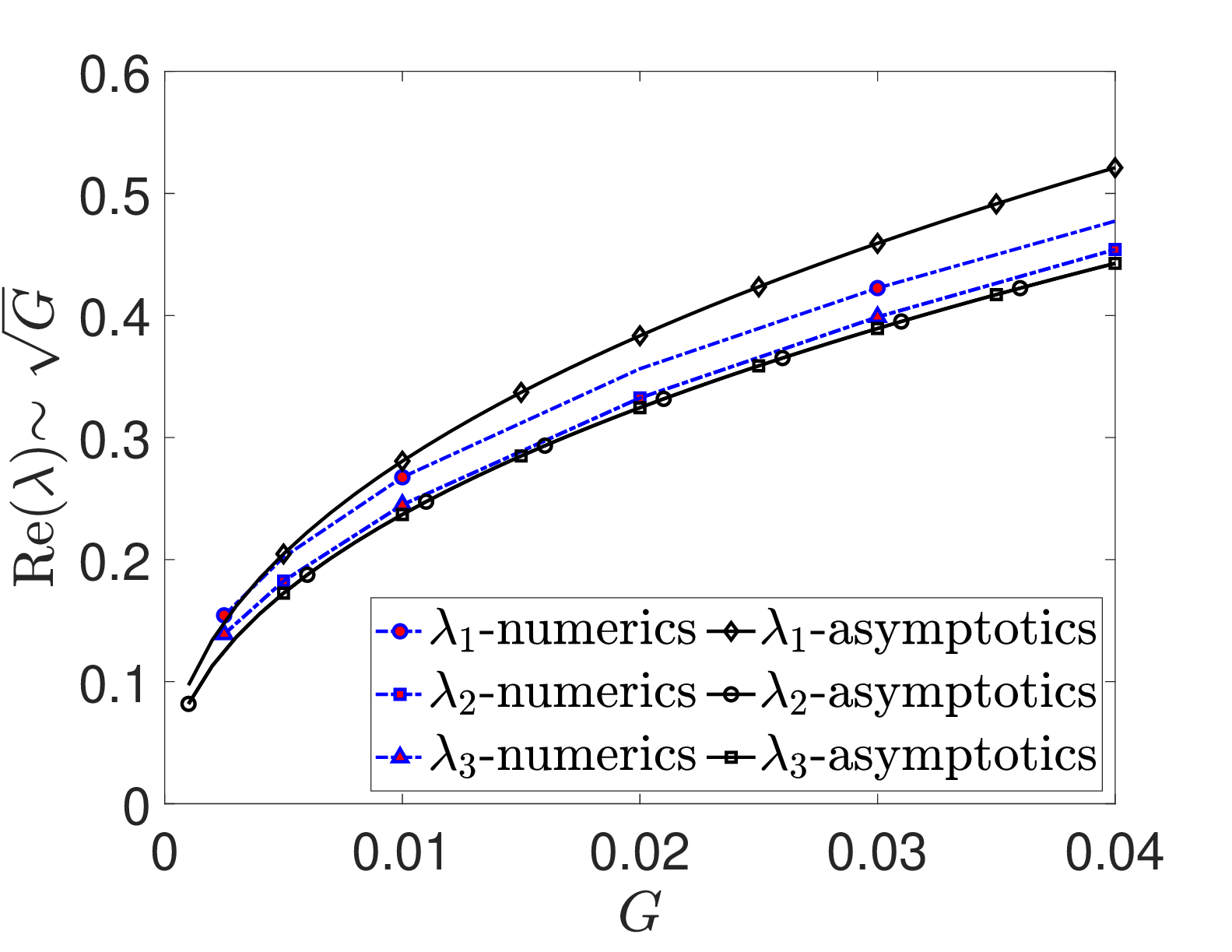}     &  \includegraphics[width=0.33\linewidth]{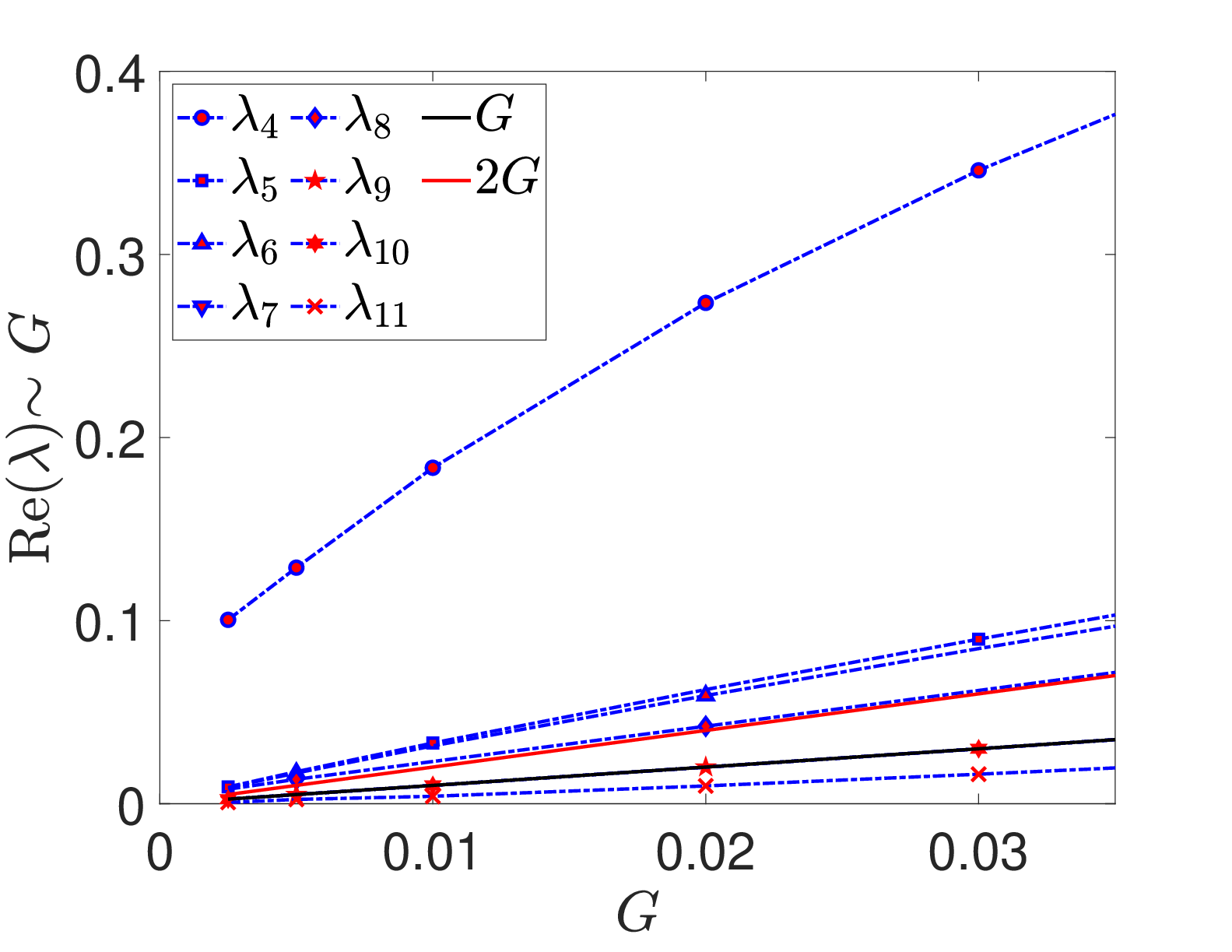} & \includegraphics[width=0.33\linewidth]{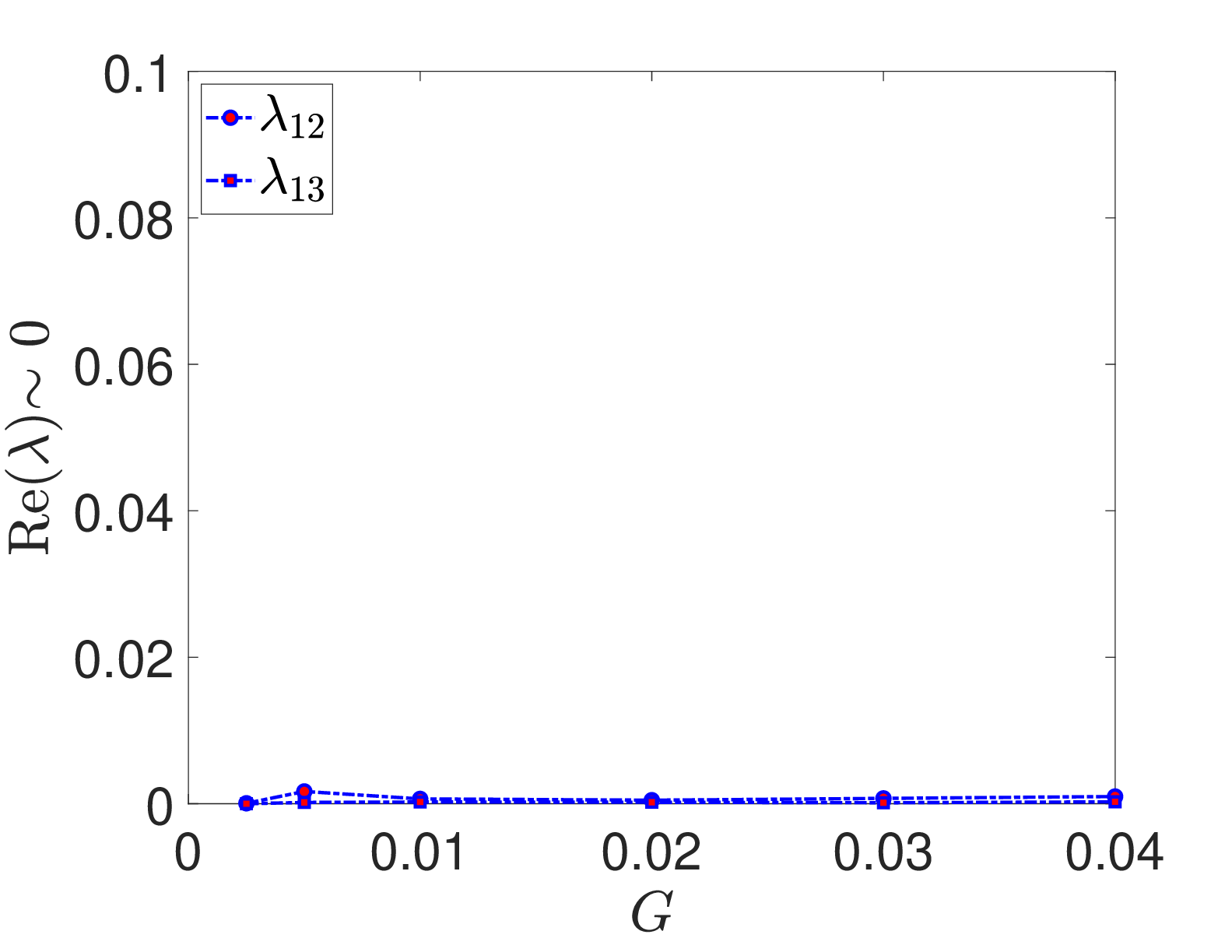}\\
     (i)    & (ii) & (iii) 
     \end{tabular}
     \begin{tabular}{c}
          \includegraphics[width=0.33\linewidth]{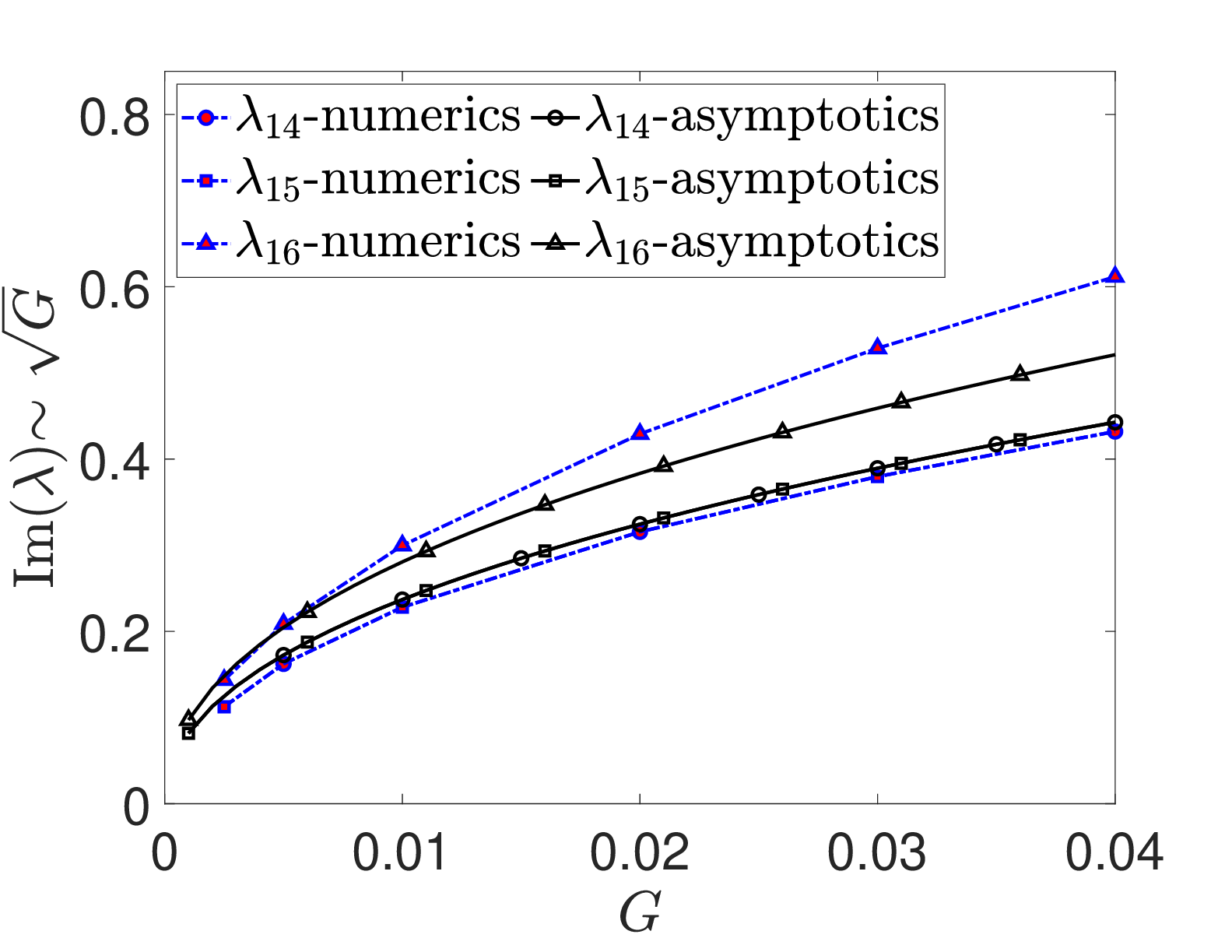} \\
     (iv)
    \end{tabular}
    \caption{Dependence of dominant eigenvalues on the blowup rate $G$ for four-peaked solutions (square configuration).
    (i) Dominant real eigenvalues with $\sqrt{G}$ scaling (black curves: asymptotic predictions).
    (ii) Dominant real eigenvalues with $G$ scaling;
    black and red lines indicate $G$ and $2G$, respectively. 
    (iii) Eigenvalues $\lambda_{12}, \lambda_{13}$ closest to the origin.
    (iv) Purely imaginary eigenvalues $\lambda_{14}$, $\lambda_{15}$ and $\lambda_{16}$ with $\sqrt{G}$ scaling (black lines: asymptotic predictions).}
    \label{fig:domeigen4peaks}
\end{figure}

\begin{figure}[ht!]
    \centering
    \begin{tabular}{ccc}
\includegraphics[width=0.33\linewidth]{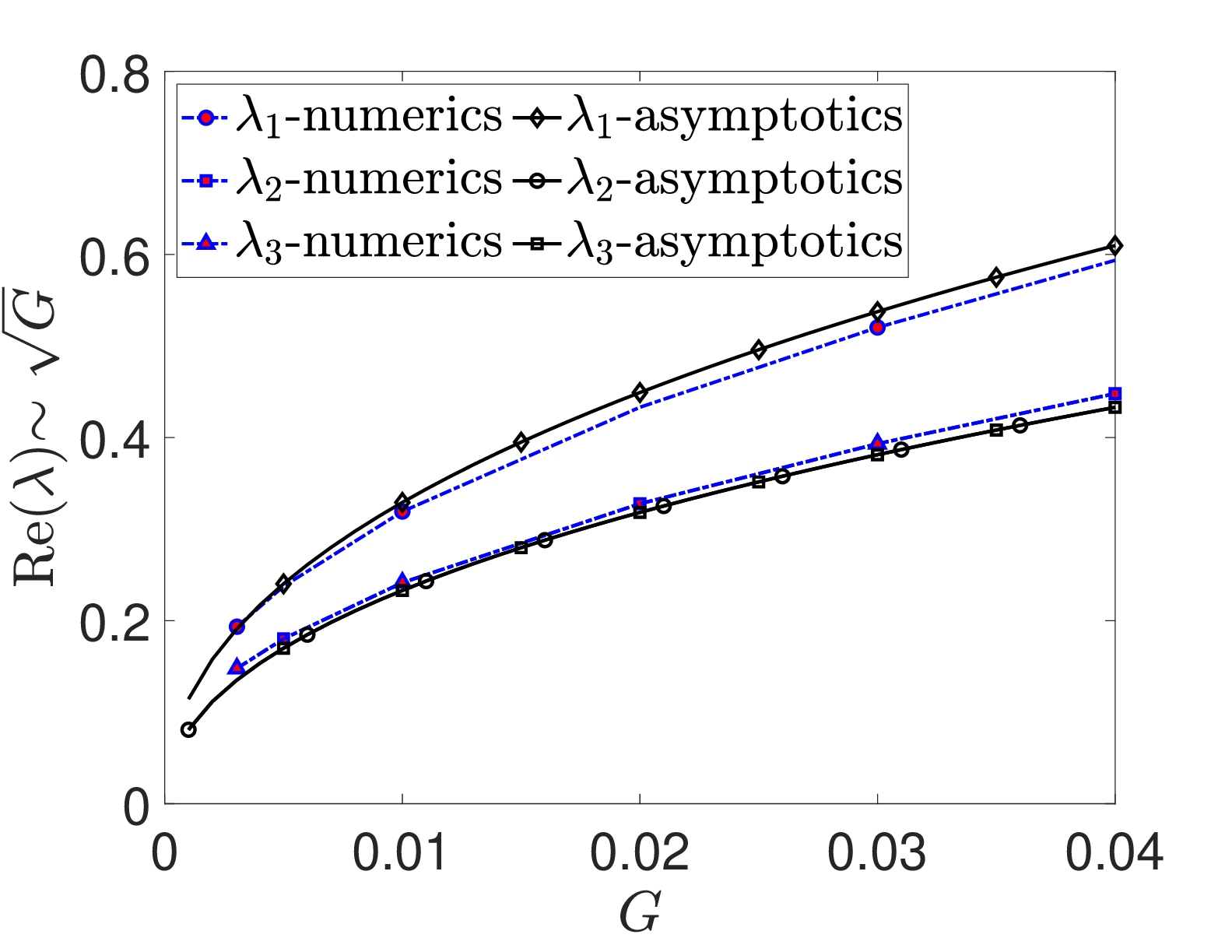}     &  \includegraphics[width=0.33\linewidth]{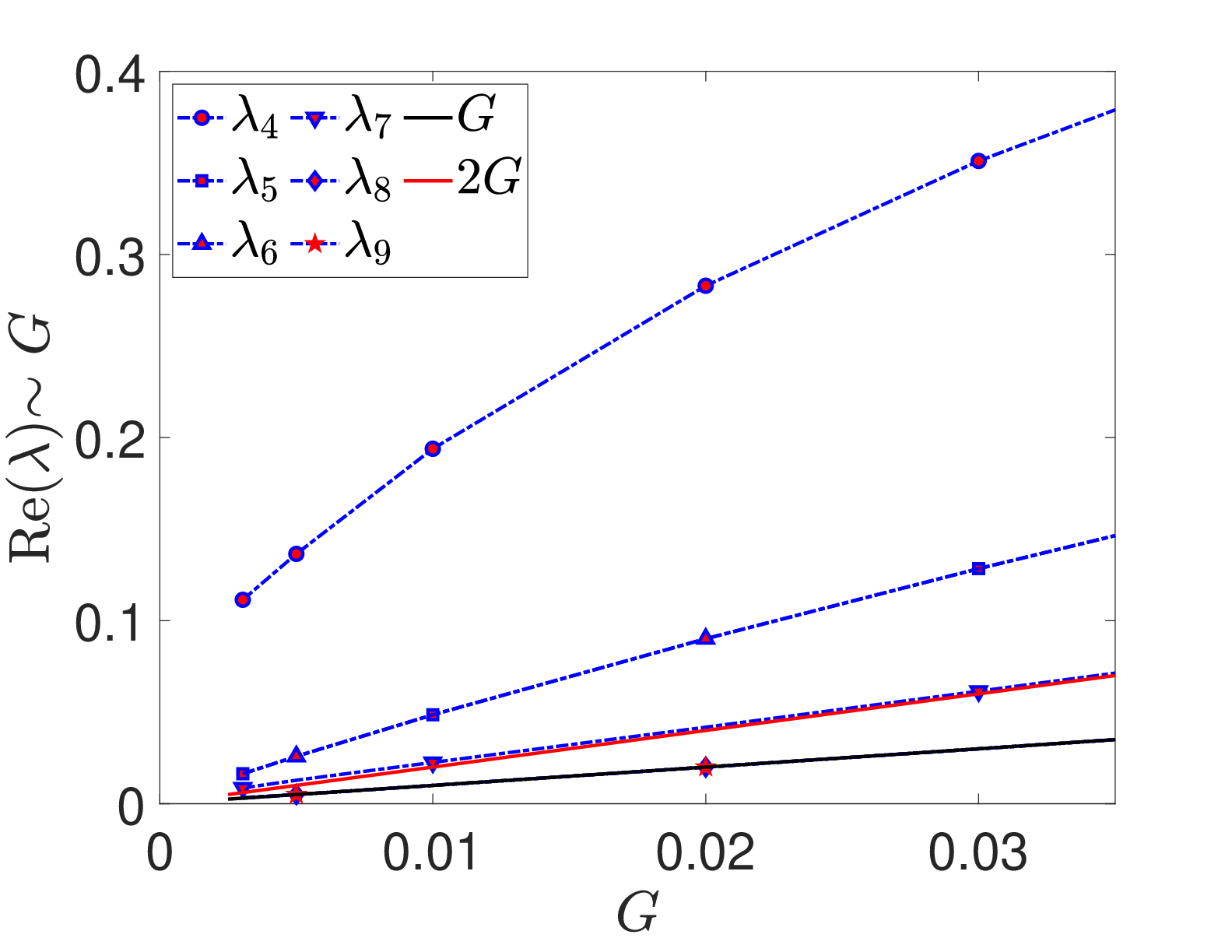} & \includegraphics[width=0.33\linewidth]{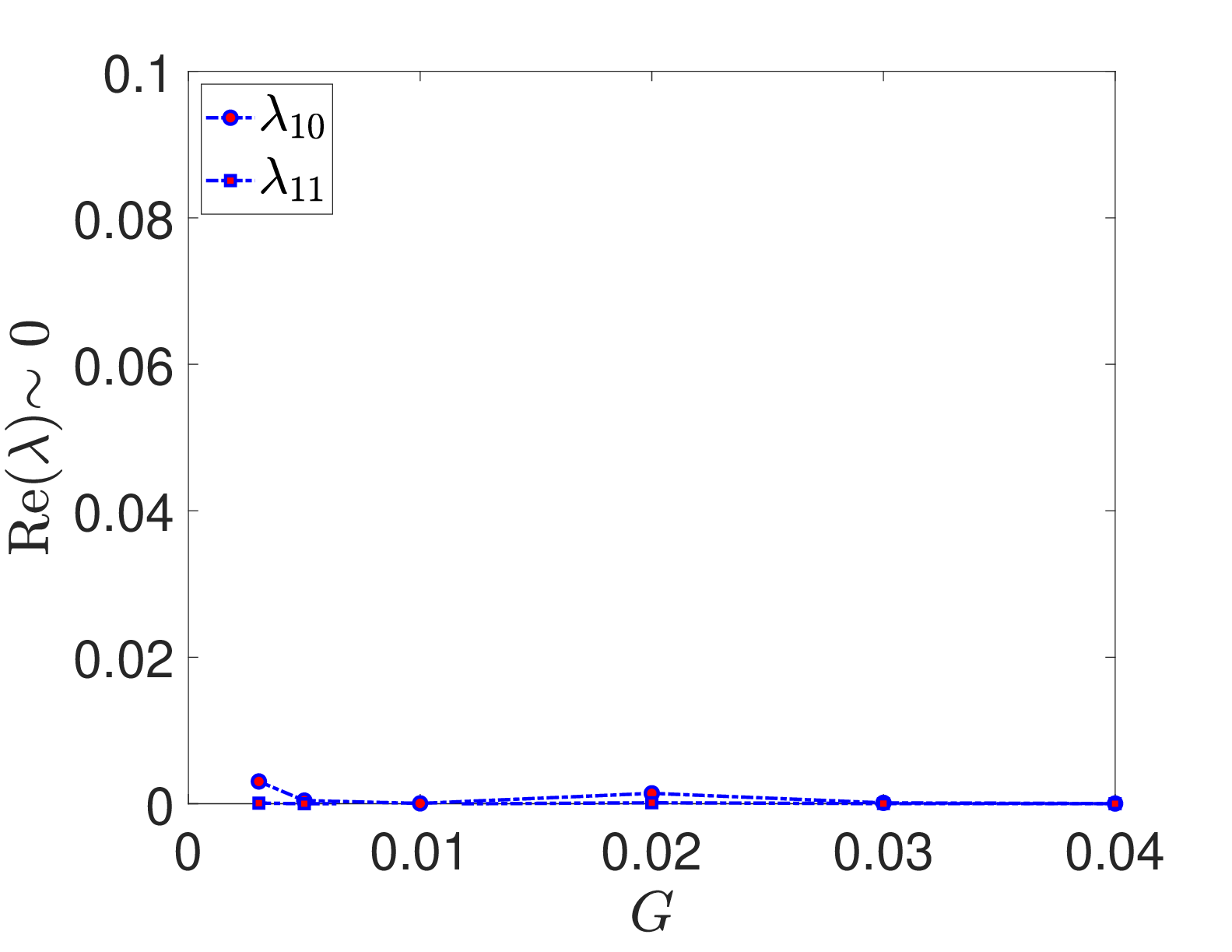}  \\
     (i)    & (ii) & (iii) \\
     \end{tabular}
     \begin{tabular}{cc}
     \includegraphics[width=0.33\linewidth]{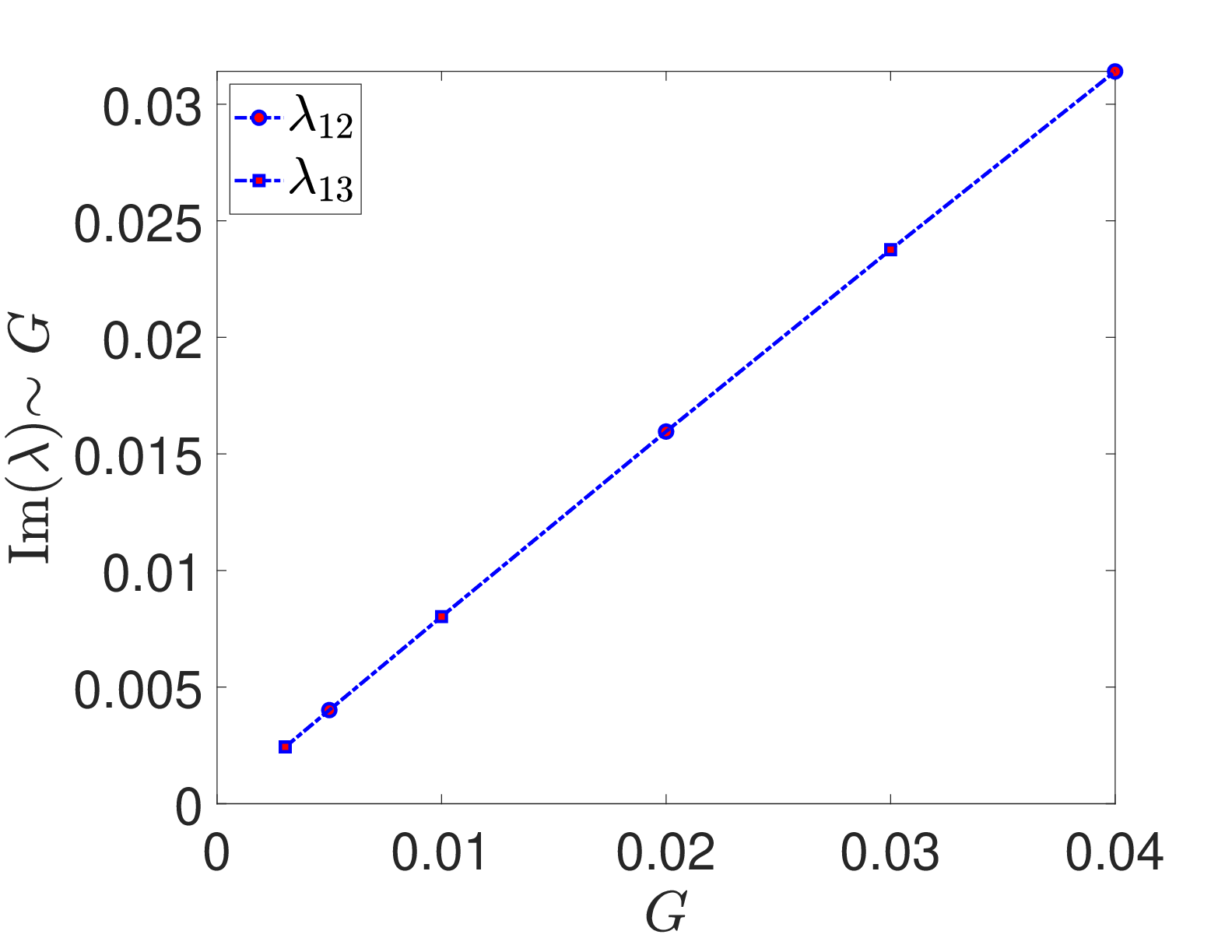} & \includegraphics[width=0.33\linewidth]{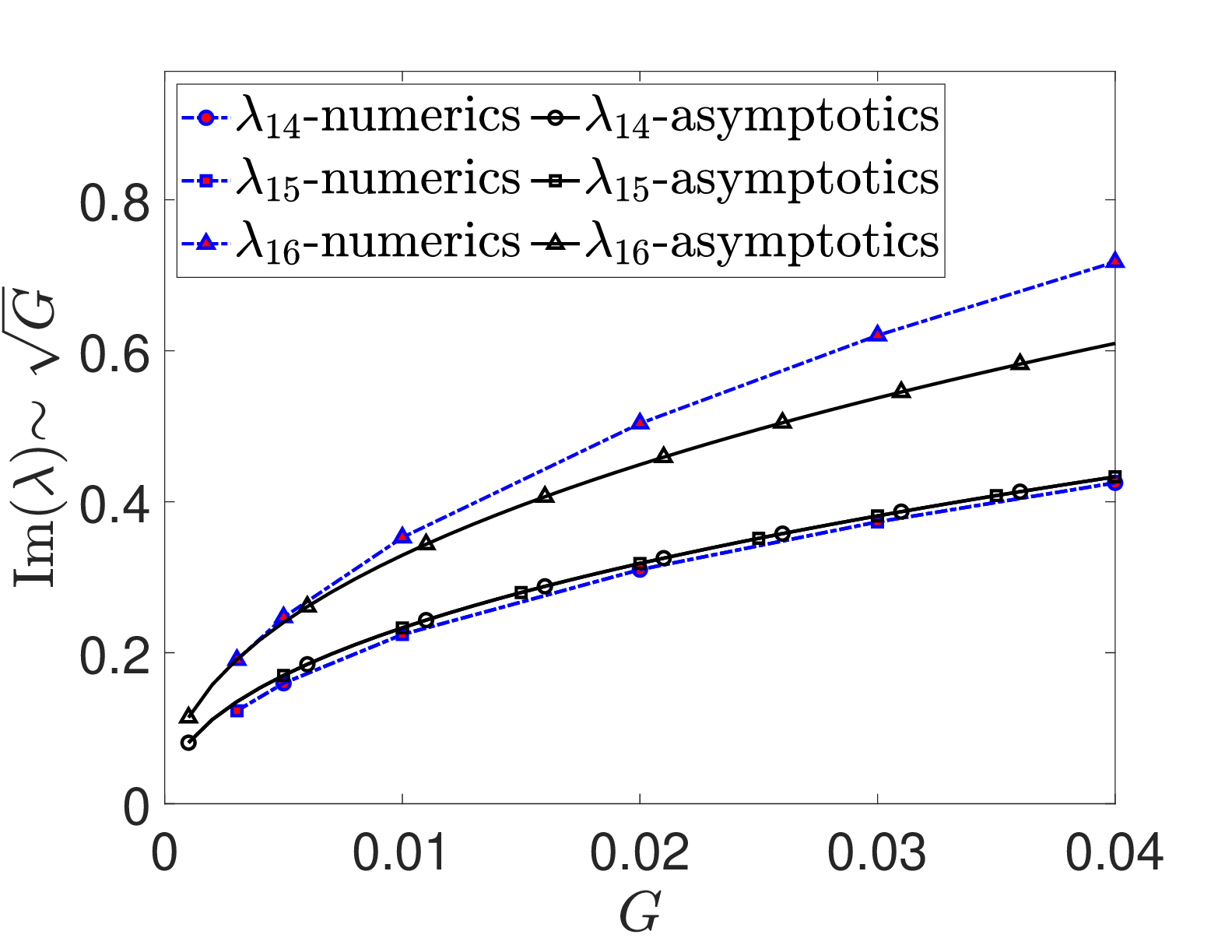} \\
     (iv)    & (v)\\     
     \end{tabular}     
    \caption{Dependence of dominant eigenvalues on the blowup rate $G$ for four-peaked solutions with one peak at the domain center and three at the vertices of an equilateral triangle.
    (i) Dominant real eigenvalues with $\sqrt{G}$ scaling (black curves: asymptotic predictions).
    (ii) Dominant real eigenvalues with $G$ scaling;
    black and red lines mark $G$ and $2G$, respectively. 
    (iii) Eigenvalues $\lambda_{10}, \lambda_{11}$ closest to the origin.
    (iv) Purely imaginary eigenvalues $\lambda_{12}$ and $\lambda_{13}$ with $G$ scaling.
    (v) Purely imaginary eigenvalues $\lambda_{14}$, $\lambda_{15}$ and $\lambda_{16}$ with $\sqrt{G}$ scaling (black lines: asymptotic predictions).}
    \label{fig:domeigen3_plus_1_peaks}
\end{figure}

\begin{figure}[ht!]
    \centering
    \begin{tabular}{ccc}
\includegraphics[width=0.33\linewidth]{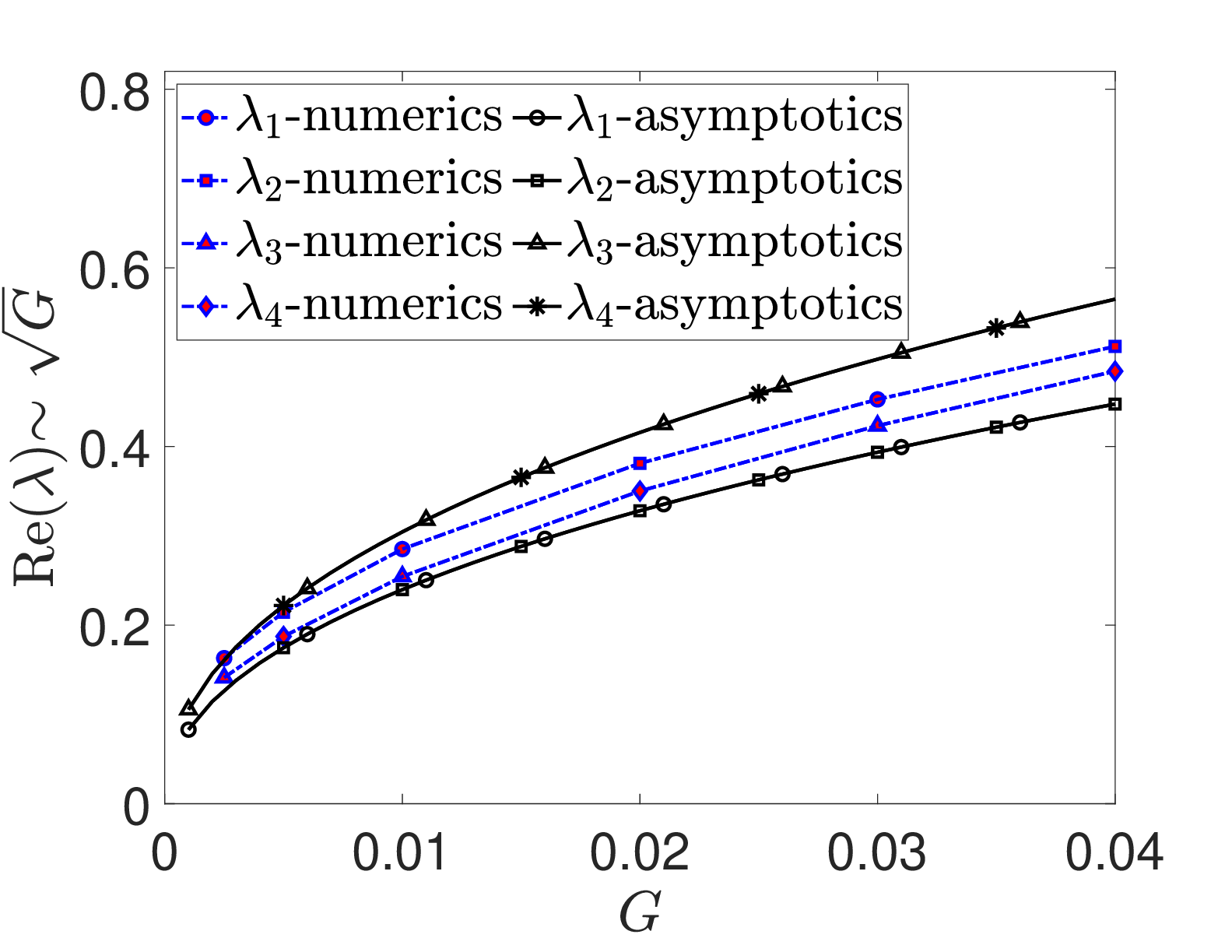}     &  \includegraphics[width=0.33\linewidth] {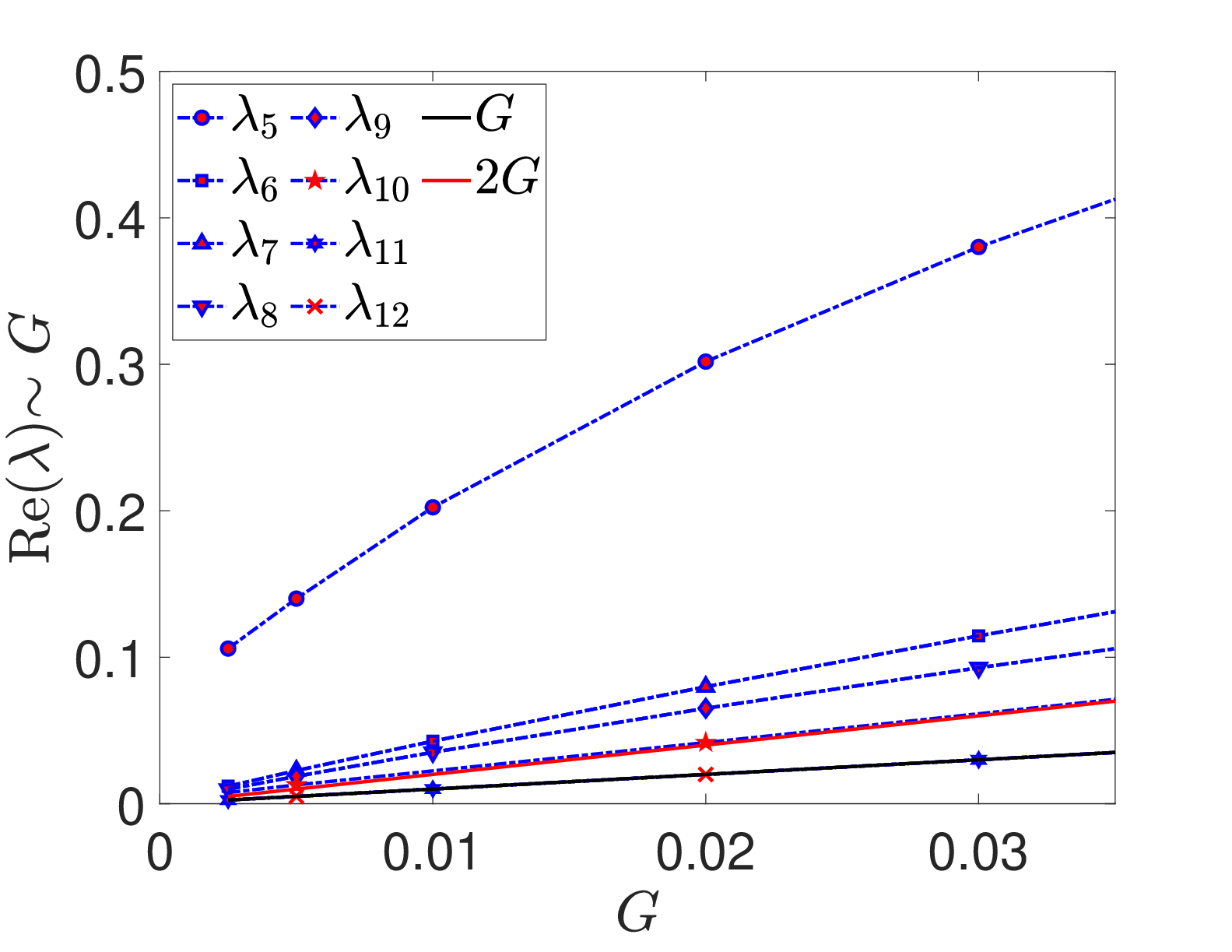} &
\includegraphics[width=0.33\linewidth]{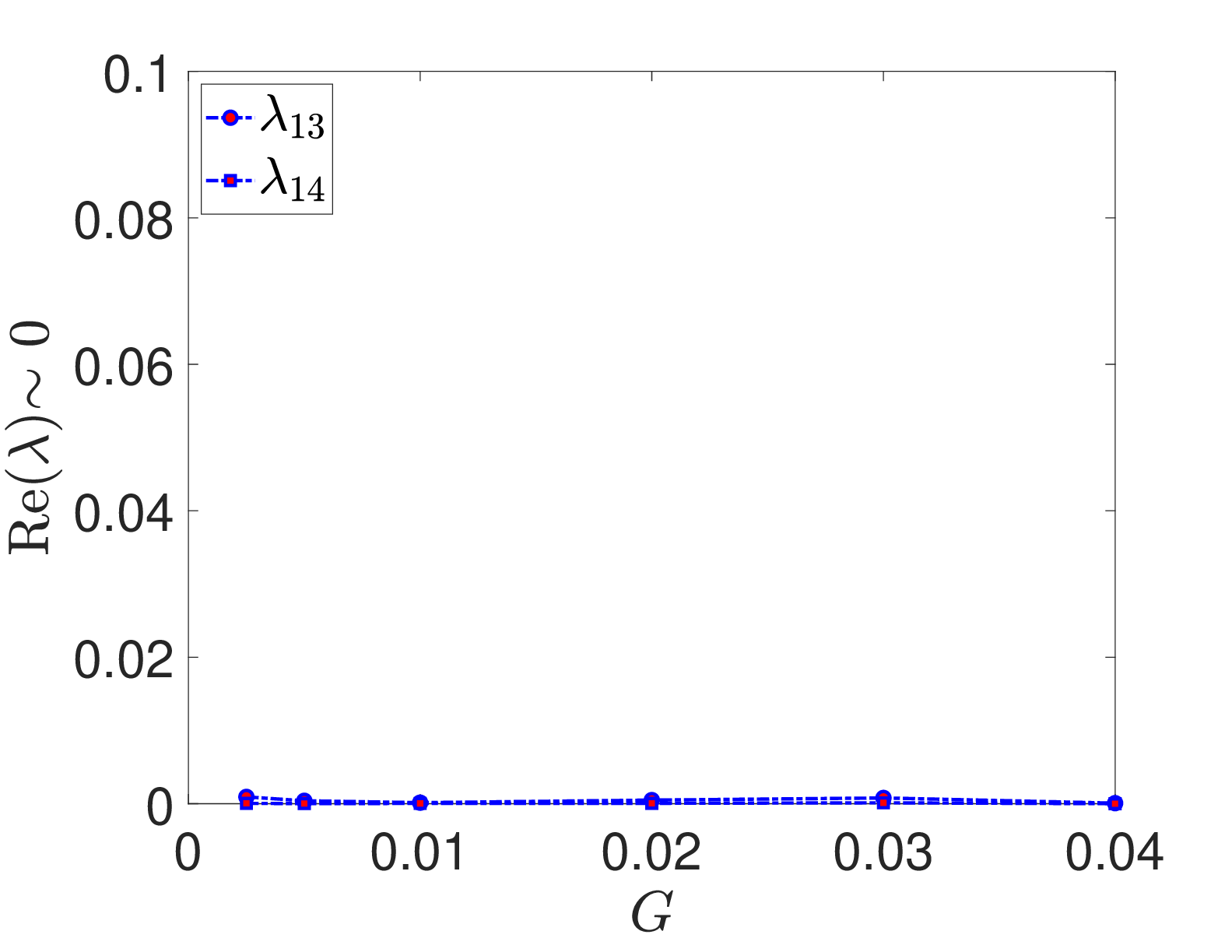} \\
     (i)    & (ii) & (iii) \\
     \end{tabular}
     \begin{tabular}{cc}
     \includegraphics[width=0.33\linewidth]{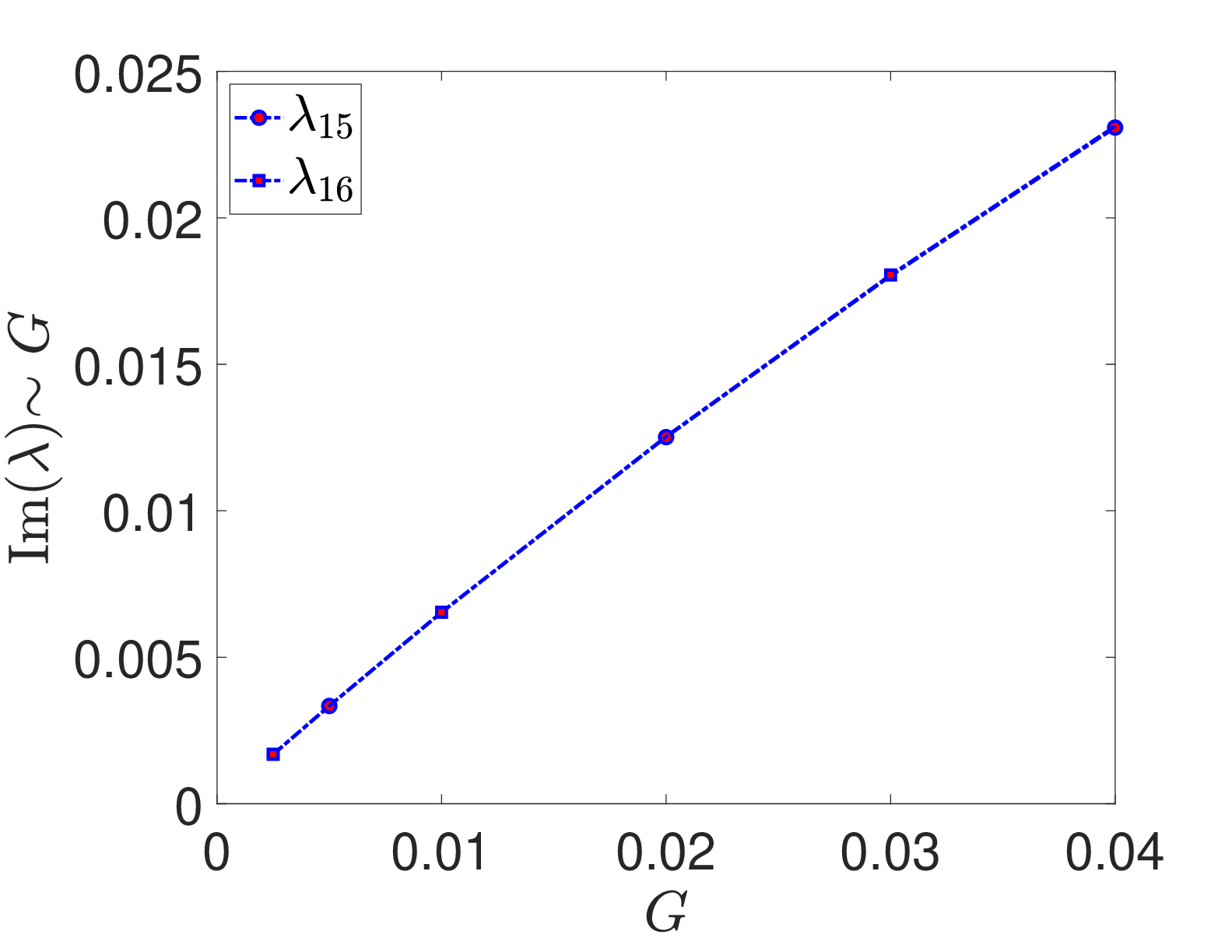} &
      \includegraphics[width=0.33\linewidth]{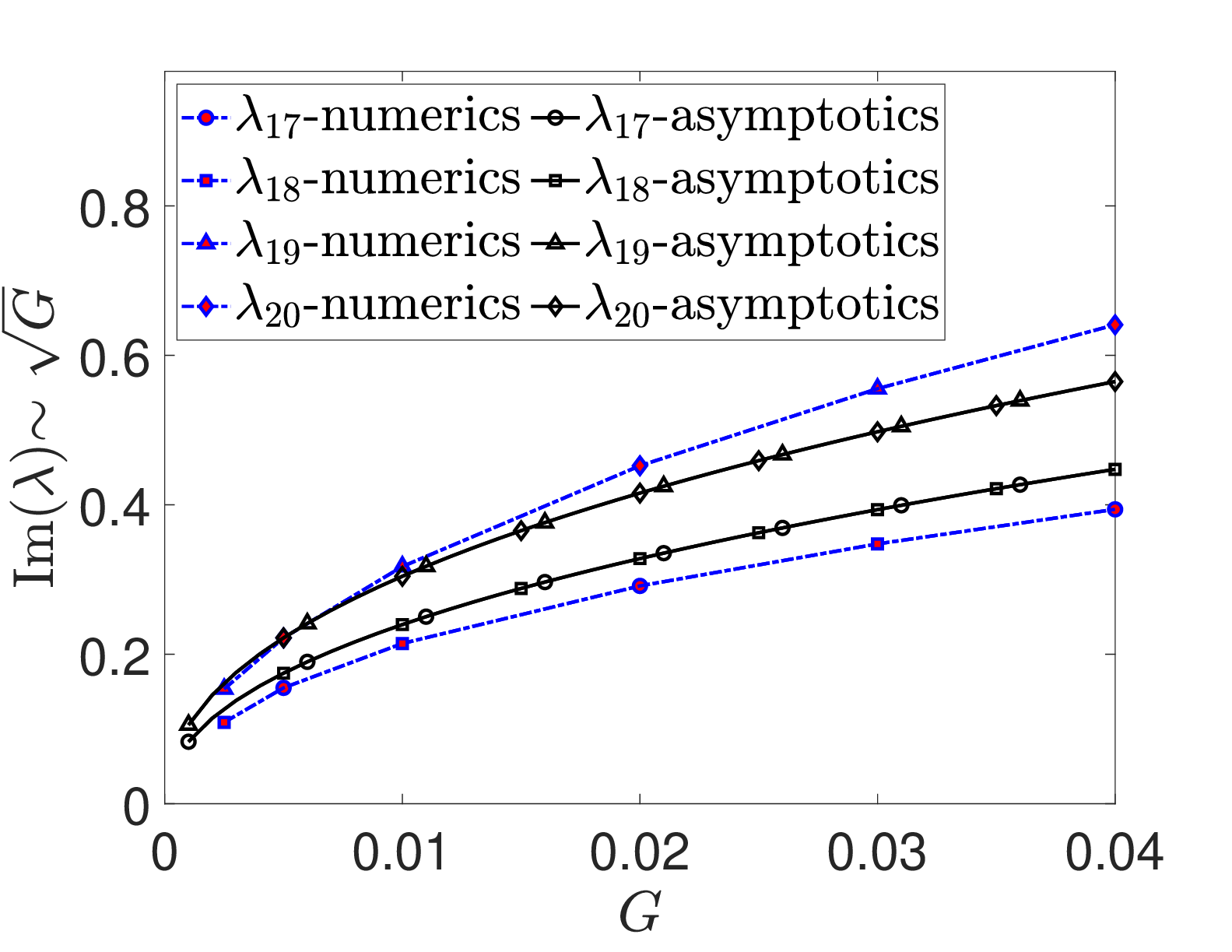} \\
     (iv) & (v) \\     
     \end{tabular}
     \caption{Dependence of dominant eigenvalues on the blowup rate $G$ for five-peaked solutions (pentagon configuration).
    (i) Dominant real eigenvalues with $\sqrt{G}$ scaling (black curves: asymptotic predictions).
    (ii) Dominant real eigenvalues with $G$ scaling;
    black and red lines indicate $G$ and $2G$, respectively. 
    (iii) Eigenvalues $\lambda_{13}, \lambda_{14}$ closest to the origin.
    (iv) Purely imaginary eigenvalues $\lambda_{15}$ and $\lambda_{16}$ with $G$ scaling.
    (v) Purely imaginary eigenvalues $\lambda_{17}$, $\lambda_{18}$, $\lambda_{19}$ and $\lambda_{20}$ with $\sqrt{G}$ scaling (black lines: asymptotic predictions).}
    \label{fig:domeigen5peaks}
\end{figure}

Similarly, we now present the dominant eigenvalues for configurations with $N=4$ and 5 peaks.
Across these cases, we consistently observe the same counting for modes with $\sqrt{G}$ and $G$ scaling. 
Figures~\ref{fig:domeigen4peaks} and \ref{fig:domeigen3_plus_1_peaks} depict results for four-peaked solutions in two configurations: peaks at the corners of a square, and a ''3+1'' configuration with  three at the vertices of an equilateral triangle and one peak at the barycenter thereof.
In both cases, we count $2\times(N=4)-2=6$ eigenvalues with $\sqrt{G}$ scaling 
(half real and half imaginary) and $2\times(N=4)=8$ with $G$ scaling, while,
as usual, two eigenvalues remain at the origin, combining for a total of
$4 \times 4=16$ modes. 
For the five-peaked solutions with peaks at the vertices of a pentagon
(Fig.~\ref{fig:domeigen5peaks}),
we count $2\times(N=5)-2=8$ dominant modes with $\sqrt{G}$ scaling (again half real and half imaginary) 
and $2\times(N=5)=10$ modes with $G$ scaling and $2$ vanishing ones for a
total of $4\times 5=20$ modes.


\section{Asymptotic Analysis of the multi-peaked steady state} 
\label{asymptotics}

In order to theoretically examine the configurations of interest,
we suppose that $\sigma$ is exponentially close to $1$, 
and look for a solution in the form
\[ \V =  \Ve
    + \Vexp,\] 
{where $\Ve$ is the regular power series expansion and} $\Vexp$ is transcendentally small in $G$. Then
\[
   \del^2\Ve+ |\Ve|^{2}\Ve - \Ve 
   + \frac{G^2 |\boldsymbol{\xi}|^2 }{4}\Ve= 0,\]
where we have dropped the subscript on $\del$.

We look for a general multi-peaked solution, with peaks located at
$\bxi =  \bX_i(G)\gg 1$, $i = 1,\ldots,N$. 
We aim to determine the $\bX_i$ along with the
dependence of $G$ on $\sigma$. Our expectation is that  $|\bX_i-\bX_j| = O(\log
G)$. 

\subsection{Near field in the vicinity of the $i$\,th peak}
\label{asynear}
We write $\bxi = \bX_i + \bx$ to give  
\beq
   \del^2\Ve+ |\Ve|^{2}\Ve - \Ve 
   + \frac{G^2 |\bx+\bX_i|^2 }{4}\Ve= 0.
   \label{inner}
 \eeq
 Expanding in powers of $G$ as
\[ \Ve =\sum_{n=0}^\infty G^{2n} \Vn_n\]
gives the leading-order equation
\[ \del^2\Vn_0+ |\Vn_0|^2\Vn_0 - \Vn_0 = 0,\]
where we have assumed $G^2|\bX_i|^2 \ll 1$.
The relevant solution is the soliton,
\beq
V_0(\bx) = S_2(|\bx|),\label{solitonI}
\eeq
say. As $|\bx|\ra \infty$,
\beq
S_2(|\bx|) \sim \frac{A_2 \ee^{-|\bx|}}{|\bx|^{1/2}}.
\label{inoutmatch}
\eeq
i.e., the self-similarly collapsing peak satisfies the well-known
asymptotics thereof~\cite{sulem,fibich2015}.

\subsection{In between the peaks}

In this region $V$ is (algebraically) small, so we can
neglect the nonlinear terms at leading order. Then
\[ \del^2V_0 - V_0 = 0.\]
The solution which decays at infinity and matches with
\eqref{inoutmatch} is
\beq
V_0 = \frac{A_2\sqrt{2}}{\sqrt{\pi}} \sum_{i=1}^N  K_0(|\bxi - \bX_i|).\label{innersum}
\eeq
As $\boldsymbol{\xi}$ approaches each peak, we see from
\eqref{innersum} that the tail from each peak acts on all the others.
This interaction produces a correction term in the inner limit of $V_0$ near $\bxi = \bX_i +
\bx$ due to the other peaks, which  is
\beqa
\frac{A_2\sqrt{2}}{\sqrt{\pi}} \sum_{j \not = i}^N  K_0(|\bX_i -
  \bX_j + \bx|) &\sim& A_2 \sum_{j \not = i}^N
 \frac{ \ee^{-|\bX_i -
  \bX_j + \bx|}}{|\bX_i -
\bX_j + \bx|^{1/2}} \non \\
& \sim &A_2 \sum_{j \not = i}^N
\frac{ \ee^{-|\bX_i - \bX_j|}
  \ee^{-(\bX_i - \bX_j)\cdot\bx/|\bX_i - \bX_j|}
 }{|\bX_i -
\bX_j + \bx|^{1/2}}
.\label{out}
\eeqa

\subsection{Next order in the near field }
Equating coefficients of $G^2$ in \eqref{inner} gives
\[ \del^2V_1 + 3 V_0^2V_1 - V_1 
   = -\frac{|\bx+\bX_i|^2 }{4}V_0.
\]
The homogeneous version is satisfied by each component of $\del
V_0$. Multiplying by $ {\bf k} \cdot\del V_0$ (for some aribtrary
constant vector ${\bf k}$), and integrating, gives, after some
manipulation,  the solvability
condition as 
\beqas
- \int_{B_R} {\bf k} \cdot \del V_0  \frac{(|\bx|^2 + 2 \bx \cdot \bX_i
  + |\bX_i|^2 )}{4}V_0\, \d V  
 &=& \int_{\dd B_R}\left[( {\bf k} \cdot \del V_0)( \del
    V_1\cdot{\bf n}) - V_1 {\bf n} \cdot
({\bf k}\cdot \del) (\del V_0) 
\right]\, \d S,
\eeqas
where $B_R$ is the disk of radius $R$.

Since $V_0 = V_0(|{\bx}|)$,  the LHS may be simplified to
\[ - {\bf k} \cdot  \int_{B_R} \frac{\del V_0^2\, (\bx \cdot
    \bX_i)}{4}\, \d V =  \int_{B_R}
  \frac{V_0^2 \del \cdot ({\bf k} (\bx \cdot
    \bX_i))}{4}\, \d V =    
( {\bf k}  \cdot
    \bX_i) \int_{B_R}\frac{V_0^2}{4}\, \d V \sim
\frac{ {\bf k}  \cdot
    \bX_i}{4} b_0,
\]
where
\[ b_0 =  \int_{\R^2} V_0^2\, \d V .\]
For the RHS we evaluate the boundary terms by matching with
(\ref{out}).
As $|\bx| \ra \infty$,
\[ V_0 = S_2(|\bx|) \sim \frac{A_2 \ee^{-|\bx|}}{|\bx|^{1/2}}, 
\qquad
G^2 V_1 \sim 
A_2 \sum_{j \not = i}^N
\frac{ \ee^{-|\bX_i - \bX_j|}
  \ee^{-(\bX_i - \bX_j)\cdot\bx/|\bX_i - \bX_j|}
 }{|\bX_i -
\bX_j|^{1/2}},
\]
so that
\beqas
{\bf n}\cdot \del ({\bf k}\cdot  \del V_0) & = & {\bf n}\cdot \del\left(
 S_2'(|\bx|) \frac{{\bf k}\cdot\bx}{|\bx|}\right) =  
 S_2''(|\bx|) \frac{{\bf k}\cdot\bx}{|\bx|},\\
({\bf k}\cdot  \del V_0)(\del V_1 \cdot {\bf n}) & = & 
\left(
 S_2'(|\bx|) \frac{{\bf k}\cdot\bx}{|\bx|}\right)
\left(-\frac{A_2}{G^2} \sum_{j \not = i}^N
\frac{ \ee^{-|\bX_i - \bX_j|}
  \ee^{-(\bX_i - \bX_j)\cdot\bx/|\bX_i - \bX_j|}
 }{|\bX_i -
\bX_j|^{3/2}}\frac{(\bX_i - \bX_j)\cdot{\bx}}{|\bx|}
\right).
\eeqas
Thus
\begin{multline*}
( {\bf k} \cdot \del V_0)( \del
    V_1\cdot{\bf n}) - V_1 {\bf n} \cdot
({\bf k}\cdot \del) (\del V_0) \\    \sim  
 -
 \frac{{\bf k}\cdot\bx}{|\bx|^{3/2}}\frac{A_2^2}{G^2} 
 \sum_{j \not = i}^N
\frac{ \ee^{-|\bX_i - \bX_j|}
  \ee^{-|\bx|-(\bX_i - \bX_j)\cdot\bx/|\bX_i - \bX_j|}
 }{|\bX_i -
\bX_j|^{1/2}}\left(1- \frac{(\bX_i - \bX_j)\cdot \bx}{|\bX_i -
\bX_j||\bx|}\right)
\end{multline*}
as $|\bx| \ra \infty$.
Thus, letting $R \ra \infty$,  the solvability condition is
\beqas
\frac{ \bX_i}{4} b_0 & =&  - \frac{A_2^2}{G^2}
\lim_{R\ra\infty} \int_{\dd B_R}  \frac{\bx}{|\bx|^{3/2}}
 \sum_{j \not = i}^N
\frac{ \ee^{-|\bX_i - \bX_j|}
  \ee^{-|\bx|-(\bX_i - \bX_j)\cdot\bx/|\bX_i - \bX_j|}
 }{|\bX_i -
\bX_j|^{1/2}}\left(1- \frac{(\bX_i - \bX_j)\cdot \bx}{|\bX_i -
\bX_j||\bx|}\right)
\, \d S\\ & =&  - \frac{A_2^2}{G^2}
\lim_{R\ra\infty} \int_{0}^{2 \pi}  (\cos \theta, \sin \theta)
 \sum_{j \not = i}^N
\frac{ \ee^{-\Delta_{ij}}
  \ee^{-R-R  \cos (\theta-\phi_{ij})}
 }{\Delta_{ij}^{1/2}}\left(1-  \cos (\theta-\phi_{ij})\right)
\, R^{1/2} \d \theta 
\eeqas
where $\bD_{ij} = \bX_i - \bX_j = \Delta_{ij} (\cos \phi_{ij},\sin \phi_{ij})$.
The integrals can be evaluated using Laplace's method.
The stationary points  are given by
\[ \sin(\theta-\phi_{ij})=0 \qquad \Rightarrow \qquad \theta =
  \phi_{ij}, \quad \phi_{ij}+\pi.\]
If $\cos(\theta-\phi_{ij})=1$ the integrand is exponentially small
in $R$, so the relevant stationary point is $\theta = \phi_{ij}+\pi$,
$\cos(\theta-\phi_{ij})=-1$. Then
\begin{multline*}
\lim_{R\ra\infty} \int_{0}^{2 \pi}  (\cos \theta, \sin \theta)
   \ee^{-R-R  \cos (\theta-\phi_{ij})}
 \left(1-  \cos (\theta-\phi_{ij})\right)
\, R^{1/2} \d \theta  \\
 \sim  -2(\cos
  \phi_{ij},\sin \phi_{ij}) \lim_{R\ra\infty} \int_{-\infty}^{\infty}  
   \ee^{-R(\theta-\phi_{ij}-\pi)^2/2}
\, R^{1/2} \d \theta   =  -2\sqrt{2 \pi}(\cos
  \phi_{ij},\sin \phi_{ij}) 
\end{multline*}
Thus the solvability condition is
\[ b_0\bX_i =     8 \sqrt{2 \pi}\frac{A_2^2}{G^2}
  \frac{\ee^{-\Delta_{ij}}}{\Delta_{ij}^{1/2}}(\cos 
  \phi_{ij},\sin \phi_{ij}) 
\]
Rewriting this in terms of $\bX_i$ gives
\beq
b_0\bX_i =    8 \sqrt{2 \pi}\frac{A_2^2}{G^2}\sum_{j
    \not = i}^{N}\ee^{-|\bX_i-\bX_j|}
  \frac{(\bX_i-\bX_j) }{|\bX_i-\bX_j|^{3/2}} .
\label{asypositions}
  \eeq

\subsection{Normal form}
We have found the positions of the peaks, but as yet we have not
determined the dependence of the blow-up rate $G$ on the parameter
$\sigma$. To do so we need to consider the behaviour in the far field,
away from all the peaks.

\subsubsection{Far field}

When $\xi$ is large the term $G^2 |\bxi|^2 V/4$ can no longer be neglected. 
Thus there is a different balance in the equation in the far field. 
We rescale $\bxi = \brho/G$ to give
\[
 G^2  \del^2V+ |V|^{2}V - V 
 + \frac{|\brho|^2}{4} V= 0.\]
Since $V$ is  exponentially small in the far field (consider $|\bxi|$
large in \eqref{innersum})  we
can neglect the nonlinear term.
We look for a Liouville-Green (WKB) solution of the form
\[ V \sim  \ee^{\phi(\brho)/G} \sum_{n=0}^\infty \A_n(\brho) G^{n+1/2}.\]
At leading order this gives
\[ |\del\phi|^2 =1-\frac{|\brho|^2}{4}.\]
Since the peaks are all clustered near the origin on this scale ($|G
\log G| \ll 1$) we anticipate a radial solution.
Note the turning point at $|\brho|=2$. For $|\brho|<2$ the relevant solution is
\[ \fdd{\phi}{\rho} = -\left(1-\frac{\rho^2}{4}\right)^{1/2},\]
where $\rho = |\brho|$, so that the solution is decreasing (and $V$ is exponentially decaying) as $\rho$ increases.
Let us fix the constant by writing
\[ \phi = -\int_0^\rho \left(1-\frac{\bar{\rho}^2}{4}\right)^{1/2}\, \d \bar{\rho}.\]
The leading-order amplitude equation is
\[ 2 \del\phi \cdot \del\A_0 + \del^2\phi\, \A_0 = 0, \]
{which, with radial symmetry, reads}
\[ 2\fdd{\phi}{\rho} \pd{A_0}{\rho} + \left(\sdd{\phi}{\rho}+\frac{1}{\rho}\fdd{\phi}{\rho}\right) \A_0 =  0, \]
so that
\[ \A_0 = \frac{a_0}{\rho^{1/2}}\left(-\fdd{\phi}{\rho}\right)^{-1/2} = \frac{2^{1/2}\, a_0}{\rho^{1/2}(4-\rho^2)^{1/4}},\]
where $a_0 = a_0(\theta)$.
As $\rho \ra 0$
\beq
\ee^{\phi(\rho)/G}A_0 \sim \frac{a_0(\theta)}{\rho^{1/2}} \ee^{ -\rho/G } .\label{in0}
\eeq
Writing (\ref{innersum}) in terms of $\rho$  and expanding for small $G$ gives 
\[
V_0 
\sim   \frac{ G^{1/2} A_2}{\rho^{1/2}}\ee^{-\rho/G}\sum_{i=1}^N 
\ee^{X_{i,1}\cos \theta + X_{i,2} \sin \theta}.
\]
Matching with \eqref{in0} gives
\[ a_0(\theta) = A_2\sum_{i=1}^N 
\ee^{X_{i,1}\cos \theta + X_{i,2} \sin \theta}.\]

\subsubsection{Solution beyond the turning point}

As $\rho \ra \infty$ only the solution in which
\[ \phi' = \ii\left(\frac{\rho^2}{4}-1\right)^{1/2}\]
has a finite Hamiltonian.
Thus for $\rho>2$, the WKB solution is
\[ V = \al \ee^{\ii \phi_2(\rho)/G} G^{1/2}\sum_{n=0}^{\infty} A_n(\brho) (\ii G)^n,
\]
for some constant $\alpha$, where
\[ \phi_2 = \int_2^\rho \left(\frac{\bar{\rho}^2}{4}-1\right)^{1/2}\, \d \bar{\rho}, \qquad A_0(\brho) =\frac{2^{1/2}a_0(\theta)}{\rho^{1/2}(\rho^2-4)^{1/4}}.
\]

\subsubsection{Turning point}
The turning point causes an exponentially small reflection back into the near field.
The azimuthal angle acts as a parameter only.
The solution in the turning point region is exactly the same as in \cite{jon1}, with the result that $\alpha = \ee^{\ii \pi/4}$ and the WKB solution in $\rho<2$ is modified to 
\beqas
V &=&  \ee^{\phi(\rho)/G}G^{1/2} \sum_{n=0}^{\infty} A_n(\brho) G^n + \gamma\ee^{- \phi(\rho)/G} G^{1/2}\sum_{n=0}^{\infty} A_n(\brho) (-G)^n \qquad \rho<2,
\eeqas
where, for an infinite domain, 
\[ \gamma = \frac{\ii}{2}\ee^{2\phi(2)/G}=\frac{\ii}{2}\ee^{-\pi/G}.\]

\subsubsection{Matching back into the near field}
\label{sec:sol}
We have found that the exponentially small correction to the far field in $0<\rho<2$ is
\[ \Vexp = \gamma\ee^{- \phi(\rho)/G}G^{1/2} \sum_{n=0}^{\infty} A_n(\brho) (-G)^n,
\]
and as $\rho \ra 0$, we have
\beq
\Vexp \sim \frac{a_0(\theta) \gamma G^{1/2}}{\rho^{1/2}} \ee^{\rho/G}. \label{match}
\eeq
Writing $V = \Ve + \Vexp$ 
where
$\Ve$
is the original (multi-peaked) algebraic expansion, Taylor
expanding and neglecting quadratic terms in $\Vexp$ but keeping now the term involving $\sigma-1$ gives, noting that $\Ve$ is real,
\[
\del^2 \Vexp+ \Ve^2 ( \Vexp^*+2 \Vexp)  
  -\Vexp
  + \frac{G^2 |\bxi|^2 }{4}\Vexp
  =  -2(\sigma-1) \Ve^3 \log \Ve+\frac{\ii (\sigma-1) G}{ \sigma} \Ve.
\]
We  separate the expression into real and imaginary parts. Writing $\Vexp = \Uexp + \ii \Wexp$ gives
\beqa
   \del^2 \Uexp+
 3 \Ve^2 \Uexp   
 -\Uexp 
  + \frac{G^2  |\bxi|^2 }{4}\Uexp&=& - 2(\sigma-1) \Ve^3 \log \Ve,\label{Ueq}\\ 
  \del^2\Wexp+
 \Ve^2 \Wexp   
 -\Wexp 
  + \frac{G^2  |\bxi|^2 }{4}\Wexp&=& 
  \frac{ (\sigma-1) G}{ \sigma} \Ve. \label{Weq}
\eeqa
Note that $\Ve$ satisfies the homogeneous version of (\ref{Weq}).
The resulting solvability condition will determine $\sigma$.
It is easiest now to consider this equation throughout the whole near-field region, rather than considering each peak separately.

Multiplying (\ref{Weq}) by $\Ve$ and integrating over a ball of radius $R$
 gives
\begin{multline}
\int_{B_R} \left(\Ve  \del^2\Wexp+
 \Ve^3 \Wexp   
 -\Ve\Wexp 
 + \frac{G^2 |\bxi|^2 }{4}\Ve\Wexp\right)\, \d V \\
  = \int_{\dd B_R}\left(\Ve \del\Wexp - \Wexp \del\Ve\right)\cdot {\bf n}\, \d S  
 =  
   \frac{ (\sigma-1) G}{\sigma} \int_{B_R}\Ve^2\, \d x.\label{sigsol}
\end{multline}
As $R \ra \infty$ we evaluate the boundary terms by matching using (\ref{match}). We find
\beqas
\left.\Ve \pd{\Wexp}{\Xi} - \Wexp \pd{\Ve}{\Xi}\right|_{\Xi = R}
& = &\frac{a_0^2}{R}  \ee^{-\pi/G} \qquad \mbox{ as }R \ra \infty,
\eeqas
where $\Xi = |\bxi|$.
Then
\beqas
\int_{\dd B_R}\left(\Ve \del\Wexp - \Wexp \del\Ve\right)\cdot {\bf n}\, \d S
& = &  \ee^{-\pi/G} \int_{0}^{2 \pi}a_0^2(\theta)\, \d \theta\\
& = &  A_2^2\ee^{-\pi/G} \int_{0}^{2 \pi}\left(\sum_{i=1}^N\ee^{X_{i,1}\cos \theta + X_{i,2} \sin \theta}\right)^2\, \d \theta\\
& = &  A_2^2\ee^{-\pi/G} \int_{0}^{2 \pi}\left(\sum_{i=1}^N\ee^{R_i \cos (\theta-\phi_i)}\right)^2\, \d \theta,
\eeqas
where $\bX_i = R_i(\cos \phi_i, \sin \phi_i)$. 
The dominant contribution to the RHS of (\ref{sigsol}) comes from each of the peaks, since $\Ve$ is $O(G^2)$ in between the peaks. 
Then
\[ \int_{B_R} V_G(x)^2\, \d x \sim N b_0.\]
Thus
\begin{equation}
\label{eq:sigmaG}
    \sigma - 1 \sim \frac{\sigma}{N b_0 G}  A_2^2\ee^{-\pi/G} \int_{0}^{2 \pi}\left(\sum_{i=1}^N\ee^{R_i \cos (\theta-\phi_i)}\right)^2\, \d \theta.
\end{equation}

Since $R_i$ is large as $G \ra 0$ we could evaluate via Laplace's
method again, although since $R_i$ is only logarithmically large the
errors would be substantial.

\section{Asymptotic Analysis of the leading order multi-peaked eigenvalues}
\label{asymptoticseigenvalues}

Denote the steady state by $\Vs$.
We linearize about the steady state by writing
\begin{eqnarray}
\V(\bxi,\tau) = \Vs(\bxi) + \eps\left( f(\bxi) \ee^{\la \tau} + g^*(\bxi) \ee^{\la^* \tau} \right)
\label{linear14}
\end{eqnarray} 
and linearizing in $\eps$, which  leads to the eigenvalue problem
\beqa
\ii\la f+\del^2 f +
\sigma |\Vs|^{2\sigma-2}\Vs^{2}g + ( \sigma+1)|\Vs|^{2 \sigma}  f 
- f
- \frac{\ii (d\sigma-2) G}{2 \sigma} f
+ \frac{G^2 |\boldsymbol{\xi}|^2  }{4} f &=& 0,\label{geqn}\\
-\ii  \la g
+ \del^2 g 
+  \sigma |\Vs|^{2\sigma-2}(\Vs^*)^{2} f 
+ ( \sigma+1)|\Vs|^{2 \sigma} g 
-  g 
+ \frac{\ii (d\sigma-2) G}{2 \sigma}  g 
+ \frac{G^2 |\boldsymbol{\xi}|^2 }{4} g  &=& 0.\label{feqn}
\eeqa
Since $\Vs$ is real to all orders, and $\sigma$ is
exponentially close to 1, we may write
\beqa
\ii\la f+ \del^2 f +
 \Vs^{2}g +2 \Vs^{2} f 
- f
+ \frac{G^2 |\bxi|^2 }{4} f &=& \tst,\label{geqn1}\\
-\ii  \la g
+ \del^2 g 
+   \Vs^{2} f 
+ 2\Vs^{2} g 
-  g 
+ \frac{G^2 |\bxi|^2 }{4} g  &=& \tst,\label{feqn1}
\eeqa
where $\tst$ stands for transcendentally small terms, i.e., terms which are
$O(G^n)$ for all $n$ as $G \ra 0$.
We aim to find asymptotic approximations to the real
eigenvalues. Since $\Vs$ is real to all orders, when $\la$ is real we
may write $g = f^* + \tst$, so that we may deal with the single
equation 
\beq
\ii\la f+\del^2 f +
 \Vs^{2}f^* +2 \Vs^{2} f 
- f
+ \frac{G^2 |\bxi|^2 }{4} f = \tst
\eeq
Anticipating that $\la$ is small, the leading-order eigenfunctions satisfy
\beq
\del^2 f_0 +
 V_0^{2}f_0^* +2 V_0^{2} f_0 
- f_0  = 0.\label{eqnf0}
\eeq
There are three solutions to this equation, 
namely
\[ f_0 = \pd{V_0}{\xi} , \qquad f_0 = \pd{V_0}{\eta}, \qquad f_0 = \ii V_0.\]
For a multi-peaked steady state we may take a different multiple of each
 at each peak. The general analysis rapidly becomes very complicated
 (analogous to the one-dimensional multi-peaked eigenvalue calculations
 in \cite{multi1dpaper}). We therefore attempt to approximate only
 the dominant family of eigenvalues, which are $O(G^{1/2})$ as $G \ra
 0$, for which the analysis is more straightfoward.
We anticipate the scaling by writing 
$\la = G^{1/2}\bar{\la}$. We will perturb in powers of $G^{1/2}$. The
analysis mirrors that in \cite{multi1dpaper} for the one-dimensional case.

Throughout,  we will encounter inhomogeneous versions of (\ref{eqnf0}) in the form
\beq
\del^2 f +
 V_0^{2}f^* +2 V_0^{2} f 
- f  = \RHS.\label{eqnf0rhs}
\eeq
Since the imaginary part of the homogeneous equation has the
nontrivial solution $V_0$, and the real part of the homogeneous
equation has the nontrivial solution $\del V_0$, there are three solvability conditions on (\ref{eqnf0rhs}), namely, 
\beqa
\im \left[ \int_{\dd B_R} \left(\del f V_0 - f \del V_0 \right)\cdot
  \mathbf{n} \, \d S\right]  &=& \int_{B_R} \im(\RHS)V_0\, \d V,\label{solv1}\\
\re \left[ \int_{\dd B_R} \left(\del f \pd{V_0}{\xi} - f \del \pd{V_0}{\xi} \right)\cdot
  \mathbf{n} \, \d S\right]  &=&\int_{B_R} \re(\RHS)\pd{V_0}{\xi}\, \d V,\label{solv2}\\
\re \left[ \int_{\dd B_R} \left(\del f \pd{V_0}{\eta} - f \del \pd{V_0}{\eta}\right)\cdot
  \mathbf{n} \, \d S\right] &=& \int_{B_R} \re(\RHS)\pd{V_0}{\eta}\, \d V.\label{solv3}
\eeqa

\subsection{Inner expansion near  the $i$th peak}
With $\bxi = \bX_i + \bx$, we expand
\[ f = \Ca_i \sum_{n=0}^\infty G^{n/2}f_n,\]
where $f_0 = \ii V_0$.
At $O(G^{1/2})$ we have
\[
\eq{f_1}{f_1^*} =\bar{\la} V_0 .
\]
The solvability conditions (\ref{solv1})-(\ref{solv3}) are automatically satisfied, so that $\bar{\la}$ is undetermined, and 
\[ f_1 = \frac{\bar{\la}}{2}\left(V_0 + \bx \cdot \del V_0\right).
\]
At $O(G)$ we find \footnote{As in \cite{multi1dpaper}, really we should expand $\la$ (and $\Ca_i$) in powers of $G$, so that there would be additional terms on the RHS due to $\la_2$. However, these terms will simply mirror the terms involving $\bar{\la}$ at the previous order, and we can ignore them if we are only interested in the leading approximation of the eigenvalue.}
\[
\eq{f_2}{f_2^*} 
=-\frac{\ii \bar{\la}^2}{2}\left( V_0 + \bx \cdot \del V_0\right).
\]
Now
\[
\int_{B_R} V_0\, \bx \cdot \del V_0\, \d V  =  \frac{1}{2}\int_{B_R}
\bx \cdot \del V_0^2\, \d V = \frac{1}{2}\int_{B_R}
\del \cdot (\bx V_0^2) - 2 V_0^2 \, \d V = -\int_{B_R}V_0^2 \, \d V .
\]
Thus, the solvability conditions (\ref{solv1})-(\ref{solv3}) are automatically satisfied, so that $\bar{\la}$ is still undetermined, and the solution is
\[
f_2  = - \frac{\ii \bar{\la}^2}{8} |\bx|^2 V_0.
\]
At $O(G^{3/2})$ we find
\[
\eq{f_3}{f_3^*} 
= - \frac{\bar{\la}^3 |\bx|^2}{8} V_0.
\]
As before, the solvability conditions are automatically satisfied
because of symmetry, and the solution is
\[
f_3  =  \frac{ \bar{\la}^3}{2}  \hat{V}_1, 
\qquad
g_3  =  \frac{\bar{\la}^3}{2 }\hat{V}_1,
\]
where $\hat{V}_1$ 
satisfies
\[
\eqr{\hat{V}_1} = -\frac{|\bx|^2 V_0}{4}, \qquad \hat{V_1} \ra 0 \mbox{ as } x \ra \pm \infty,
\]
that is, $\hat{V_1}$ is the next term in the expansion of the standard single-peaked solution (i.e. not
accounting for the shift in origin).
At $O(G^{2})$ we find 
\beq
\eq{f_4}{f_4^*} 
= - \frac{\ii  \bar{\la}^4}{2}  \hat{V}_1 - \frac{\ii |\bx+\bX_i|^2}{4}V_0 - 2 \ii V_1 V_0^2.\label{iV4}
\eeq
Here at last the solvability condition (\ref{solv1}) is not automatically satisfied, and  will determine $\bar{\la}$. Note that $V_1$ grows exponentially as $x \ra \pm\infty$ (in contrast
to $\hat{V}_1$).
The equation for $V_1$ is
\beq
\del^2 V_1 + 3 V_0^2 V_1 - V_1 = -\frac{|\bx+\bX_i|^2}{4}V_0.\label{inner1}
\eeq
Multiplying (\ref{inner1}) by $V_0$ and integrating by parts gives
\beqas
-\int_{B_R} \frac{(\bx+\bX_i)^2}{4}V_0^2\, \d V
& = & \int_{B_R}
2 V_0^3V_1\, \d V +
\int_{\dd B_R} \left(V_0 \del V_1 - V_1 \del V_0 \right)\cdot \bn \,
\d S.
\eeqas
Thus, the solvability conditions on (\ref{iV4}) are
\begin{align}
\im &\left[ \int_{\dd B_R} \left(\del f_4 V_0 - f_4 \del V_0 \right)\cdot
  \mathbf{n} \, \d S\right]   = 
 -   \frac{\bar{\la}^4}{2}  \int_{B_R} 
  \hat{V}_1 V_0\, \d V  
+
\int_{\dd B_R} \left(V_0 \del V_1 - V_1 \del V_0 \right)\cdot \bn \,
\d S,\label{sol1}\\
\re &\left[ \int_{\dd B_R} \left(\del f_4 \pd{V_0}{\xi} - f_4 \del \pd{V_0}{\xi} \right)\cdot
  \mathbf{n} \, \d S\right]  = 
0
\label{sol2},\\
\re &\left[ \int_{\dd B_R} \left(\del f_4 \pd{V_0}{\eta} - f_4 \del \pd{V_0}{\eta}\right)\cdot
  \mathbf{n} \, \d S\right]=
0.
\label{sol3}
\end{align}
The boundary terms on the RHS can be evaluated using the steady state analysis.
To evaluate the boundary terms on the LHS we need to consider $f$ in the region between peaks.

\subsection{In between the peaks}
Away from the peaks  $\Vs = O(G^2 \log G)$ and accordingly we can neglect the
nonlinear term. 
Thus,
\beq
\ii G^{1/2}\bar{\la} f+\del^2  f
- f
+ \frac{G^2 |\bxi|^2 }{4} f = O(G^8 \log^4 G f),
\eeq
Expanding
\[ f = \sum_{k=0}^\infty G^{k/2} f_k,\] 
gives
\beq
\del^2  f_0 -  f_0 = 0,
\eeq
at leading order. Since $f_0 \sim \ii \Ca_i V_0$ near the $i$\,th peak,
matching gives the solution as
\[ f_0 = \frac{A_2 \sqrt{2}\,\ii}{\sqrt{\pi}} \sum_{j=1}^N \Ca_j
  K_0(|\bxi - \bX_j|).
\]
Evaluating close to the $i$\,th peak by writing $\bxi = \bX_i + \bx$ gives
\beqa
 f_0 \sim  &\sim&  A_2 \ii\sum_{j \not = i}^N
 \frac{\Ca_j \ee^{-|\bX_i -
  \bX_j + \bx|}}{|\bX_i -
\bX_j + \bx|^{1/2}} \non \\
& \sim &A_2\ii \sum_{j \not = i}^N
\frac{\Ca_j \ee^{-|\bX_i - \bX_j|}
  \ee^{-(\bX_i - \bX_j)\cdot\bx/|\bX_i - \bX_j|}
 }{|\bX_i -
\bX_j + \bx|^{1/2}}
.\label{outA}
\eeqa
Thus, the matching condition on the inner expansion near the  $i$\,th peak is
\beq
 \Ca_{i} f_4G ^2 \sim A_2 \ii\sum_{j \not = i}^N
\frac{\Ca_j \ee^{-|\bX_i - \bX_j|}
  \ee^{-(\bX_i - \bX_j)\cdot\bx/|\bX_i - \bX_j|}
 }{|\bX_i -
\bX_j + \bx|^{1/2}},\label{f4inf}
\eeq
as $|\bx| \ra \infty$.

\subsection{Solvability condition}
Using \eqref{f4inf}  gives
\beqa
\lim_{R \ra \infty}\re \left[ \int_{\dd B_R} \left(\del f_4 \pd{V_0}{\xi} - f_4 \del \pd{V_0}{\xi} \right)\cdot
  \mathbf{n} \, \d S\right]  &=& 
0
,\\
\lim_{R \ra \infty}\re \left[ \int_{\dd B_R} \left(\del f_4 \pd{V_0}{\eta} - f_4 \del \pd{V_0}{\eta}\right)\cdot
  \mathbf{n} \, \d S\right]&=& 
0.
\eeqa
and 
\beqas
\lefteqn{
\lim_{R \ra \infty}\im \left[ \int_{\dd B_R} \Ca_i \left(\del f_4 V_0 - f_4 \del V_0 \right)\cdot
  \mathbf{n} \, \d S\right]}\qquad && \\ &=&
\frac{A_2^2 }{G^2}\lim_{R \ra \infty}\int_{\dd B_R} \frac{1}{|\bx|^{1/2}}
 \sum_{j \not = i}^N \Ca_j\frac{ \ee^{-|\bX_i - \bX_j|- |\bx|-(\bX_i - \bX_j)\cdot\bx/|\bX_i - \bX_j|} }{|\bX_i -
\bX_j|^{1/2}}\left(1 - \frac{(\bX_i - \bX_j)\cdot{\bx}}{|\bX_i -
  \bX_j||\bx|}\right)\, \d S\\
& = &\lim_{R \ra \infty}\frac{A_2^2 R^{1/2}}{G^2}\int_{0}^{2 \pi} 
 \sum_{j \not = i}^N \Ca_j\frac{ \ee^{-\Delta_{ij} - R- R
     \cos(\theta-\phi_{ij}) }}
   {\Delta_{ij}^{1/2}}\left(1 -\cos(\theta-\phi_{ij})\right)\, \d
   \theta\\
   & = &\frac{2 \sqrt{2 \pi}\,A_2^2}{G^2}
 \sum_{j \not = i}^N \frac{\Ca_j \ee^{-\Delta_{ij} }}{ \Delta_{ij}^{1/2}}
 \eeqas
 using Laplace's method. Similarly,  the steady state analysis shows
 \beqas
\lefteqn{\lim_{R \ra \infty}\int_{\dd B_R} \left(V_0 \del V_1 - V_1 \del V_0 \right)\cdot \bn \,
  \d S }\qquad && \\
& = &
\frac{A_2^2 }{G^2}\lim_{R \ra \infty}\int_{\dd B_R} \frac{1}{|\bx|^{1/2}}
 \sum_{j \not = i}^N \frac{ \ee^{-|\bX_i - \bX_j|- |\bx|-(\bX_i - \bX_j)\cdot\bx/|\bX_i - \bX_j|} }{|\bX_i -
\bX_j|^{1/2}}\left(1 - \frac{(\bX_i - \bX_j)\cdot{\bx}}{|\bX_i -
  \bX_j||\bx|}\right)\, \d S\\
   & = &\frac{ 2\sqrt{2 \pi}\,A_2^2}{G^2}
 \sum_{j \not = i}^N \frac{ \ee^{-\Delta_{ij} }}{ \Delta_{ij}^{1/2}}.
\eeqas
Thus, letting $R\ra\infty$,  (\ref{sol2})-(\ref{sol3}) are satisfied, and  
the remaining condition (\ref{sol1})  is
\beq
  \Ca_i b_1\frac{\bar{\la}^4}{2}   =\frac{2 \sqrt{2 \pi}\,A_2^2}{G^2}
\sum_{j \not = i}^N (\Ca_i-\Ca_j)\frac{ \ee^{-\Delta_{ij} }}{ \Delta_{ij}^{1/2}},\label{Ghalfeigs}
\eeq
where
\[ b_1 = \int_{B_R} 
\hat{V}_1 V_0\, \d V.\]
Equation (\ref{Ghalfeigs}) is a set of $N$ homogeneous linear equations for the $N$ coefficients $\Ca_i$, $i = 1,\ldots,N$. For a nontrivial solution the determinant must vanish, and it is this condition that determines the eigenvalue $\bar{\la}$.
Note that for $\Ca_i =\Ca$ for all $i$, $\bar{\la}=0$ is a solution for any $N$. This corresponds to  the global phase invariance, with eigenvalue $\la=0$. 
Thus we expect that there are $N-1$ eigenvalues which are $O(G^{1/2})$.
Note also that the eigenvalues appear in pairs---for each positive
eigenvalue there is a corresponding negative eigenvalue with the same
(leading order) weights $\Ca_i$. Although not covered by the
preceeding analysis (since we assumed that $\la$ was real), a similar
analysis shows that there are also (pairs of complex conjugate) purely imaginary
eigenvalues with the same modulus.

Numerically we find: $b_1 \approx 1.786.$

\subsection{Examples}

We now provide a number of concrete case examples of the eigenvalue
calculations for our multi-peak solutions that have been used in 
our earlier numerical comparisons.

\subsubsection{Two peaks}
With $-\bX_1=\bX_2 = (X,0)$, where
\begin{equation}
\label{eq:twohumpspos}
   b_0 X^{3/2} = \frac{8 \sqrt{\pi} A_2^2}{G^2}
\ee^{-2X}, 
\end{equation}

Eq.~(\ref{Ghalfeigs}) in this case yields:
\[
 \bar{\la}^4   =\frac{4 \sqrt{\pi}\,A_2^2}{b_1G^2}
 \left(1-\frac{\Ca_2}{\Ca_1}\right)\frac{ \ee^{-2X }}{ X^{1/2}} =\frac{4 \sqrt{\pi}\,A_2^2}{b_1G^2}
 \left(1-\frac{\Ca_1}{\Ca_2}\right)\frac{ \ee^{-2X }}{ X^{1/2}}.
\]
Thus $\Ca_2^2=\Ca_1^2$ so that $\Ca_2 = \pm \Ca_1$. The additional solution
$\Ca_2 = - \Ca_1$ has positive eigenvalue
\[ \bar{\la}^4=\frac{8 \sqrt{\pi}\,A_2^2}{b_1G^2}
  \frac{2 \ee^{-2X }}{ X^{1/2}} =
  \frac{Xb_0}{b_1}
  ,\]
i.e.,
\[ \la = \left(\frac{Xb_0}{b_1}\right)^{1/4}G^{1/2} .\]

\subsubsection{Three Aligned Peaks}

With $-\bX_1=\bX_3 = (X,0)$, $\bX_2 = (0,0)$,
we have
\[
X^{3/2} b_0  \frac{G^2}{8 \sqrt{\pi} A_2^2}=   \left(
 \sqrt{2}\, \ee^{-X}
  + \ee^{-2X}
  \right).
\]

Equation (\ref{Ghalfeigs}) yields in this case
\beqas
  \frac{ b_1 G^2\bar{\la}^4}{4\sqrt{2 \pi}\,A_2^2 }  & =&
 \left(1-\frac{\Ca_2}{\Ca_1}\right)\frac{ \ee^{-X }}{
   X^{1/2}}
 +   \left(1-\frac{\Ca_3}{\Ca_1}\right)\frac{ \ee^{-2X }}{ (2X)^{1/2}},\\
 \frac{ b_1 G^2\bar{\la}^4}{4\sqrt{2 \pi}\,A_2^2 }  & =&
 \left(1-\frac{\Ca_1}{\Ca_2}\right)\frac{ \ee^{-X }}{
   X^{1/2}}
 +   \left(1-\frac{\Ca_3}{\Ca_2}\right)\frac{ \ee^{-X}}{X^{1/2}},\\
 \frac{ b_1 G^2\bar{\la}^4}{4\sqrt{2 \pi}\,A_2^2 }  & =&
 \left(1-\frac{\Ca_1}{\Ca_3}\right)\frac{ \ee^{-2X }}{
   (2X)^{1/2}}
 +   \left(1-\frac{\Ca_2}{\Ca_3}\right)\frac{ \ee^{-X}}{
   X^{1/2}}.
 \eeqas
The solutions are $(\Ca_1,\Ca_2,\Ca_3) = (1,1,1)$ with $\bar{\la} = 0$ as
expected, along with
\[ (\Ca_1,\Ca_2,\Ca_3) = (1,-2,1), \qquad (\Ca_1,\Ca_2,\Ca_3) =
  (1,0,-1),\]
with eigenvalues
\[  \frac{ b_1 G^2\bar{\la}^4}{4\sqrt{2 \pi}\,A_2^2 }  =  \frac{3\ee^{-X}}{X^{1/2}},
\]
and
\[  \frac{ b_1 G^2\bar{\la}^4}{4\sqrt{2 \pi}\,A_2^2 }  = \frac{\ee^{-X} +
    \sqrt{2} \ee^{-2X}}{X^{1/2}},
\]
respectively.

\subsubsection{Three peaks in an equilateral triangle arrangment}
With $\bX_j= X (\cos(2 \pi j/3),\sin(2 \pi j/3))$, we have that $\bX_3 - \bX_1 = (3/2,\sqrt{3}/2)X$, and $\bX_3 - \bX_2 = (3/2,-\sqrt{3}/2)X$.
~In this case, the 
equilibrium
position is
\[  b_0 X  = 3^{1/4}8 \sqrt{2 \pi} \frac{A_2^2}{G^2}
    \frac{\ee^{-\sqrt{3}X}}{X^{1/2}}.
    \]
Then Eq.~(\ref{Ghalfeigs}) is 
\beqas
 \frac{b_1G^2\bar{\la}^4}{4\sqrt{2 \pi} A_2^2} &=&  
 \left(1-\frac{\Ca_{2}}{\Ca_1}\right)\frac{\ee^{-\sqrt{3}X}}{3^{1/4}X^{1/2}}+
 \left(1-\frac{\Ca_{3}}{\Ca_1}\right)\frac{\ee^{-\sqrt{3}X}}{3^{1/4}X^{1/2}}\\
 &=& \left(1-\frac{\Ca_{1}}{\Ca_2}\right)\frac{\ee^{-\sqrt{3}X}}{3^{1/4}X^{1/2}}+
 \left(1-\frac{\Ca_{3}}{\Ca_2}\right)\frac{\ee^{-\sqrt{3}X}}{3^{1/4}X^{1/2}}\\
 &=& \left(1-\frac{\Ca_{1}}{\Ca_3}\right)\frac{\ee^{-\sqrt{3}X}}{3^{1/4}X^{1/2}}+
 \left(1-\frac{\Ca_{2}}{\Ca_3}\right)\frac{\ee^{-\sqrt{3}X}}{3^{1/4}X^{1/2}},
 \eeqas
with solution $(\Ca_1,\Ca_2,\Ca_3) =(1,1,1)$ as expected along with the additional solutions 
\[ (\Ca_1,\Ca_2,\Ca_3) =  (1,\Ca,-1-\Ca), 
\]
which has the corresponding positive eigenvalue
\[ \bar{\la}^4= \frac{4\sqrt{2 \pi} A_2^2} {b_1G^2}
\frac{3^{3/4}\ee^{-\sqrt{3}X}}{X^{1/2}} = \frac{3^{1/2} b_0
  X}{2b_1},\]
i.e., finally we obtain
\[ \la = \left(\frac{3^{1/2} b_0
      X}{2b_1}\right)^{1/4}G^{1/2} 
  .\]
The coefficient $a$ is not determined at this order.
{If we continue we will find a solvability condition on $a$
  at a higher order. Generically we might expect to find a quadratic
  equation for $a$ (the simplest nonlinearity), so that this
  eigenvalue is approximately a double eigenvalue, splitting at
  $O(G)$. This is consistent with the numerical results presented earlier.}

\subsubsection{Four peaks}
With $\bX_j= X (\cos(\pi j/2),\sin(\pi j/2))$,
\[ \bX_4 - \bX_1 = (X,-X), \qquad  \bX_4 - \bX_2 = (2X,0),\qquad
\bX_4-\bX_3 = (X,X),\]
and the equilibrium
position is
\[  b_0 X^{3/2}  =8 \sqrt{2\pi} \frac{A_2^2}{G^2}
  \left(2^{1/4}\ee^{-\sqrt{2}X}+
    \frac{\ee^{-2X}}{2^{1/2}}\right).
    \]
Then      
(\ref{Ghalfeigs}) yields
\beqas
 \frac{b_1G^2\bar{\la}^4}{4\sqrt{2 \pi} A_2^2} &=&  
 \left(1-\frac{\Ca_{2}}{\Ca_1}\right)\frac{\ee^{-\sqrt{2}X}}{2^{1/4}X^{1/2}}+
 \left(1-\frac{\Ca_{3}}{\Ca_1}\right)\frac{\ee^{-2X}}{2^{1/2}X^{1/2}}+ \left(1-\frac{\Ca_{4}}{\Ca_1}\right)\frac{\ee^{-\sqrt{2}X}}{2^{1/4}X^{1/2}}\\
&=&  
 \left(1-\frac{\Ca_{1}}{\Ca_2}\right)\frac{\ee^{-\sqrt{2}X}}{2^{1/4}X^{1/2}}+
 \left(1-\frac{\Ca_{4}}{\Ca_2}\right)\frac{\ee^{-2X}}{2^{1/2}X^{1/2}}+ \left(1-\frac{\Ca_{3}}{\Ca_2}\right)\frac{\ee^{-\sqrt{2}X}}{2^{1/4}X^{1/2}}\\
&=&  
 \left(1-\frac{\Ca_{2}}{\Ca_3}\right)\frac{\ee^{-\sqrt{2}X}}{2^{1/4}X^{1/2}}+
 \left(1-\frac{\Ca_{1}}{\Ca_3}\right)\frac{\ee^{-2X}}{2^{1/2}X^{1/2}}+ \left(1-\frac{\Ca_{4}}{\Ca_3}\right)\frac{\ee^{-\sqrt{2}X}}{2^{1/4}X^{1/2}}\\
&=&  
 \left(1-\frac{\Ca_{1}}{\Ca_4}\right)\frac{\ee^{-\sqrt{2}X}}{2^{1/4}X^{1/2}}+
 \left(1-\frac{\Ca_{2}}{\Ca_4}\right)\frac{\ee^{-2X}}{2^{1/2}X^{1/2}}+ \left(1-\frac{\Ca_{3}}{\Ca_4}\right)\frac{\ee^{-\sqrt{2}X}}{2^{1/4}X^{1/2}}.
 \eeqas
with solution $(\Ca_1,\Ca_2,\Ca_3,\Ca_4) =(1,1,1,1)$ as expected along with the additional solutions 
\[ (\Ca_1,\Ca_2,\Ca_3,\Ca_4) =  (1,\Ca,-1,-\Ca), 
\]
which has corresponding positive eigenvalue
\[  \bar{\la}^4 =
\frac{4\sqrt{2 \pi} A_2^2}{b_1G^2} \frac{(2^{1/2}\ee^{-2X}+
  2^{3/4}\ee^{-\sqrt{2}X})}{X^{1/2}} ,\] 
(again, $a$ is not determined and we expect to find two possible values of $a$
at next order) and
\[ (\Ca_1,\Ca_2,\Ca_3,\Ca_4) =  (1,-1,1,-1), 
\]
which has corresponding positive eigenvalue
\[  \bar{\la}^4 =
\frac{16\sqrt{\pi} A_2^2}{b_1G^2} \frac{ 2^{1/4}\ee^{-\sqrt{2}X}}{X^{1/2}}.\]

As before, we can use the equation connecting $X$ and $G$ above to
express $\bar{\lambda}$ purely as a function of $X$, and multiplying $\bar{\lambda}$
by $G^{1/2}$ we obtain the final form of the relevant eigenvalues,
which has been used for our comparison with numerical computations
in Fig.~\ref{fig:domeigen4peaks}.
 
\subsubsection{Four peaks: equilateral triangle with a fourth peak at the barycenter}

With $\bX_j= X (\cos(2 \pi j/3),\sin(2 \pi j/3))$, $\bX_4 = (0,0)$, 
\[ \Delta_{12} =\Delta_{13}=\Delta_{23} = \sqrt{3}\, X,\qquad
  \Delta_{14} =\Delta_{24}=\Delta_{34} = X,\]
we have
\[
b_0 \frac{G^2}{ 8 \sqrt{2 \pi}A_2^2} X^{3/2} =   
  3^{1/4}\ee^{-\sqrt{3}\,X} +\ee^{-X}.
\]
Equation (\ref{Ghalfeigs}) in this case reads
\beq
  \Ca_i b_1\frac{\bar{\la}^4}{2}   =\frac{2 \sqrt{2 \pi}\,A_2^2}{G^2}
\sum_{j \not = i}^N (\Ca_i-\Ca_j)\frac{ \ee^{-\Delta_{ij} }}{ \Delta_{ij}^{1/2}}.
\eeq
\beqas
  \frac{ b_1 G^2\bar{\la}^4}{4\sqrt{2 \pi}\,A_2^2 }  & =&
 \left(1-\frac{\Ca_2}{\Ca_1}\right)\frac{ \ee^{-\sqrt{3}\,X }}{
   3^{1/4}X^{1/2}}
 + \left(1-\frac{\Ca_3}{\Ca_1}\right)\frac{ \ee^{-\sqrt{3}\,X }}{
   3^{1/4}X^{1/2}}
 +   \left(1-\frac{\Ca_4}{\Ca_1}\right)\frac{ \ee^{-X }}{ X^{1/2}},\\
  \frac{ b_1 G^2\bar{\la}^4}{4\sqrt{2 \pi}\,A_2^2 }  & =&
 \left(1-\frac{\Ca_1}{\Ca_2}\right)\frac{ \ee^{-\sqrt{3}\,X }}{
   3^{1/4}X^{1/2}}
 + \left(1-\frac{\Ca_3}{\Ca_2}\right)\frac{ \ee^{-\sqrt{3}\,X }}{
   3^{1/4}X^{1/2}}
 +   \left(1-\frac{\Ca_4}{\Ca_2}\right)\frac{ \ee^{-X }}{ X^{1/2}},\\
 \frac{ b_1 G^2\bar{\la}^4}{4\sqrt{2 \pi}\,A_2^2 }  & =&
 \left(1-\frac{\Ca_1}{\Ca_3}\right)\frac{ \ee^{-\sqrt{3}\,X }}{
   3^{1/4}X^{1/2}}
 + \left(1-\frac{\Ca_2}{\Ca_3}\right)\frac{ \ee^{-\sqrt{3}\,X }}{
   3^{1/4}X^{1/2}}
 +   \left(1-\frac{\Ca_4}{\Ca_3}\right)\frac{ \ee^{-X }}{ X^{1/2}},\\
 \frac{ b_1 G^2\bar{\la}^4}{4\sqrt{2 \pi}\,A_2^2 }  & =&
 \left(1-\frac{\Ca_1}{\Ca_4}\right)\frac{ \ee^{-X }}{
   X^{1/2}}
 + \left(1-\frac{\Ca_2}{\Ca_4}\right)\frac{ \ee^{-X }}{
   X^{1/2}}
 +   \left(1-\frac{\Ca_3}{\Ca_4}\right)\frac{ \ee^{-X }}{ X^{1/2}}.
 \eeqas
 In addition to $(\Ca_1,\Ca_2,\Ca_3,\Ca_4) = (1,1,1,1)$ we have solutions
 \[ (\Ca_1,\Ca_2,\Ca_3,\Ca_4) = (1,1,1,-3),\]
 with eigenvalue
 \[  \frac{ b_1 G^2\bar{\la}^4}{4\sqrt{2 \pi}\,A_2^2 } = \frac{4
     \ee^{-X}}{X^{1/2}},\] 
 and
 \[ (\Ca_1,\Ca_2,\Ca_3,\Ca_4) = (1,a,-1-a,0),\]
 with (approximately double) eigenvalue
 \[  \frac{ b_1 G^2\bar{\la}^4}{4\sqrt{2 \pi}\,A_2^2 } = \frac{3^{3/4} \ee^{-\sqrt{3}\,X}+\ee^{-X}}{X^{1/2}}.\]
 As usual, multiplying $\bar{\lambda}$ by $G^{1/2}$ gives
the final form of the relevant eigenvalues.

\subsubsection{Five peaks}
With $\bX_j= X (\cos(2\pi j/5),\sin(2\pi j/5))$,
the equilibrium
position is
\[  b_0 X^{3/2}  =8 \sqrt{\pi} \frac{A_2^2}{G^2}
\frac{1}{2^{1/4}}  \left(   (5-\sqrt{5})^{1/4} \ee^{-\sqrt{(5-\sqrt{5})/2}\, X} +
    (5+\sqrt{5})^{1/4} \ee^{-\sqrt{(5+\sqrt{5})/2}\, X} \right).
    \]
The solutions of \eqref{Ghalfeigs} are $(\Ca_1,\Ca_2,\Ca_3,\Ca_4,\Ca_5) = (1,1,1,1,1)$ as
expected, along with the additional solutions
\beqas
\left(1,a,-1-\frac{(1+\sqrt{5})a}{2},(1+a)\frac{(1+\sqrt{5})}{2},-a
  -\frac{(1+\sqrt{5})}{2}\right),\\ 
\left(1,a,-1-\frac{(1-\sqrt{5})a}{2},(1+a)\frac{(1-\sqrt{5})}{2},-a
  -\frac{(1-\sqrt{5})}{2}\right),
\eeqas
with eigenvalues given by
\beqas
\frac{b_1G^2\bar{\la}^4}{4\sqrt{2 \pi} A_2^2} X^{1/2}&=&
\frac{1}{2}\left( (5+\sqrt{5})\ee^{-\sqrt{(5-\sqrt{5})/2}X} +
  (5-\sqrt{5})\ee^{-\sqrt{(5+\sqrt{5})/2}X}\right),\\
  \frac{b_1G^2\bar{\la}^4}{4\sqrt{2 \pi} A_2^2} X^{1/2}&=&\frac{1}{2}\left( (5-\sqrt{5})\ee^{-\sqrt{(5-\sqrt{5})/2}X} +
  (5+\sqrt{5})\ee^{-\sqrt{(5+\sqrt{5})/2}X}\right),
\eeqas
respectively. We expect each of these to be a double eigenvalue (to
leading order).

\section{Self-similar structures of increased complexity}
\begin{figure}
    \centering
    \includegraphics[width=0.5\linewidth]{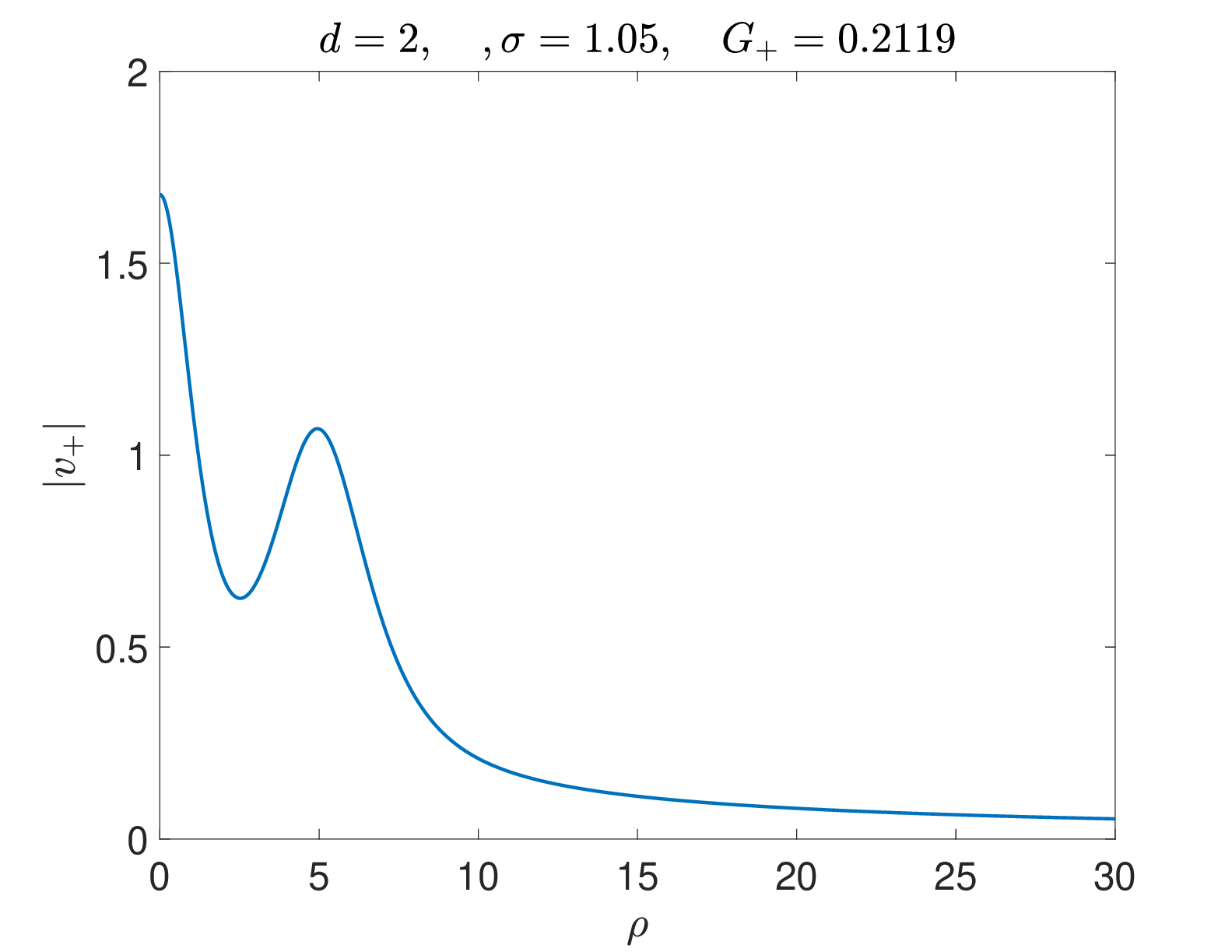}
    \caption{Amplitude $|v_+|$ of the radially symmetric solution to the boundary value problem of Eq.~(\ref{eq:formulti})-(\ref{eq:multibc}) for $\sigma=1.05$.
    The solution features one peak at the origin and a ring at $\rho \approx 4.96$.
    }
    \label{fig:three_humped_example}
\end{figure}

In Secs.~\ref{sec:multiradial} and \ref{sec:nonradial}, we described two approaches for numerically computing non-radially symmetric, multi-peaked self-similar solutions. 
In the first approach, we start from a  radially symmetric ring solution (see Fig.~\ref{fig:two_humped_example}), compute the corresponding 2D ring solution in Cartesian coordinates (see Fig.~\ref{fig:two_humped_2D}), assess its stability, and perturb it along eigenvectors associated with the dominant unstable modes. This produces an initial guess for damped Newton iterations that can converge to multi-peaked solutions as shown in Fig.~\ref{fig:perturbexample}.
Following the same procedure, we can construct initial guesses for more complex structures and converge to non-trivial self-similar configurations. 
For instance, if we initialize suitably chosen values of $\re{[v_+(0)]}$ and $G_+$ to solve the boundary value problem Eqs.~(\ref{eq:formulti})-(\ref{eq:multibc}), we can compute a radially symmetric solution including a peak at the origin and a ring. 
More specifically, using the initial guesses: $\re{[v_+(0)]}=1.5, G_+=0.2$ at $\sigma=1.05$, we converge to the solution shown in Fig.~\ref{fig:three_humped_example}. 
The converged parameters are: $\re{[v_+(0)]}=1.6790, G_+=0.2119$ for $\sigma=1.05$, and the solution features two localized peaks: one at the origin, and a ring at $\rho \approx 4.96$.

\begin{figure}[ht!]
\centering
\begin{tabular}{cc}
\includegraphics[width=0.49\linewidth]{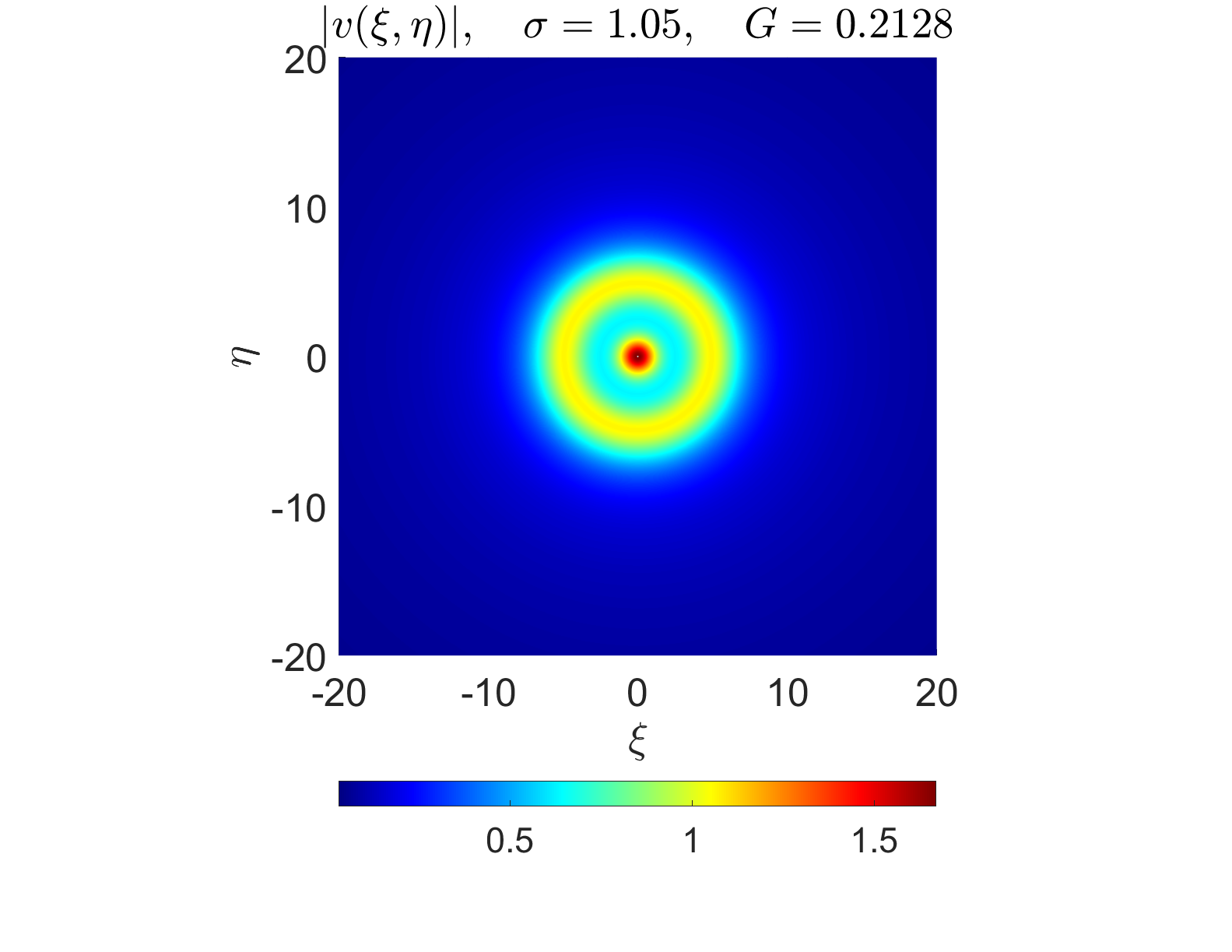}     &  \includegraphics[width=0.49\linewidth]{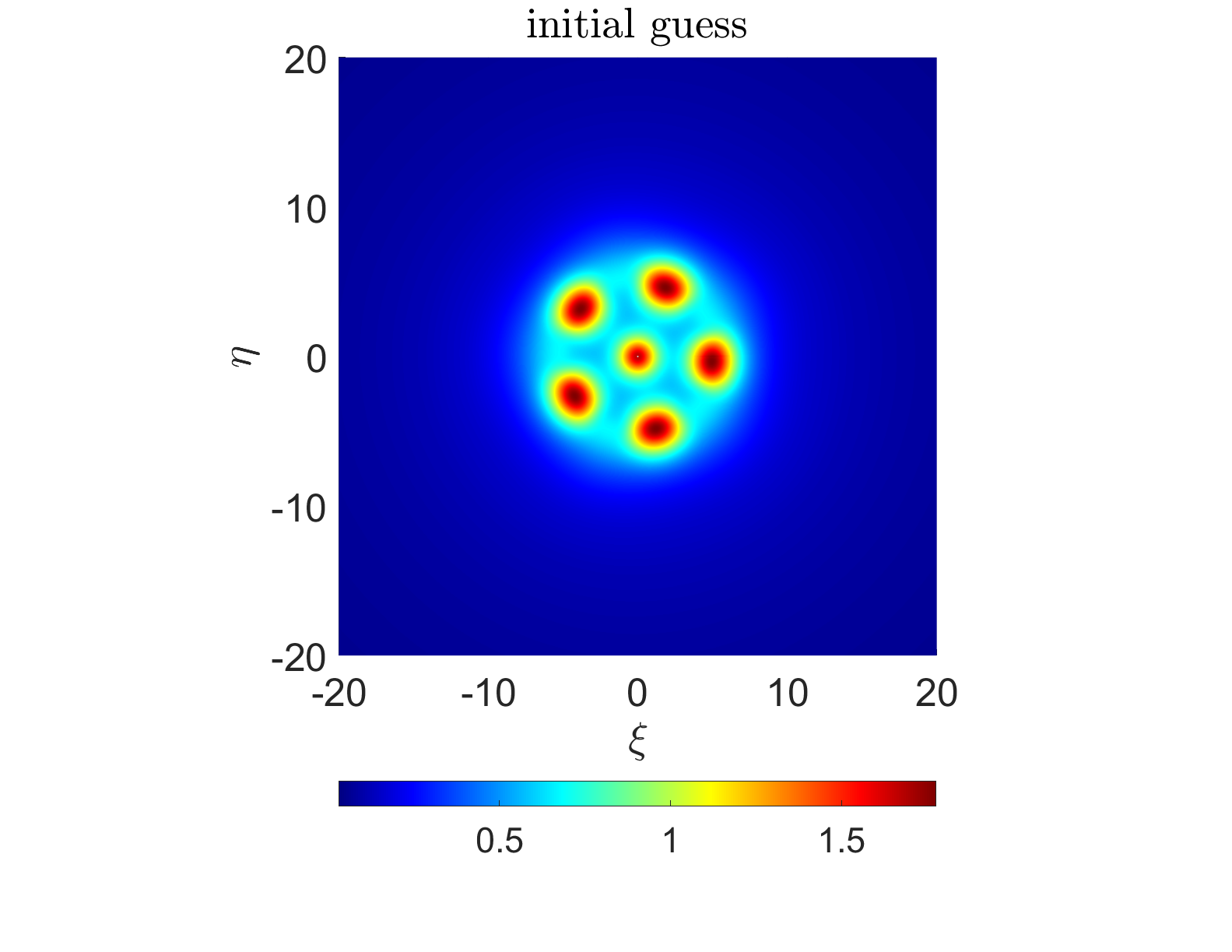}\\
 (a)    & (b) \\
 \end{tabular}
\begin{tabular}{c}
 \includegraphics[width=0.49\linewidth]{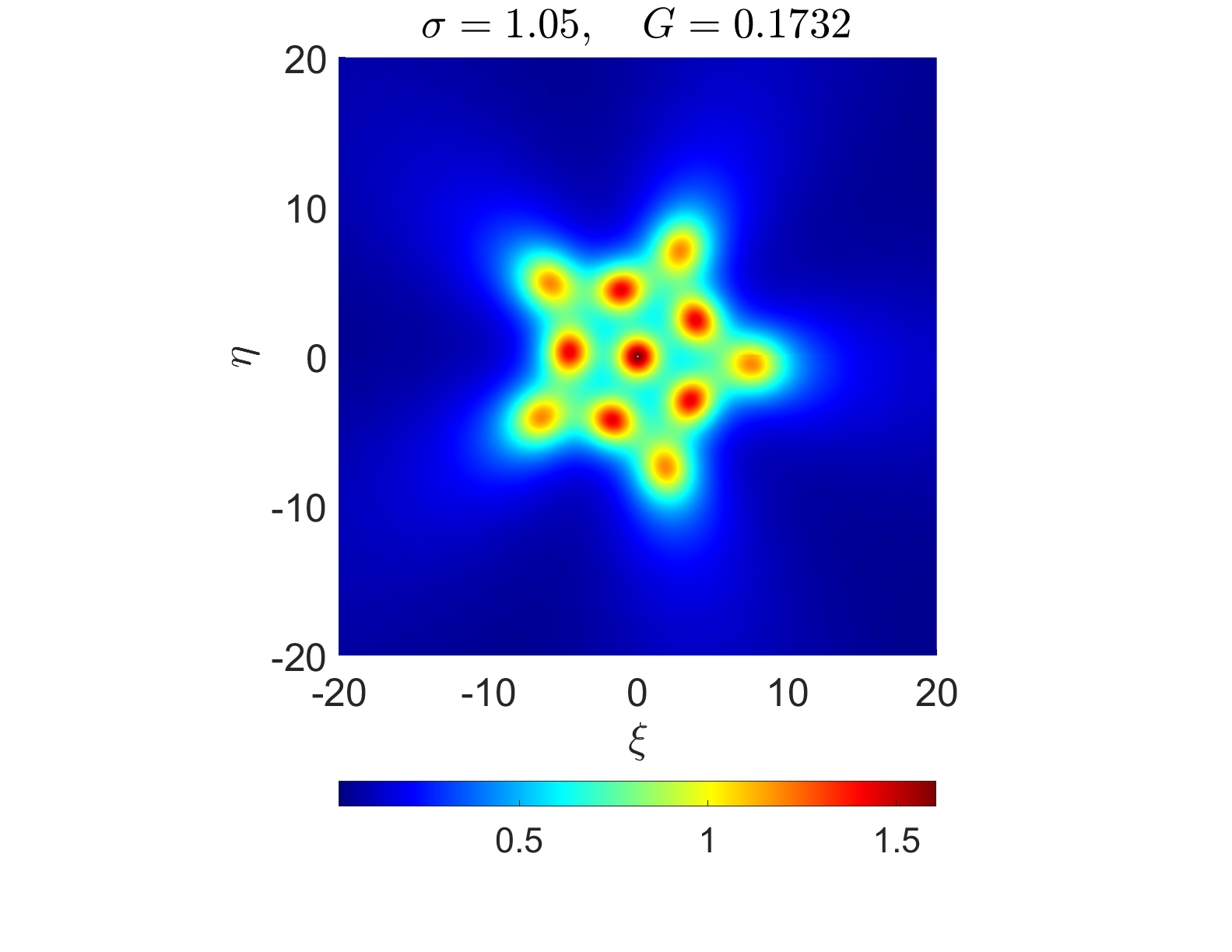}\\
 (c)
\end{tabular} 
\caption{
\label{fig:three_humped_2D}
(a) Unstable self-similar ring solution with an additional center peak (here we show the amplitude) of the 2D NLS equation at $\sigma=1.05$.
(b) Initial guess for damped Newton iterations, obtained by perturbing the  ring-with-center-bump solution at $\sigma=1.05$ along the eigenvector associated with $\lambda=0.93$. 
(c) Converged eleven-peaked solution; the blow-up rate is $G=0.1732$. 
All panels illustrate the amplitude $|v|$ of the solution.
}
\end{figure}

This solution is then used as input to the Newton solver for Eq.~(\ref{eq:nlscoexp}); upon convergence, we ``lift'' the radially symmetric profile to Cartesian coordinates.
The resulting configuration is --as expected-- a radially symmetric ring solution with an additional central bump (see Fig.~\ref{fig:three_humped_2D}(a)).

Spectral analysis of this solution reveals several dominant eigenmodes. As before, we identify symmetry-related eigenvalues ($\lambda=2G$, $\lambda=G$ and $\lambda=0$), as well as additional unstable modes that can break the radial symmetry and trigger the formation of multi-peaked solutions.
As an illustration, perturbing the ring-with-center-hump solution along the eigenvector associated with $\lambda=0.93$ yields the initial configuration shown in Fig.~\ref{fig:three_humped_2D}(b), which we use as an initial guess for our damped Newton iterations. 
After a few Newton steps, the method converges to a star-shaped profile with eleven peaks (Fig.\ref{fig:three_humped_2D}(c)) and blow-up rate $G=0.1732$. 

\begin{figure}[ht!]
\centering
\begin{tabular}{cc}
\includegraphics[width=0.49\linewidth]{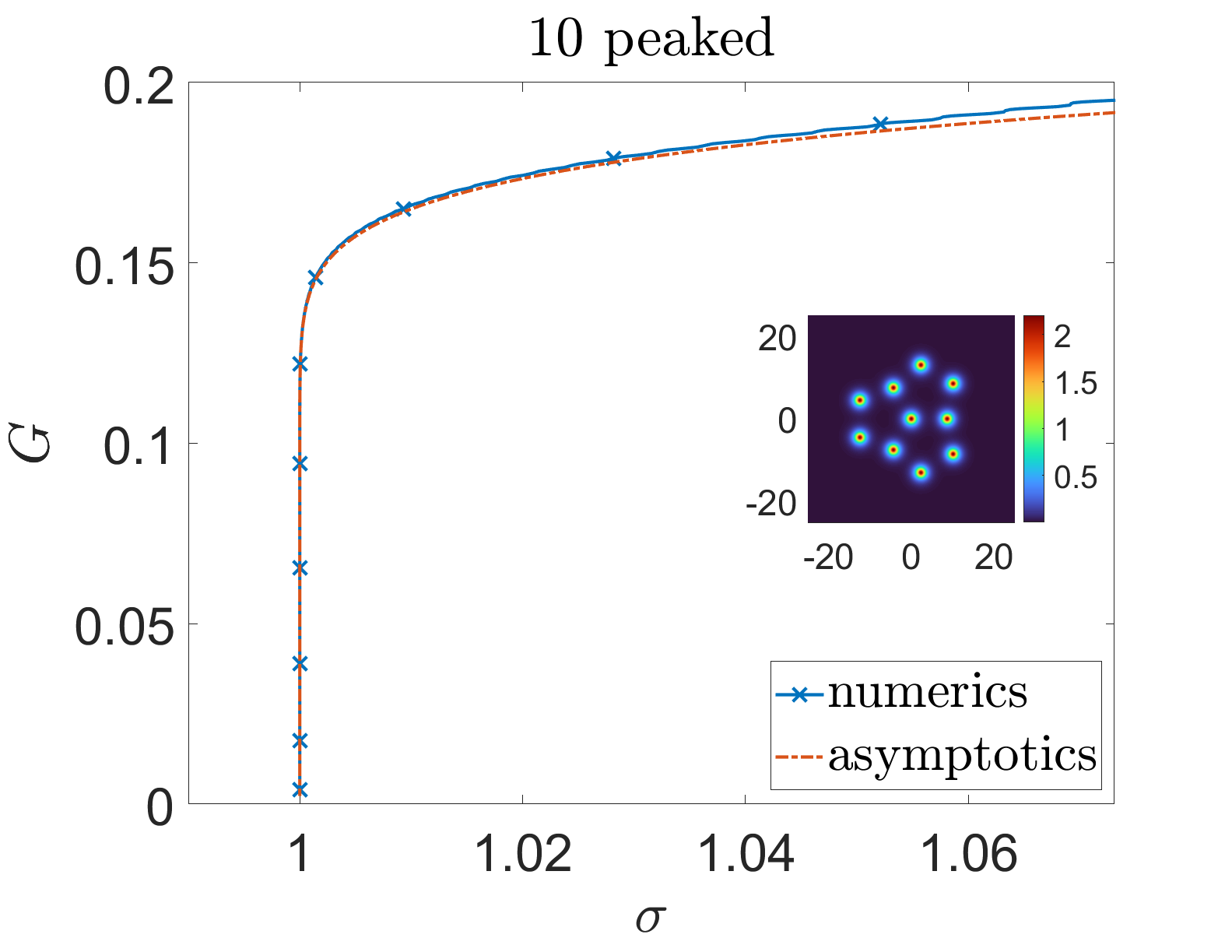}     &  \includegraphics[width=0.49\linewidth]{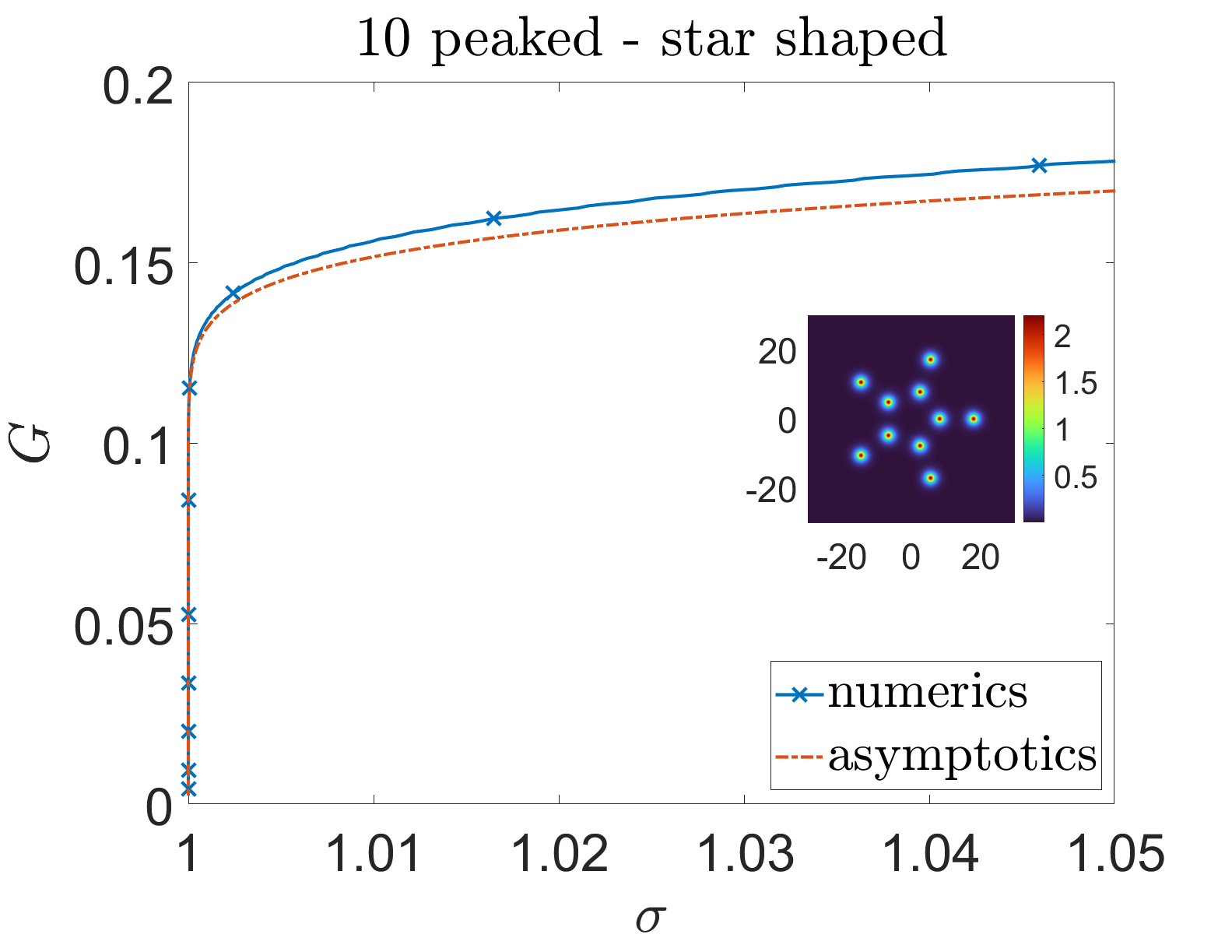}\\
 (a)    & (b)
\end{tabular} 
\caption{
\label{fig:exotic1}
Bifurcation diagrams of self-similar solution branches with higher structural complexity (solutions with 10 peaks). 
The dash-dotted red lines correspond to asymptotic predictions showing very good agreement for small $G$ values.
The insets illustrate in (a) a solution with blow-up rate $G=2.2 \times 10^{-3}$, and in (b) a star-shaped solution with blow-up rate $G=2.3 \times 10^{-3}$ .
}
\end{figure}

Using the same methodology, we can compute a variety of configurations with multiple peaks and track entire solution branches via parameter continuation. Representative examples are shown in Fig.~\ref{fig:exotic1}, where we trace self-similar solution branches with ten peaks and configurations of increased complexity; insets display the corresponding multi-peaked profiles. While we do not pursue
this avenue further, given also the instability of such multi-peak solutions,
we use this opportunity to highlight the feature that the model
does possess a wide range of such multi-peak configurations, 
reminiscent in a way of the wide range of multi-vortex configurations
that exist in point vortex and similar models in two spatial dimensions~\cite{Aref_2007}. 
It is also noteworthy that recent experiments such as the one of~\cite{banerjee2024collapse}
suggest that such configurations may arise in a transient form in
physical experiments, e.g., in atomic BECs, nonlinear optics and
elsewhere. Hence, the study of such multi-peak states
is of interest for practical reasons,
beyond the intrinsic mathematical appeal of the relevant
two-dimensional Toda lattice with a mass system.

\section{Conclusions \& Future Challenges}
\label{sec:conclusions}

In the present work we have revisited the potential of 
nonlinear Schr{\"o}dinger models to bear solutions
beyond ones featuring a single collapsing peak~\cite{sulem,fibich2015}
in the supercritical regime of a two-dimensional model with nonlinearity
exponent beyond the cubic critical one. While such examples
 for two-peaks had been previously established
theoretically~\cite{Nawa1998} and illustrated in
particular cases numerically~\cite{REN2000246} (see, e.g.,
Figs.~17-18 therein) ---including more recently in 
particular examples via rigorous numerics~\cite{dahne2024selfsimilarsingularsolutionsnonlinear}---, 
to the best of
our knowledge, no systematic theory existed characterizing
either the existence or the stability of such multi-peaked 
states, although early works had suggested their more
general existence in one spatial dimension in~\cite{budd:1999}
and more recently their (in)stability in~\cite{multi1dpaper}.
We find a wide range of possible configurations and corroborate
their existence numerically. 

The competition that enables the existence and identification of
such delicate states stems from the equilibrium configurations
of an interacting particle system of the ``collapse spots'',
arising through the balance of the inter-pulse interaction
force and the ``onsite force'' emerging from the system's
phase structure (and complex pattern). This is a remarkable (and unexpected)
two-dimensional generalization of a Toda lattice with a mass that
has also been previously encountered, e.g., in the one-dimensional
context of dark solitons in a parabolic
trap~\cite{Coles_2010}.
This analysis provides us with excellent initial guesses that
led our fixed-point iteration scheme to converge to a wide
range of such configurations and makes it possible to continue them
over parameters (such as the nonlinearity exponent).
We then turned our attention to the stability of the relevant
states, bearing in mind the multiplicity of eigenvalue pairs
that are arising due to the multi-peaked nature of the configuration
(and of each peak bearing $4$ associated pairs in isolation). 
While we did not provide the full functional characterization
 obtained in the one-dimensional setting as concerns the
eigenvalues, we nevertheless described the full qualitative 
situation of the eigenvalue dependence on the characteristic
parameter of the system, namely the blowup rate $G$. 
In particular, we identified $N-1$ real and $N-1$ (positive)
imaginary eigenvalues of ``size'' $G^{1/2}$, $2 N$ eigenvalues
of size $G$ and 2 vanishing eigenvalues reflecting symmetries.
For the dominant real and imaginary eigenvalues, we were also
able to provide quantitative characterizations, while the
ones of O$(G)$ turned out to be more technically involved 
(without necessarily being more insightful, as the latter
concern sub-dominant terms in the resulting dynamics).

Naturally, the progressively increased tractability of not
only one-dimensional, but also two-dimensional collapsing
settings of the nonlinear Schr{\"o}dinger equation
(and of the Korteweg-de Vries equation~\cite{Chapman_2024})
paves the way towards future studies and interesting additional
possibilities. To mark just a few of the associated 
considerations we note the following: it would be interesting
to connect the profiles considered herein more firmly with the 
(in)stability analysis of the singular ring (and multi-ring generalization) solutions
considered in~\cite{FIBICH2005193,FIBICH2005193}. It is anticipated
---also per our fixed-point iteration results above---
that the transverse, azimuthal instability of the latter 
leads spontaneously to the former, a feature of interest in its own
right~\cite{CAPLAN20091399}. 
This is even more so the case in the context of vortical
structures, as the latter would emulate the recent experiment
of~\cite{banerjee2024collapse}. In the latter setting, a vortex
was placed at the center of an atomic Bose-Einstein condensate,
precluding the breakup of the ring toward the center, but rather
seeding its destabilization azimuthally along the ring's periphery.
Extending this to multiple vortices or multiple rings would seem 
particularly exciting in the way of generating higher order
multi-peaked waveforms. A more complete understanding of the 
two-dimensional realm ---facilitated in part by the results herein---
is, in itself, a further motivation for exploring more complex
settings.
~Those include three-dimensional, single-component 
ones~\cite{fibich2015,sulem}, as well as potentially multi-component
ones that are of considerable interest in their own right and
which have been used as a tool for building effective single
component attractive states, as in the experimental work of~\cite{Bakkali-Hassani2021,Bakkali-Hassani2022}. Such studies are currently in 
progress and will be reported in future publications.

\begin{acknowledgments}
This research was supported by the U.S. National Science Foundation under the awards 
DMS-2204782 \& DMS-2527314 (E.G.C.), and PHY-2408988 (P.G.K.), as well as by the U.S. Army Research Office (ARO) under award number W911NF-26-1-A043 (E.G.C.).~This research was partly conducted while P.G.K. was  visiting the Okinawa Institute of Science and
Technology (OIST) through the Theoretical Sciences Visiting Program (TSVP), the University of
Sydney through the visitor program of the Sydney Mathematical Research Institute (SMRI) and the Department of Mechanical Engineering at Seoul National 
University through a Fulbright Fellowship. Their support is gratefully acknowledged.
Finally, this work was also  supported by a grant from the Simons Foundation [SFI-MPS-SFM-00011048, P.G.K]. 
\end{acknowledgments}

\bibliographystyle{unsrt}

\appendix

\renewcommand{\theequation}{\Alph{section}.\arabic{equation}}
\renewcommand{\thefigure}{\Alph{section}.\arabic{figure}}
\setcounter{equation}{0}
\setcounter{figure}{0}

\section{Looking Ahead: State-Of-The-Art Self-Similar Computations in \texttt{FreeFem\,$++$}}

\begin{figure}[pt!]
\begin{center}
\begin{overpic}
[width=0.32\linewidth]{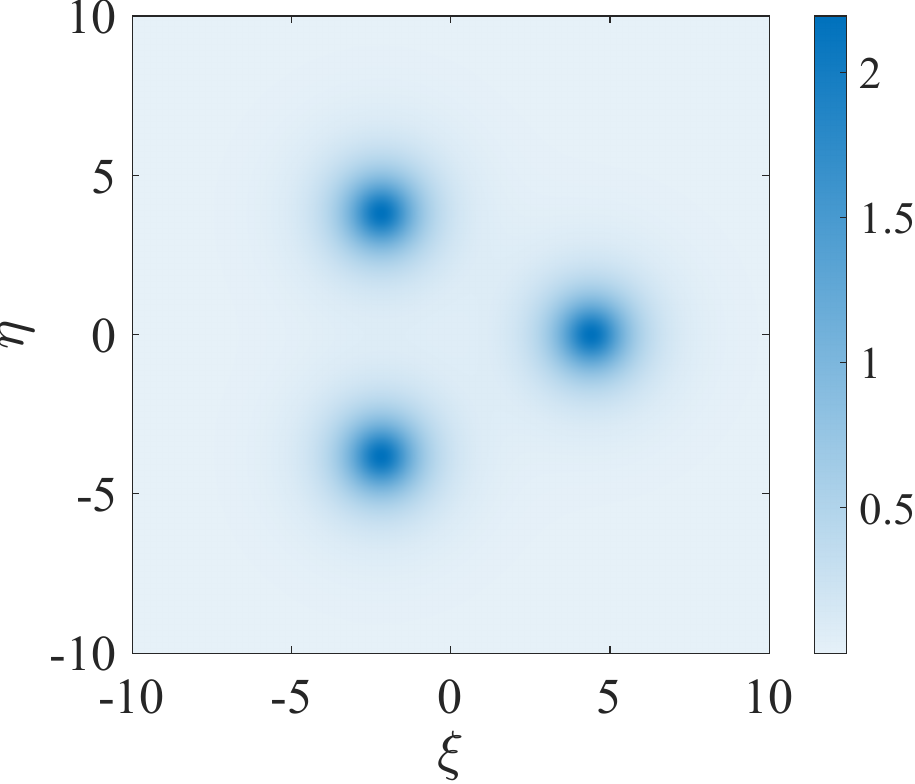}
\put(16,75){\small $\mathrm{\textbf{(a)}}$}
\end{overpic}
\begin{overpic}
[width=0.28\linewidth]{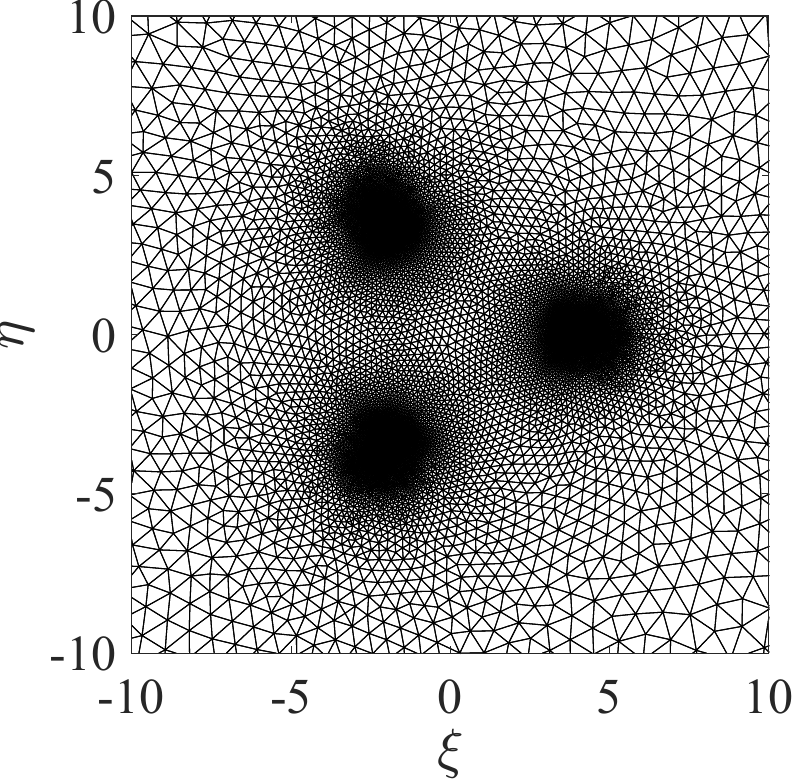}
\put(18,87){\small $\mathrm{\textbf{(b)}}$}
\end{overpic}
\begin{overpic}
[width=0.35\linewidth]{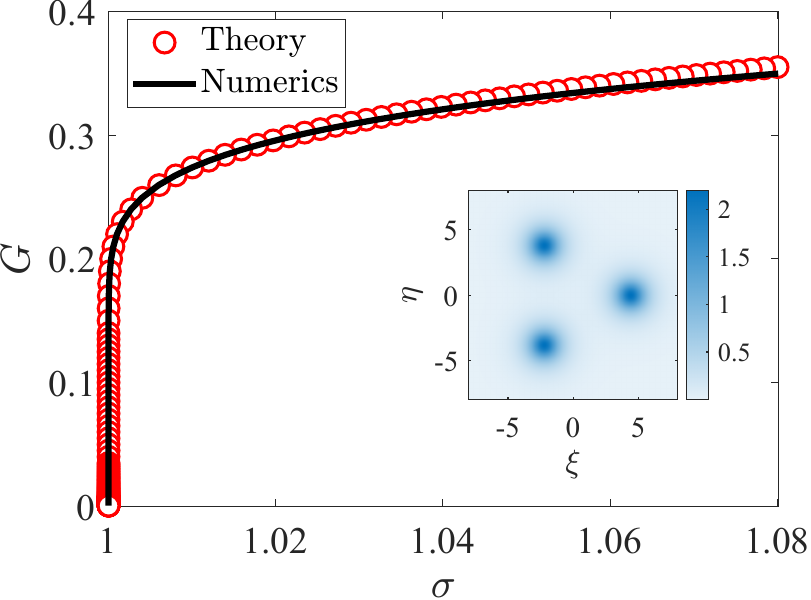}
\put(15,55.5){\small $\mathrm{\textbf{(c)}}$}
\end{overpic}\\
\vskip 0.3cm
\begin{overpic}
[width=0.32\linewidth]{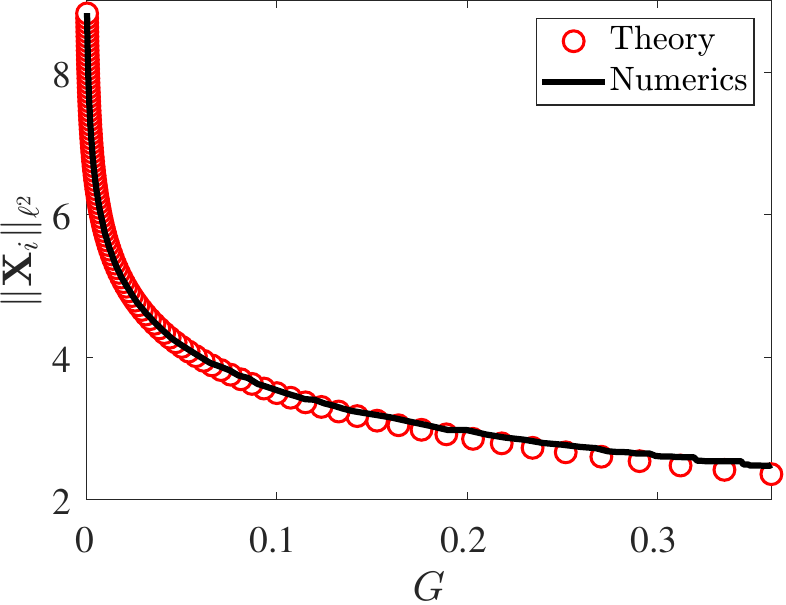}
\put(16,70){\small $\mathrm{\textbf{(d)}}$}
\end{overpic}
\begin{overpic}
[width=0.32\linewidth]{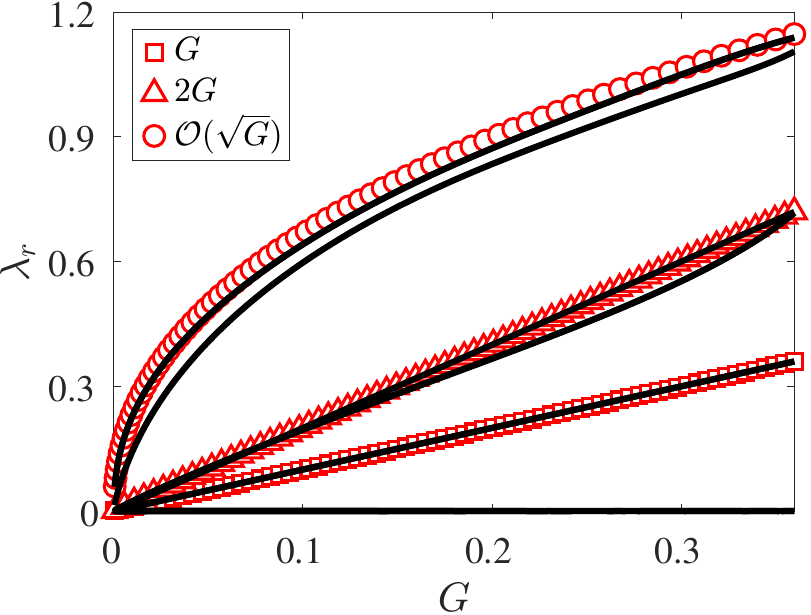}
\put(17.3,48){\small $\mathrm{\textbf{(e)}}$}
\end{overpic}
\begin{overpic}
[width=0.32\linewidth]{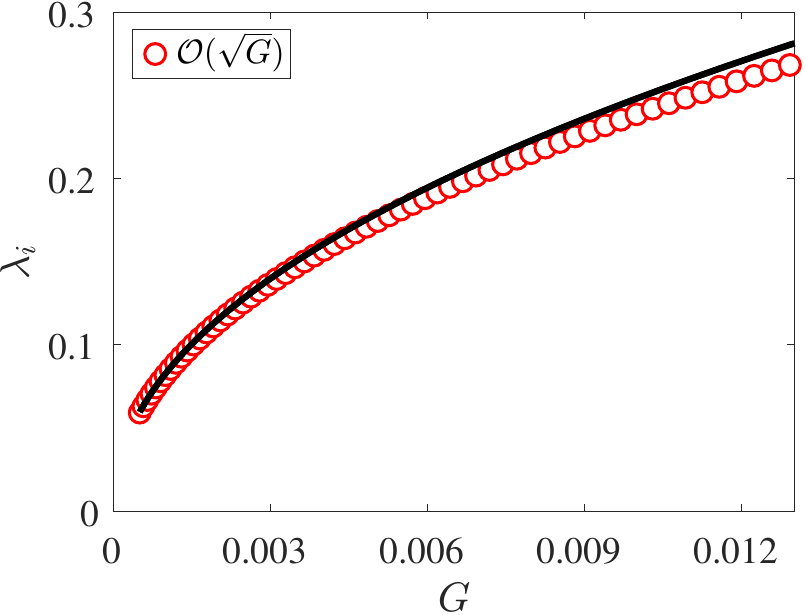}
\put(17.3,58){\small $\mathrm{\textbf{(f)}}$}
\end{overpic}
\end{center}
\caption{
\label{ff_3_peaks_equilateral}
Summary of numerical results on the self-similarly blowing-up three-peak branch (see, Figs.~\ref{fig:bifdiagrammulti}(iii),~\ref{fig:peak_comparison}(i) and \ref{fig:domeigen3peaks}) in the main text).~In panels \textbf{(a)} and \textbf{(b)} the $|v|$ and the mesh is shown for $\sigma=1$ (and computed $G=0.04$), and the $G$ as a function of $\sigma$ is shown in panel \textbf{(c)}.~The distance of the peaks from the origin, i.e., $\|\mathbf{X}_{i}\|_{\ell^{2}}$ is shown in panel \textbf{(d)} as a function
of $G$.~Panel \textbf{(e)} showcases the dependence of all positive real eigenvalues on $G$ with solid black lines where the theoretical predictions are included with red markers (see, the legend therein).~Finally, the imaginary eigenvalue close to $\mathcal{O}(\sqrt{G})$ is presented as a function of $G$ together with the asymptotic prediction (red circles)
in panel \textbf{(f)}.~The computations herein were performed on $\Omega=(-100,100)^{2}$ with $\frac{\partial v}{\partial {\boldsymbol{n}}}\Big|_{\partial \Omega}=0$ using $P^{4}$ elements and $\approx 1\textrm{M}$ degrees of freedom.
}
\end{figure}

In this appendix, we briefly draw a comparison between
the theoretical findings of Sec.~\ref{asymptotics} with numerical
computations performed in the finite element software \texttt{FreeFem\,$++$}~\cite{freefem}.~The backbone of the developed computational framework employs overlapping Schwarz domain decomposition methods (DDMs)~\cite{ddm_book} and leverages the interfacing of \texttt{FreeFem\,$++$}
with the scalable PETSc and SLEPc high-performance computing libraries~\cite{petsc} (see, also~\cite{SADAKA2025109378} for a recent yet concrete application in BECs).~Robust convergence of Newton's method to self-similar states is restored by penalizing the invariances of the PDE through the introduction of Lagrange multipliers and suitable phase conditions, see~\cite{uecker_book} (and references therein) for the relevant theoretical details.~Within this framework, the pseudo-arclength continuation method~\cite{kuznetsov_book} was developed in \texttt{FreeFem\,$++$}, and employed in order to trace branches of self-similar solutions (see, Fig.~\ref{ff_3_peaks_equilateral}(a) below).~Simultaneously, the underlying augmented linear systems (including the phase condition for computing the blowup rate $G$) in Newton's method were solved by using the bordering algorithm/Schur complement method~\cite{doi:10.1137/0904039}.~For these computations, $P^{4}$ finite elements were employed and solutions were sought in $V^{2}=V\times V\ni\left(v_{r},v_{i}\right)$ where $V=H^{1}(\Omega)$.

In addition to the penalization of symmetries, another key strength of this computational framework is the use of adaptive mesh refinement (AMR) as implemented in \texttt{FreeFem\,$++$}, where we corroborated our findings using the \texttt{mmg} tools~\cite{Dapogny2014} that are already interfaced therein (see, Fig.~\ref{ff_3_peaks_equilateral}(b), for a concrete example of an adapted mesh).~This allowed us to pose the problem on larger (yet finite) computational domain $\Omega$ while keeping the number of degrees of freedom small, hence providing not only a \textit{reliable}, but also \textit{an efficient computational approach} for obtaining these 2D self-similar branches and studying their spectral characteristics.~We also note in passing that the eigenvalue computations that were performed in SLEPc were further corroborated by using the FEAST~\cite{Polizzi2009} eigenvalue solver (see, Fig.~\ref{ff_3_peaks_equilateral}(d)-(f)).~Indicatively, Fig.~\ref{ff_3_peaks_equilateral} presents numerical results on the self-similarly blowing-up three-peak branch (in an equilateral triangle arrangement) by using this computational framework.~The initial guess fed to the solver was prepared by utilizing the theoretical prediction of Eq.~\eqref{eq:peakasymptotics}.~An excellent agreement between asymptotic analysis and numerical results is clearly evident.~We have computed many other states discussed in this present work using this framework, and its relevant details will be reported elsewhere.

\end{document}